\documentclass[preprint,12pt,3p]{elsarticle}

\usepackage{amssymb}
\usepackage{subcaption}
\usepackage{graphicx}
\usepackage[export]{adjustbox}
\usepackage{placeins}
\usepackage{array}
\usepackage{tabu}
\usepackage{xcolor}

\journal{XXX}

\begin{document}

\begin{frontmatter}

\title{Fluid forces and vortex patterns of an oscillating cylinder pair in still water with both side-by-side and tandem configurations}

\author[lable1]{Ang Li}
\author[lable2]{Shengmin Shi}
\author[label3]{Dixia Fan\corref{cor1}}
\address[lable1]{School of Naval Architecture, Ocean and Civil Engineering, Shanghai Jiao Tong University, Shanghai 200240, China}
\address[lable2]{SAIC Innovation Center, San Jose, California, 95134, USA}
\address[label3]{Department of Mechanical and Materials Engineering, Queen's University, Kingston, Ontario, K7L 3N6, Canada}
\ead{dixia.fan@queensu.ca}
\cortext[cor1]{Corresponding author}

\begin{abstract}
Models of cylinders in the oscillatory flow can be found virtually everywhere in the marine industry, such as pump towers experiencing sloshing load in a LNG ship liquid tank. However, compared to the problem of a cylinder in the uniform flow, a cylinder in the oscillatory flow is less studied, let alone multiple cylinders. Therefore, we experimentally and numerically studied two identical circular cylinders oscillating in the still water with either a side-by-side or a tandem configuration for a wide range of Keulegan-Carpenter number and Stokes number $\beta$. The experiment result shows that the hydrodynamic performance of an oscillating cylinder pair in the still water is greatly altered due to the interference between the multiple structures with different configurations. In specific, compared to the single-cylinder case, the drag coefficient is greatly enhanced when two cylinders are placed side-by-side at a small gap ratio, while dual cylinders in a tandem configuration obtain a smaller drag coefficient and oscillating lift coefficient. In order to reveal the detailed flow physics that results in significant fluid forces alternations, the detailed flow visualization is provided by the numerical simulation: the small gap between two cylinders in a side-by-side configuration will result in a strong gap jet that enhances the energy dissipation and increase the drag, while due to the flow blocking effect for two cylinders in a tandem configuration, the drag coefficient decreases.

\end{abstract}


\end{frontmatter}


\section{Introduction}
\label{sec1}

Compared to the fluid-structure interaction (FSI) problem of cylinders open to the uniform flow \cite{wang2020review} that has been widely investigated, cylinders relative to oscillatory flow has attracted less attention. However, its significance cannot be undermined for its rich physics as well as its existence in all kinds of engineering projects, especially in the ocean engineering field. Fan \cite{fan2017drag} listed such examples that its role could be found in the offshore field such as the wave-induced oscillatory flow around the risers, mooring lines and point wave energy generators, underwater structures, such as blow-out preventers (BOP) \cite{fan2016hydrodynamic} forced oscillatory motion close to the seabed forced to vibrate under the influence of the upper riser vortex-induced vibration (VIV) \cite{xu2013experimental,fan2019mapping,wang2021large} as well as pump towers in the liquefied natural gas (LNG) ship experiencing sloshing load induced by ship rolling. For all the examples mentioned above, the hydrodynamic model of the problems can be simplified as cylinders having an oscillatory motion with respect to the surrounding fluid.

Wang \cite{wang1968high} in 1968, using the methods of inner and outer expansions, first theoretically studied this problem of a fixed circular cylinder in oscillatory flow to $O(\beta^{-\frac{3}{2}})$. Based on this theory, Bearman et al. \cite{bearman1985forces} experimentally measured and reported the in-line force (along with the oscillation direction) and its decomposed hydrodynamic coefficients on a circular cylinder for different $KC$ up to 10 and $\beta$ from 196 to 1665. The results showed an accurate prediction of the cylinder forces for $KC$ below 3 and for moderately high $\beta$. Further experiments performed by Sarpakaya \cite{sarpkaya1986force} confirmed that the drag and added mass coefficient $C_d$ and $C_m$ are in tune with the theory under the critical $KC$ number at which flow transited into unstable and depends on different $\beta$ for the smooth cylinders. Apart from the in-line forces, Williamson \cite{williamson1985sinusoidal} studied lift forces (perpendicular to the oscillation direction) by oscillating a circular cylinder in still fluid. Together with the flow visualization, the results reveal that the appearance of the different harmonics of the oscillation frequency in the lift force depended on $KC$, as flow pattern changed with increasing $KC$ from pair attached vortices around cylinder ($0<KC<7$) to single-pair shedding vortices ($7<KC<15$), double-pair shedding vortices ($15<KC<24$), three-pair shedding vortices ($24<KC<32$) and four-pair shedding vortices ($32<KC<40$). A more comprehensive flow visualization study was carried out by Tatsuno et al. \cite{tatsuno1990visual} for a broad range of $KC$ from 1.6 to 15 and various $\beta$ between 5 and 160, and a comprehensive flow regimes map was established where eight different flow regimes were identified (A$^*$, A, B, C, D, E, F, and G).

Several numerical studies have also been performed on this problem. Lin et al. \cite{lin1996numerical} first performed simulation on oscillatory flow around a circular cylinder at $\beta=76$, using 2D discrete vortex method. In the results, the flow pattern was reported to be well replicated by numerical methods compared to the experimental results. D{\"u}tsch et al.\cite{dutsch1998low} numerically investigated oscillation of a circular cylinder in still fluid for a fixed $\beta=35$ and a large range of $KC$ up to 20, in-line force, as well as mean velocity distribution predicted by the simulation, showed a good agreement with their experiments \cite{dutsch1998low}. Zhao et al. \cite{zhao2014two} using Petrov-Galerkin finite-element method (PG-FEM) simulated a large range of $KC$ and $\beta$ number and successfully reenacted the 8 different flow pattern observed by Tatsuno et al \cite{tatsuno1990visual}. At the same time, though a majority of the simulation was two-dimensional, some of which has also been performed to understand the three-dimensional character of the flow around cylinder. Nehari et al. \cite{nehari2004three} and An et al. \cite{an2015two} performed both two and three-dimensional numerical study for comparison and results revealed that the major physics, such as the flow regime and in-line forces, were not affected by the three-dimensional flow characters as the two-dimensional simulation was able to capture the key phenomena.

Compared to the study of the single cylinder in uniform/oscillatory flow or the single oscillating cylinder in still fluid, the case of multiple cylinders, or even just two cylinders \cite{lin2020dynamic,tan2018numerical}, was relatively rare and far more complicated, as an additional cylinder can cause strong hydrodynamic interaction among cylinders, similar to the case in the flow \cite{wu2017kill,fan2019vortex,lin2020dynamic}. Williamson \cite{williamson1985sinusoidal} experimentally studied two identical cylinders in side-by-side configuration oscillating in the still fluid. Results showed that two cylinders would behave as a single cylinder when gap ratio (Distance between two cylinders divided by cylinder diameter, $G/d$) was larger than 1.5 and the vortex shedding from the two cylinders achieved synchronization either in the phase of anti-phase based on $KC$ and $G/d$. In Williamson's further study \cite{williamson1985fluid} on two cylinders of different diameters in a side-by-side configuration, it revealed that there were mean attraction and repulsion lift force between two cylinders, depending on $KC$. Chern et al. \cite{chern2010cfd} performed two-dimensional numerical simulations and focused on the effect of the gap ratio on the flow pattern for identical side-by-side cylinders. They found a small gap had a significant effect on the vortex shedding pattern from the cylinders that it kept asymmetrical vortex shedding pattern for the two cylinders when $KC$ was smaller than 10. Zhao et al. \cite{zhao2014two} undertook 2D simulation on both side-by-side and tandem cylinder arrangement, they found, compared to the single cylinder case, there was new flow regime when an additional cylinder was introduced (such as new GVS flow regime of gap vortex shedding from the cylinders in side-by-side configuration). Furthermore, Tong et al. \cite{tong2015oscillatory} carried out simulations about four cylinders of square arrangement in oscillatory flow, and quite different flow characters were found when compared to those in the single and two cylinders.

In the current study, both experiments and numerical simulations have been carried out to investigate the hydrodynamics of the identical dual cylinders in both tandem and side-by-side configurations oscillating in still water. Hydrodynamic coefficients of added mass coefficient $C_m$ and drag coefficient $C_d$ are reported experimentally for both single cylinder and dual cylinders covering a large number of $KC$ up to 20 and several $\beta$ from 350 to 3100. 2D numerical results, though at a smaller $\beta=20$, capture the major phenomena observed in the experiments, such as the drag enhancement of side-by-side dual cylinders and provides a critical insight into the flow pattern around single and dual cylinders of different gap ratios and arrangements, and this helps to explain the hydrodynamic force difference between the single cylinders and dual ones. It is worthwhile to point out that compared to the previous researches\cite{zhao2014two}, the current paper not only provides the experimental result on force over a large range of $\beta$ number but also emphasizes on the link between the flow pattern change and the fluid force variation due to the existence of the dual cylinder hydrodynamic interaction.

\section{Experimental Description and Numerical Method}
\label{sec2}
\subsection{Experimental Setup and Data Processing}
\label{sec2_1}
\subsubsection{Experimental Setup}
\label{sec2_1_1}

The experimental facility \cite{fan2019robotic} allows a prescribed sinusoidal oscillation of the cylinder model in the still water with different amplitudes and frequencies, hence covering a large range of $KC$ from 1 to 20 and $\beta$ from 350 to 3100, using cylinder models of three different diameters. Five different gap ratios are chosen as 0.5, 1.0, 2.0, 3.0 and $\infty$ (the single cylinder case) respectively, and meanwhile two configurations of tandem ($\theta = 0^o$) \cite{zhang2017numerical} and side-by-side ($\theta = 90^o$) \cite{fan2017drag} are tested in the current dual cylinder experiment. Detailed parameters are listed in Table \ref{tab:expcond}. The force data were recorded for two cylinders by two ATI Gamma sensors independently while the prescribed motion information was collected simultaneously. This allows the calculation of the added mass coefficient $C_m$ (force component in the phase of acceleration) and the drag coefficient $C_d$ (the force component in the phase of velocity).

\begin{table}[htbp]

   \centering
   
   \begin{tabular}{ccc}
     \hline
     \textbf{Model Parameters} & \textbf{} \\
     \hline
     Diameter $d$ & 2.54cm, 3.81cm, 5.08cm \\
     Test Length $L$ (Immersed) & 58.42cm\\
     \hline
     \textbf{Experiment Parameters} & \textbf{} \\
     \hline
     KC Number ($2\pi \frac{A}{D}$) & 1.0 to 20.0\\
     Stokes  Number $\beta$ ($Re$/$KC$) & 350, 700, 1190, 1580, 2120, 3100 \\
     Gap ratio ($G/d$) & 0.5, 1.0, 2.0, 3.0, $\infty$\\
     Configuration & Side-by-side, Tandem \\
     \hline
    \end{tabular}
    \caption{Model and experiment parameters}
    \label{tab:expcond}
\end{table}

\subsubsection{Hydrodynamic Coefficient}
\label{sec2_1_2}

The inline force (along with the direction of the motion) on the oscillating cylinders can be modeled, using Morison equation \cite{morison1950force} as follows, 

\begin{equation}
    F_x = \frac{1}{2}\rho SC_{d}\left|\dot{X}\right|\dot{X} + \rho \nabla C_{m}\ddot{X},
\end{equation}
in which $\rho$ is the fluid density, $S$ is the projected area of the model along with the motion, $\nabla$ is the model displacement and $X$ is the motion of the model. In the current experiment, the motion is prescribed as a sinusoidal motion of $X = Asin(\omega t)$. Hence the added mass coefficient $C_{m}$ and the drag coefficient $C_{d}$ can be obtained with the measured in-line force $F_x(t)$, as the following equations,
\begin{equation}
    C_d = \frac{\frac{1}{T}\int^{T}_{0}\hat{F}_{x}(t)\dot{X}(t)t}{\frac{2}{3\pi}\rho S(A\omega)^3},
\end{equation}

\begin{equation}
    C_m = \frac{\frac{1}{T}\int^{T}_{0}\hat{F}_{x}(t)\ddot{X}(t)dt}{\frac{1}{2}\rho \nabla A^2\omega^4},
\end{equation}
where $\hat{F}_x(t)$ is $F_x(t)$ after applying non-casual band-filter (non-causality avoids to introduce artificial phase shift) to obtain the force component with the same frequency as the imposed oscillation motion.

The cross flow force (perpendicular to the direction of the motion) is referred as the lift force. In the current experiment, the mean and root mean square of the lift coefficients are calculated in the following equations,
\begin{equation}
    \overline{C_l} = \frac{\frac{1}{T}\int^{T}_{0}F_y(t)dt}{\frac{1}{2}\rho S(A\omega)^2},
\end{equation}

\begin{equation}
    C^{rms}_l = \frac{\sqrt{\frac{1}{T}\int^{T}_{0}(F_y(t)-\overline{F_y(t)})^2dt}}{\frac{1}{2}\rho S(A\omega)^2},
\end{equation}
in which $F_y(t)$ is the measured cross flow force. 

\subsection{Numerical Method}
A Boundary Data Immersed Method (BDIM) based solver is used in this paper for the numerical simulation \cite{weymouth2011boundary,fan2020reinforcementArXiv,fan2020reinforcement,fan2020deep,wang2020active} to solve the problems of immersion of two oscillating cylinders in the fluid. BDIM is based on a general integration kernel formulation which combines the field equations of each domain and the inter-facial conditions analytically. The resulting governing equation for the complete domain preserves the behavior of the original system in an efficient Cartesian-grid method, including stable and accurate pressure values on the solid boundary.

The numerical simulation setup is quite similar to the experimental arrangement, except that the simulation is in 2-dimension. The calculation domain is chosen to be $40D \times 40D$ to erase the boundary and block effect, which is quite influential in this study. Unlike other traditional CFD methods, BDIM based solver does not need a complicated meshing technique for complex body shapes as well as significant structural motions. The resolution of the current simulation is based on the mesh number per cylinder diameter. Furthermore, a mesh dependent study will be presented in the later sections of a single cylinder case, together with a comparison between the current and existing studies. 

What is worth mentioning is that $\beta$ in the current experiment and numerical simulation is chosen to be different. beta number is fixed at $\beta = 20$ in the simulation for different $KC$ numbers (hence numerical simulations have a smaller $Re = KC\times \beta$ than experiments) to guarantee a 2D flow. However, despite $\beta$ difference, numerical simulation in the current research is still able to capture the major physical phenomena observed in the experiment. Hence, a flow visualization provided by the simulation is still a valuable source for understanding and explaining the experimental results.

\section{Single Cylinder Oscillating in still Water}
The single cylinder in oscillatory flow or the single oscillating cylinder in the still fluid has been studied before both experimentally and numerically. In this section, the current experimental and numerical results of a single cylinder oscillating in still water will be reported.
   
\label{sec3}
\subsection{Experimental Results: Hydrodynamic Coefficients}
\label{sec3_1}

\begin{figure}[!ht]
\centering
\begin{subfigure}{0.4\columnwidth}
\includegraphics[width=\columnwidth,keepaspectratio]{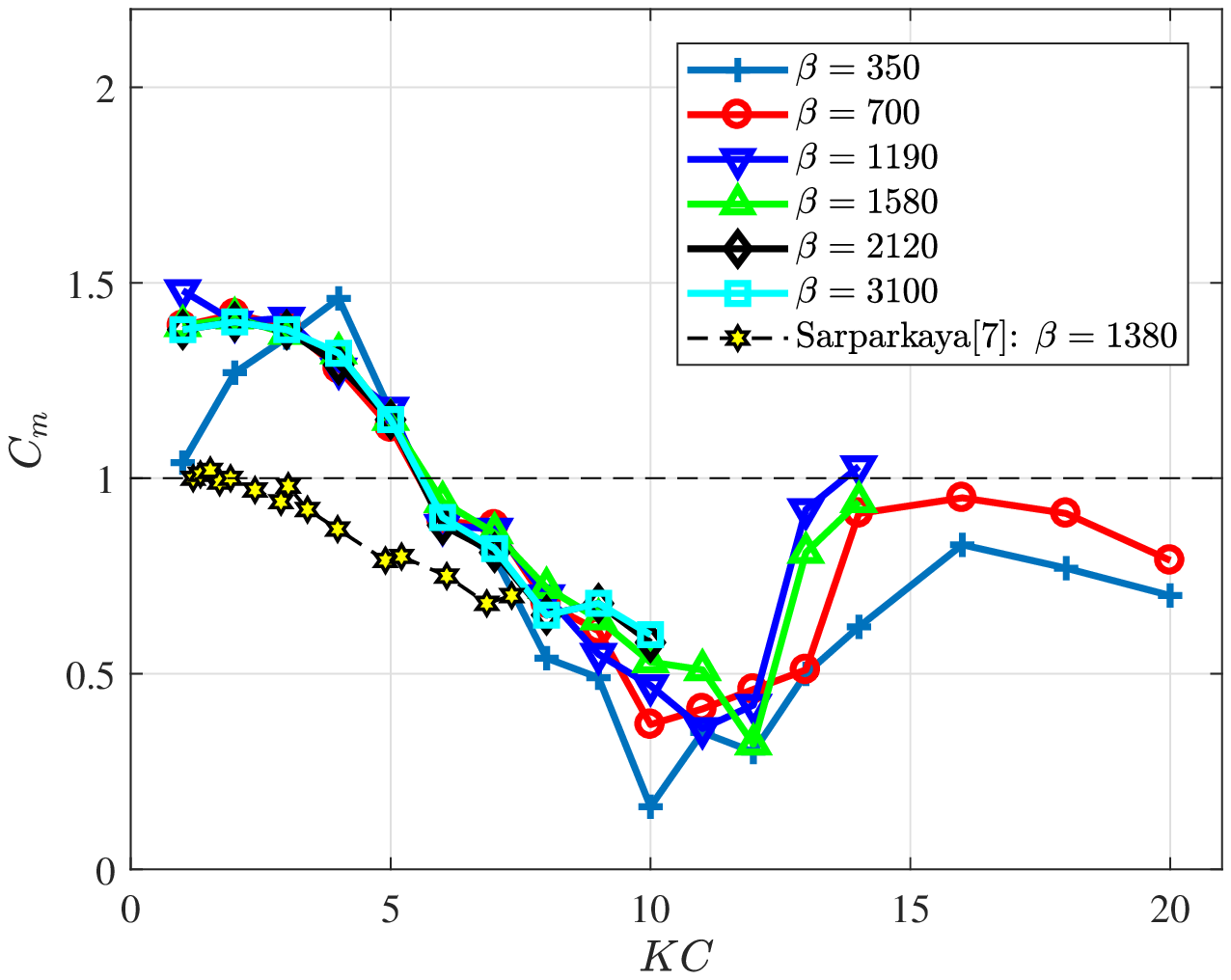}
\end{subfigure}
\begin{subfigure}{0.4\columnwidth}
\includegraphics[width=\columnwidth,keepaspectratio]{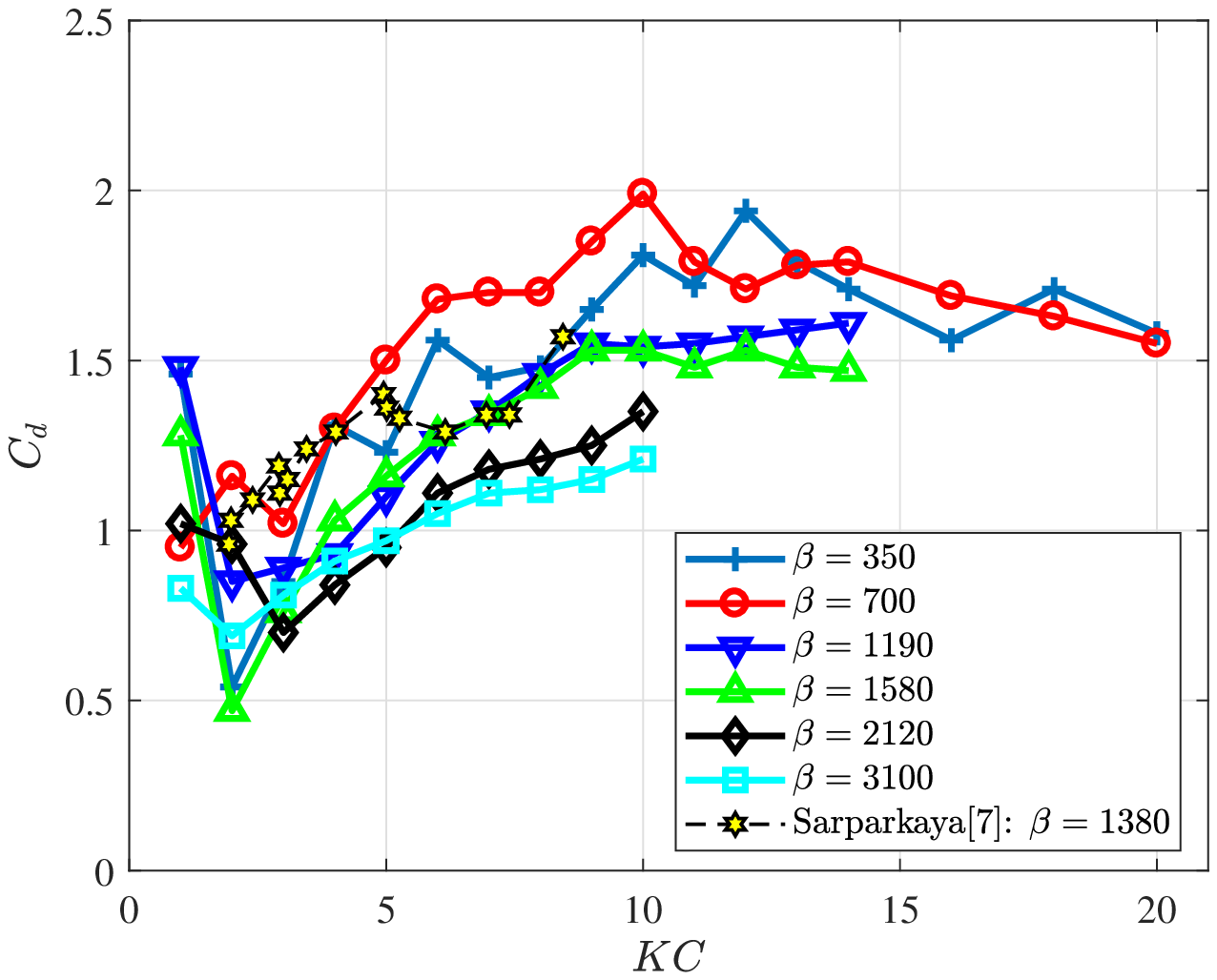}
\end{subfigure}
\caption{Single cylinder: trend of $C_m$ (left) and $C_d$ (right) with respect to $KC$. Marker \textcolor{yellow}{$\bigstar$} shows the result from Sarparkaya \cite{sarpkaya1986force}. It is noteworthy that in Sarparkaya's experiment, due to the experimental setup of a stationary cylinder in the oscillatory fluid, the added mass coefficient $C_m$ reported in \cite{sarpkaya1986force} includes Froude–Krylov force term, and hence is equal to $1+C_m$ reported in the current experiment.}
\label{fig:CSingle}
\end{figure}

\begin{figure}[!ht]
\centering
\begin{subfigure}{0.4\columnwidth}
\includegraphics[width=\columnwidth,keepaspectratio]{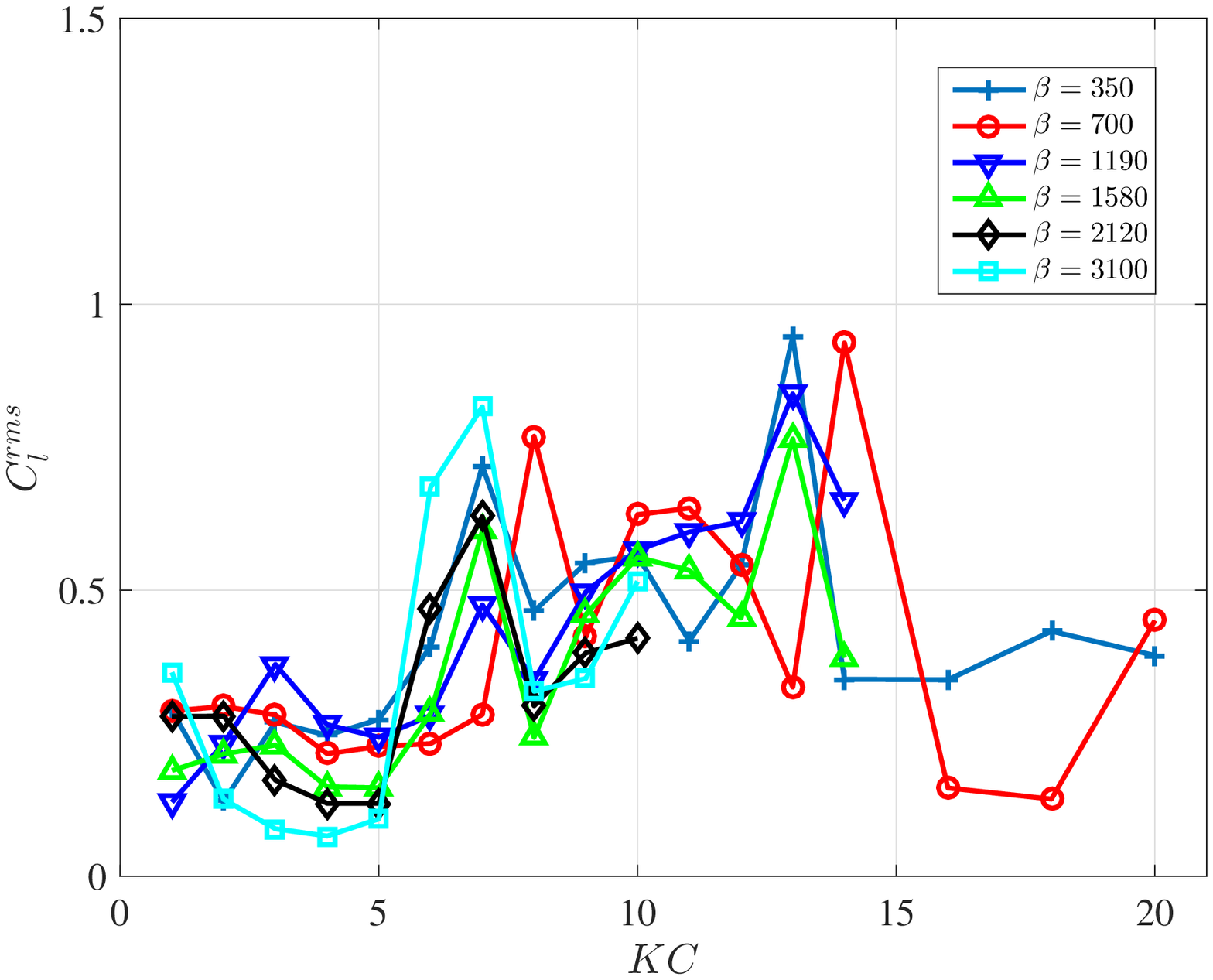}
\end{subfigure}
\begin{subfigure}{0.4\columnwidth}
\includegraphics[width=\columnwidth,keepaspectratio]{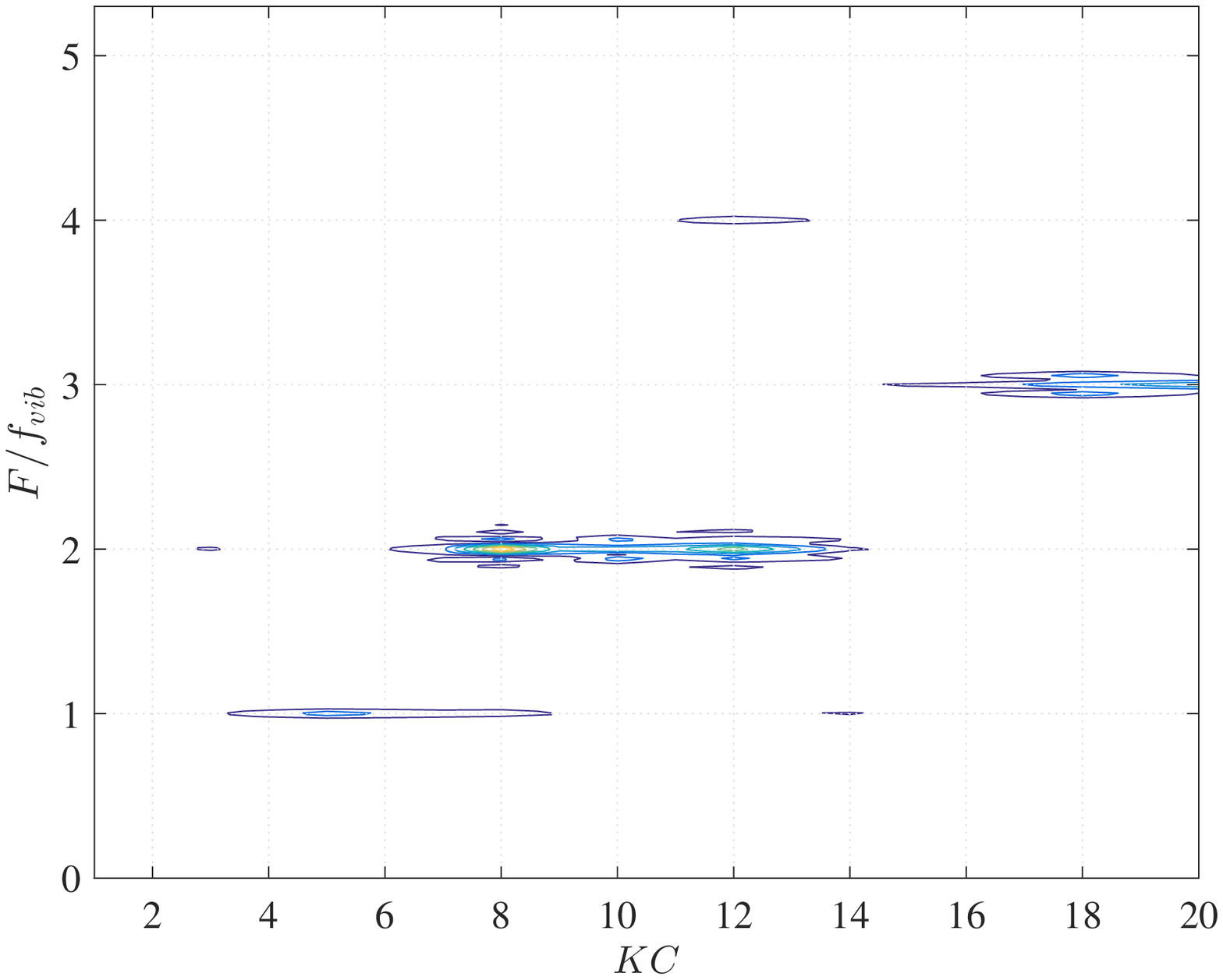}
\end{subfigure}
\caption{Single cylinder: trend of $C^{rms}_{l}$ along with $KC$ (left); PSD  of $C_l$ along with $KC$ for $\beta = 350$ (right)}
\label{fig:ClSingle}
\end{figure}

The added mass coefficient $C_m$ of the single cylinder ($G/d = \infty$) is presented in Fig. \ref{fig:CSingle}(left) with the increase of $KC$ for various $\beta$. We observe that $C_m$ decreases with an increasing $KC$ and reaches a minimum of around $KC = 11$ and then increases again to a constant value of $C_m = 1.0$. Besides, it is found that in the current $\beta$ ranging from 350 to 3100, $C_m$ is weakly related to $\beta$ variation. Meanwhile, the drag coefficient of $C_d$ is plotted in Fig. \ref{fig:CSingle}(right). It is found that $C_d$ depends on both $KC$ and $\beta$ that $C_d$ generally will jump quickly to a small value for $KC$ from 1.0 to 2.0 and then increase to a steady value with increasing $KC$. Furthermore, it is observed that with the increase of $\beta$, $C_d$ for different $KC$ will decrease accordingly. Such similar trends were reported by Sarpkaya \cite{sarpkaya1986force}. The hydrodynamic coefficients $C_m$ and $C_d$ reported in \cite{sarpkaya1986force} at $\beta = 1380$ is plotted in Fig. \ref{fig:CSingle} to compare with the those obtained in the current experiment. It  is  noteworthy  that  due to the different experimental setup between the current research (cylinder oscillating in the still water) and \cite{sarpkaya1986force} (a stationary cylinder in the oscillatory fluid), the added mass coefficient $C_m$ reported in \cite{sarpkaya1986force}  includes Froude–Krylov force term, and hence is equal to $1 + C_m$ reported in the current experiment. It is found that compared to the case of $\beta = 1580$, the two sets of experimental results show the similar trend of decreasing $C_m$ and increasing $C_d$ in the current $KC$ range. However, difference in value between the current result and \cite{sarpkaya1986force} can be observed at low $KC$ number, which may be attributed to the different experimental setup, including turbulence rate, free surface effect, model 3D effect and so forth.

The result of the lift coefficient RMS $C^{rms}_l$ is plotted in Fig. \ref{fig:ClSingle}(left) for different $\beta$ along with $KC$. It is observed that the lift coefficient RMS will have a sudden jump at $KC$ around seven and the second one at $KC$ around 14 for all $\beta$ cases in the current experiment. Moreover, the FFT results of $\beta = 350$ for all $KC$ are plotted in Fig. \ref{fig:ClSingle}(right), and it is observed the non-dimensional frequency of the lift coefficient $C_l$ ($F/f_{vib}$) changes with the increase of $KC$. The single cylinder will first have the same frequency in the lift direction as the in-line oscillation frequency when $KC$ is smaller than 6. Then a strong component of double oscillation frequency will dominate the lift response for $KC$ from 7 to 14 and is followed by the appearance of the third harmonics. These phenomena were also observed by Williamson \cite{williamson1985sinusoidal} and explained that the single oscillating cylinder in the still fluid would experience flow pattern change within a different KC range via flow visualization. When $KC <7$, vortices generated from the cylinder will attach to the moving cylinder when $7 < KC <15$, a pair of vortices will be shed away from each period of the cylinder oscillation, and this is followed by the double pair shedding vortices pattern when $KC >15$.

\subsection{Numerical Simulation: Flow Visualization}
\label{sec3_2}
Five types of meshes have been used in the current numerical study from coarse to dense, characterized by the number of nodes per cylinder diameter. Case of single cylinder of $KC=5$ and $\beta=20$ ($Re=100$) is chosen for the mesh dependence study, shown in Table \ref{tab:mesh}. The comparison of the force coefficients in Table \ref{tab:mesh} shows the convergence of the hydrodynamic coefficients from coarse to dense mesh, and mesh case 3, case 4 and case 5 achieve almost the same results and hence mesh of 100 nodes along the cylinder diameter is used in the current study. Compared with the existing numerical result, shown in Table \ref{tab:comp}, a good agreement can be observed for both $C_m$ and $C_d$.
\begin{table}[htbp]
   \centering
   \begin{tabular}{cccc}
     \hline
     \textbf{Mesh Case} & \textbf{$N_c$} & \textbf{$C_m$} & \textbf{$C_d$}\\
     \hline
     Mesh 1 & 50 & 1.28 & 2.25 \\
     Mesh 2 & 75 & 1.37 & 2.12 \\
     Mesh 3 & 100 & 1.41 & 2.02 \\
     Mesh 4 & 130 & 1.42 & 2.02 \\
     Mesh 5 & 200 & 1.40 & 2.04 \\
     \hline
    \end{tabular}
    \caption{Comparison of the hydrodynamic coefficients of single cylinder of $KC=5$ and $\beta=20$. $N_c$ is the node number along cylinder diameter}
    \label{tab:mesh}
\end{table}

\begin{table}[htbp]
   \centering
   \begin{tabular}{ccc}
     \hline
     \textbf{Source ($KC = 5, \beta= 20$)} & \textbf{$C_m$} & \textbf{$C_d$}\\
     \hline
     Current (Mesh 3) & 1.41 & 2.02 \\
     Zhao et al. (2D Simulation \cite{zhao2014two}) & 1.48 & 2.04\\
     Uzunoglu et al. (2D Simulation \cite{uzunouglu2001low}) & 1.45 & 2.10\\
     Nehari et al. (2D Simulation \cite{nehari2004three}) & 1.43 & 2.10\\
     Nehari et al. (3D Simulation \cite{nehari2004three}) & 1.47 & 2.04\\
     \hline
    \end{tabular}
    \caption{Comparison of the hydrodynamic coefficients between current and existing numerical results}
    \label{tab:comp}
\end{table}

\begin{figure}[!ht]
\centering
\begin{subfigure}{0.32\columnwidth}
\includegraphics[width=\columnwidth,keepaspectratio]{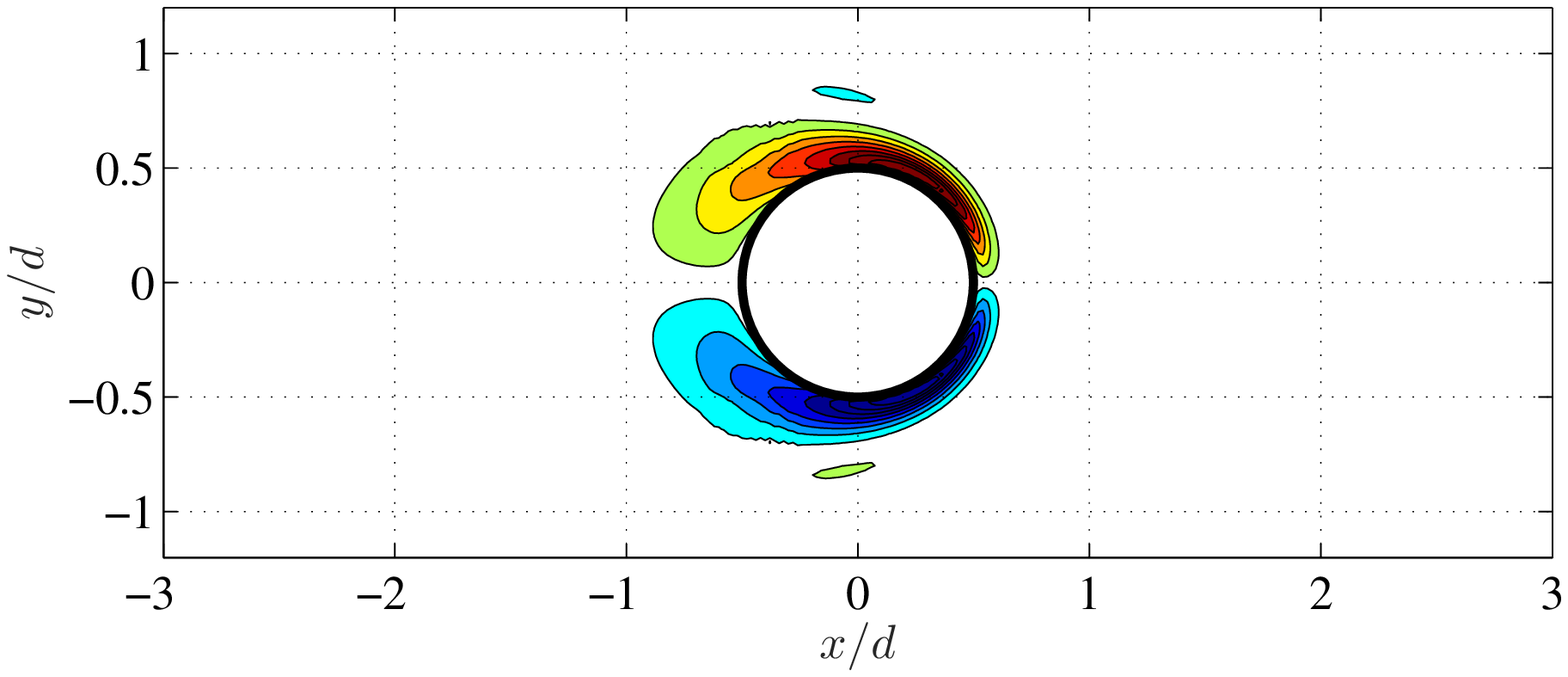}
\end{subfigure}
\begin{subfigure}{0.32\columnwidth}
\includegraphics[width=\columnwidth,keepaspectratio]{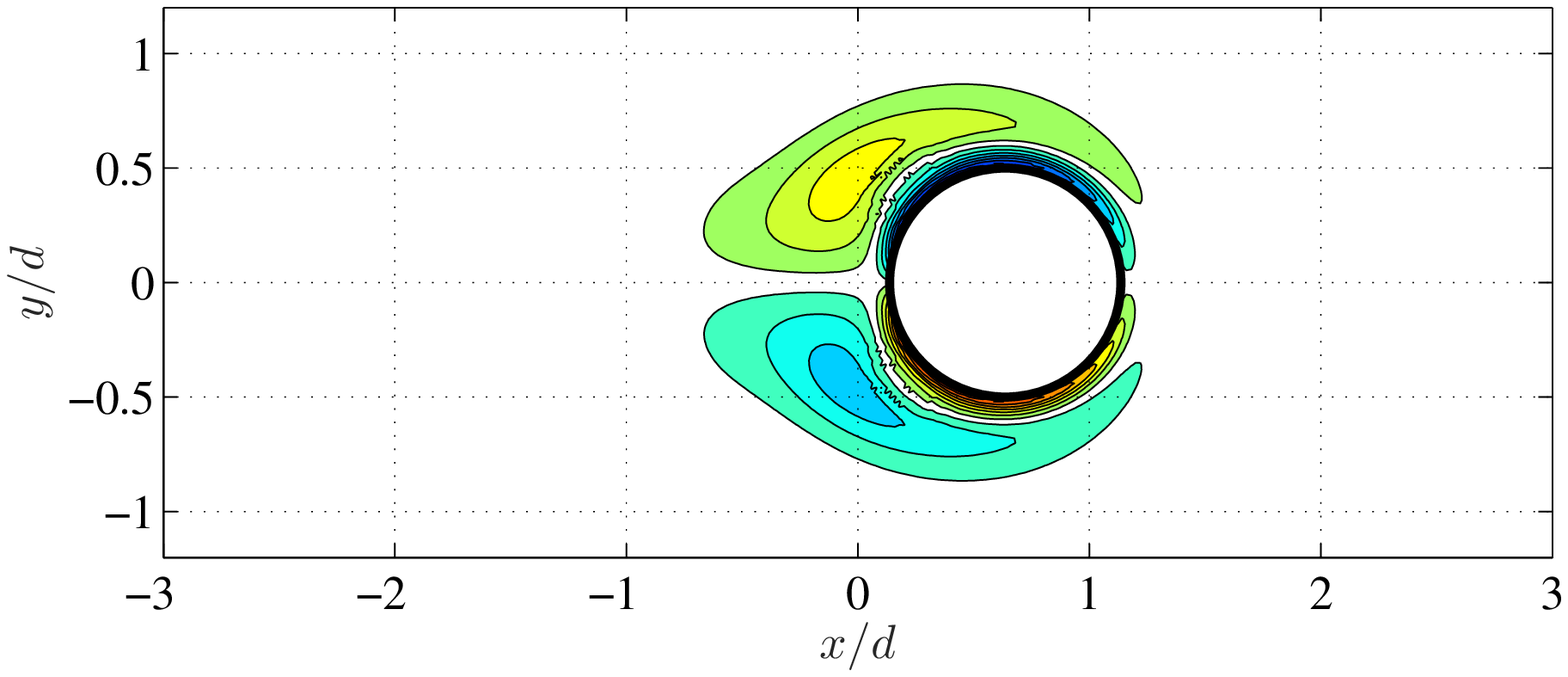}
\end{subfigure}
\begin{subfigure}{0.32\columnwidth}
\includegraphics[width=\columnwidth,keepaspectratio]{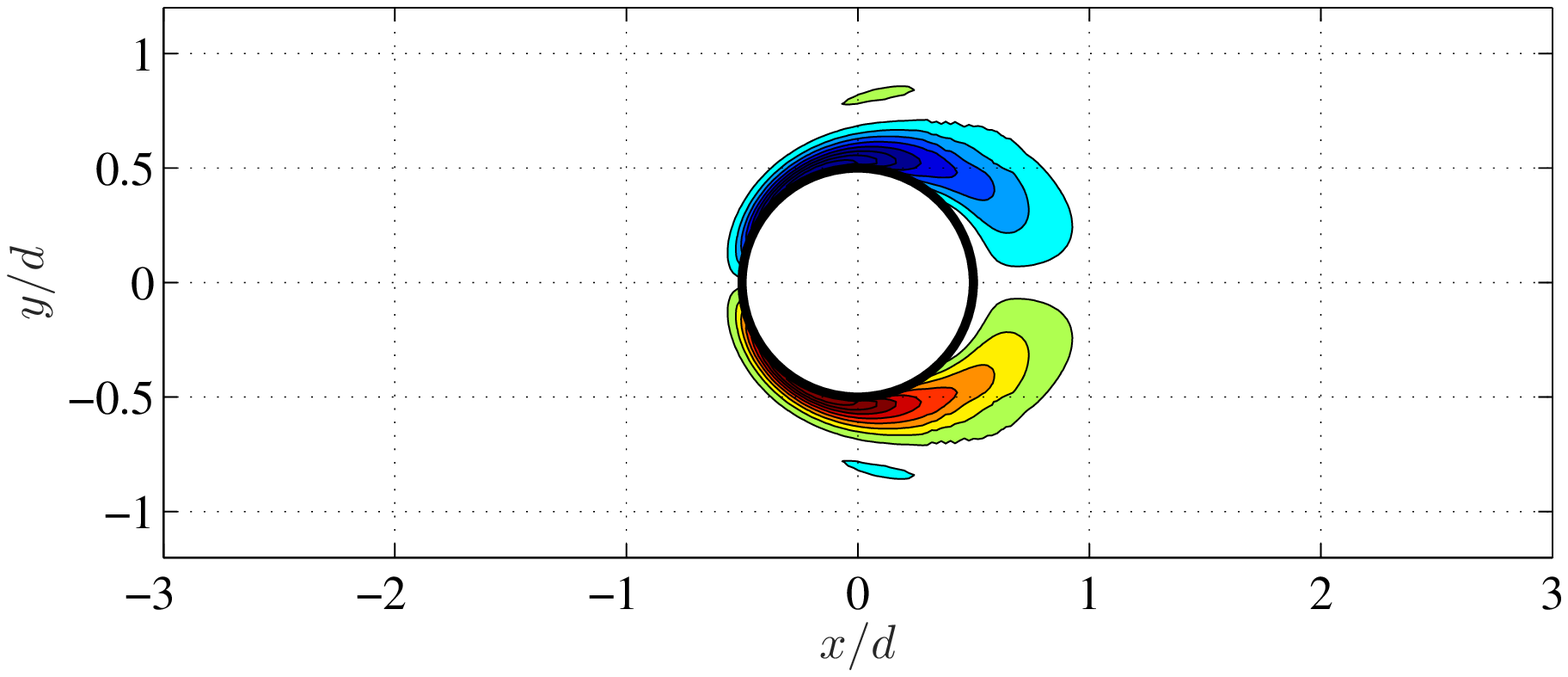}
\end{subfigure}
\begin{subfigure}{0.32\columnwidth}
\includegraphics[width=\columnwidth,keepaspectratio]{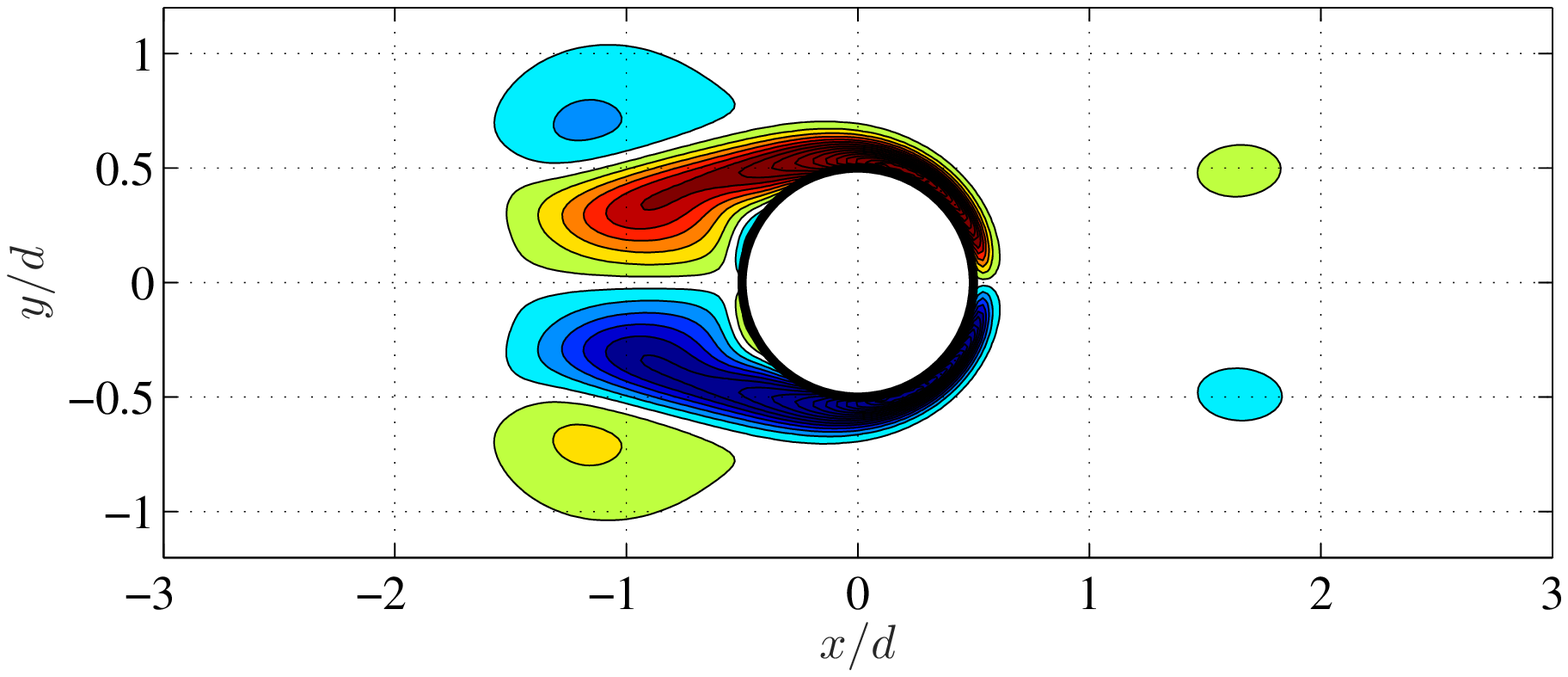}
\end{subfigure}
\begin{subfigure}{0.32\columnwidth}
\includegraphics[width=\columnwidth,keepaspectratio]{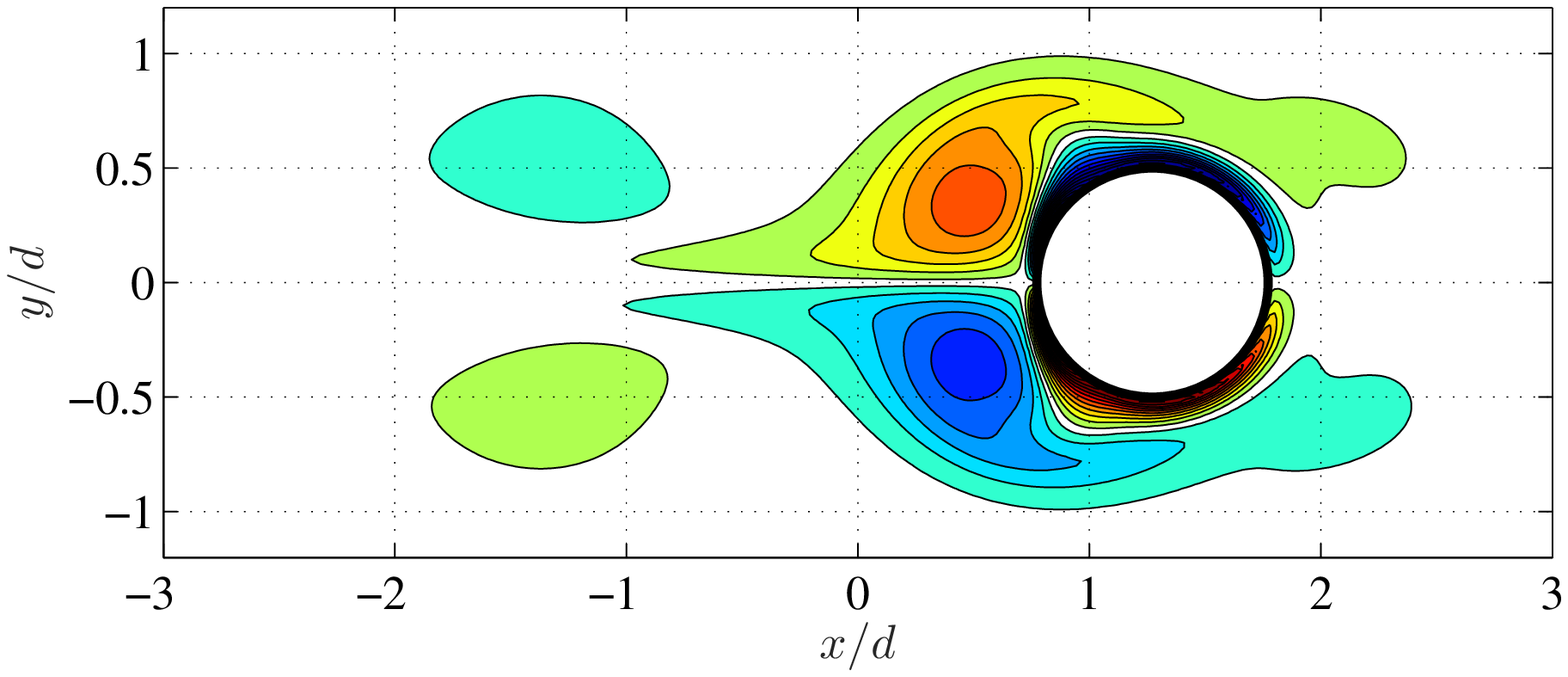}
\end{subfigure}
\begin{subfigure}{0.32\columnwidth}
\includegraphics[width=\columnwidth,keepaspectratio]{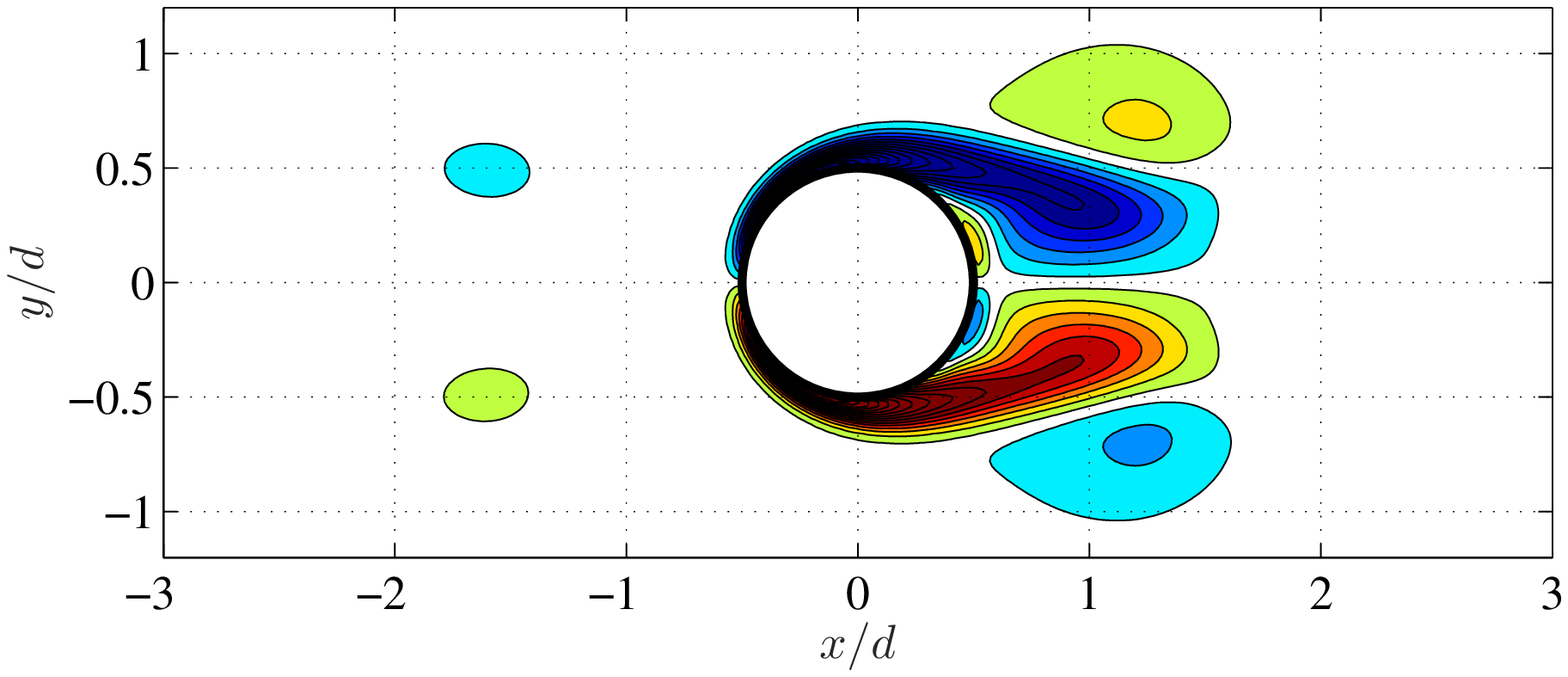}
\end{subfigure}
\begin{subfigure}{0.55\columnwidth}
\includegraphics[width=\columnwidth,keepaspectratio]{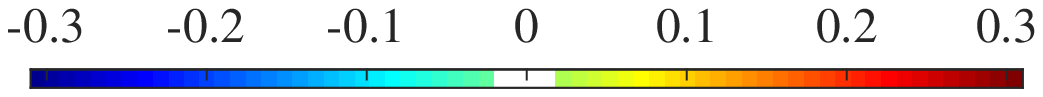}
\end{subfigure}
\caption{Snapshot of vorticity plot for single cylinder ($\beta = 20$) at $KC=4$ (first row, $C_m = 1.52$, $C_d = 2.11$) and at $KC=8$ (second row, $C_m = 1.08$, $C_d = 1.78$) at $t/\tau$ of 0 (left), $\frac{1}{4}$ (middle) and $\frac{1}{2}$ (right).}
\label{fig:vortSingle}
\end{figure}

Simulations at a constant $\beta=20$ were performed for $KC$ from 1 to 12, and the time snapshots of the vorticity around the single cylinder in a half oscillation periodfor $KC=4$ and $KC=8$ at $t/\tau$ of 0, $\frac{1}{4}$ and $\frac{1}{2}$ is displayed in Fig. \ref{fig:vortSingle} ($\tau$ is the oscillation period). By comparison, two flow regimes can be observed between $KC = 4$ and $KC = 8$. Similar to what was visualized by Williamson \cite{williamson1985sinusoidal}, when $KC <7$, the flow regime of attached vortex pairs around the cylinder is found, while it changes to a new regime of shedding a single pair of vortices in one period of oscillation when $7 < KC <15$. It is this change of flow pattern that gives rise to the dominating twice of the oscillating frequency in the lift direction.

\section{Dual Cylinder in Side-by-Side Configuration}
\label{sec4}
Experimental and numerical results of the oscillating side-by-side cylinders in still water will be presented in this section with a comparison with the single cylinder case.
\subsection{Experimental Results: Hydrodynamic Coefficients}

\label{sec4_1}
\begin{figure}[!ht]
\centering
\begin{subfigure}{0.4\columnwidth}
\includegraphics[width=\columnwidth,keepaspectratio]{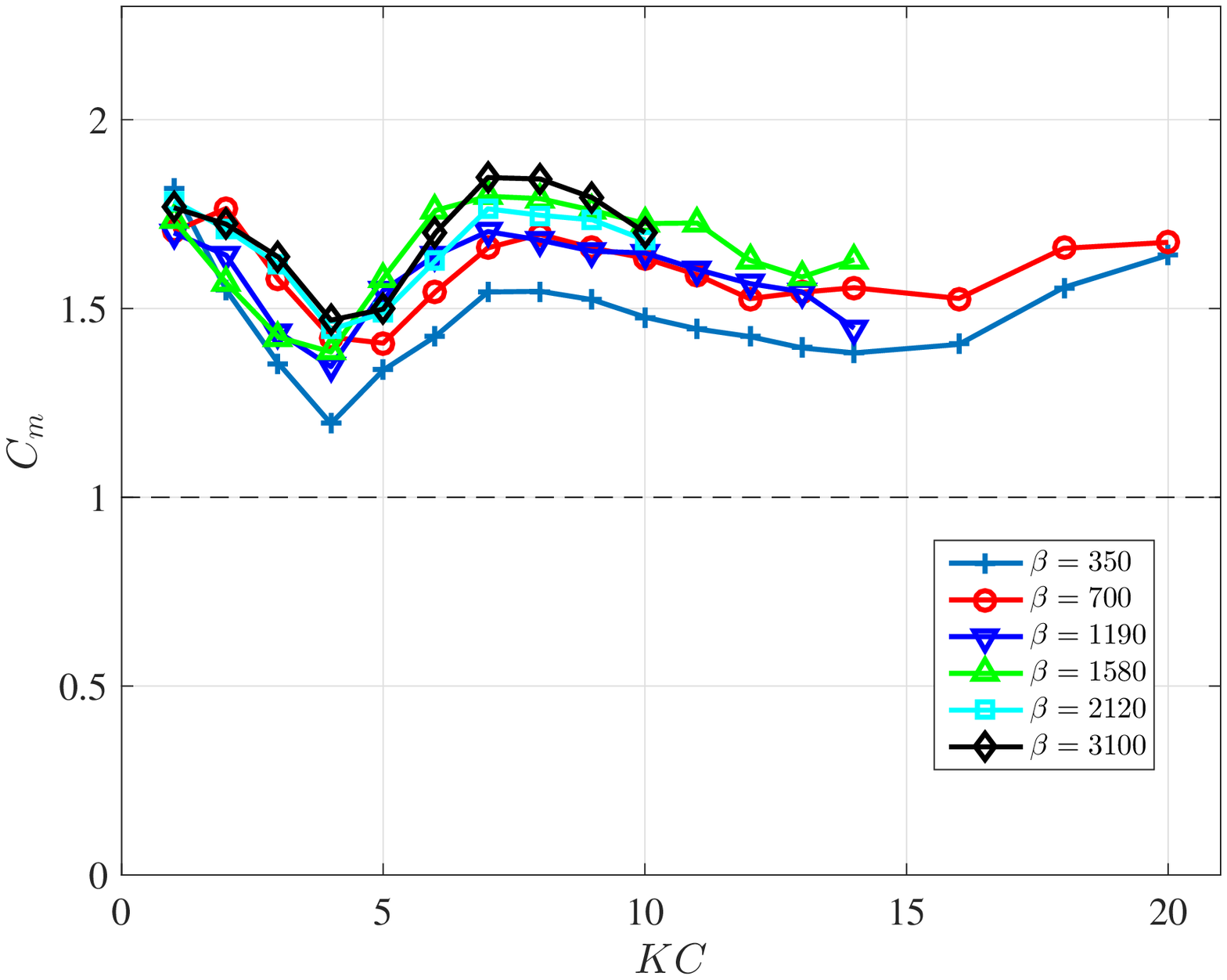}
\end{subfigure}
\begin{subfigure}{0.4\columnwidth}
\includegraphics[width=\columnwidth,keepaspectratio]{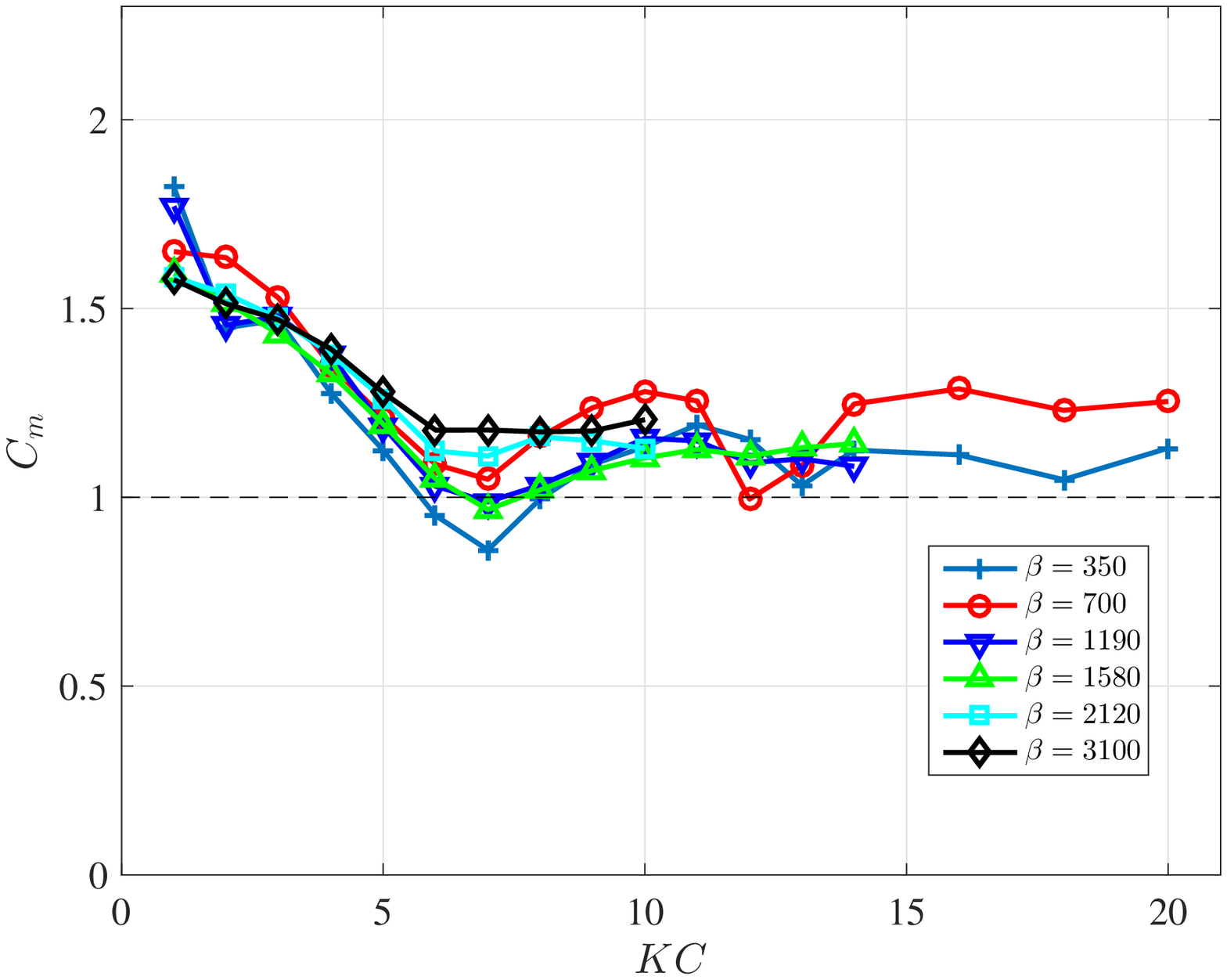}
\end{subfigure}
\begin{subfigure}{0.4\columnwidth}
\includegraphics[width=\columnwidth,keepaspectratio]{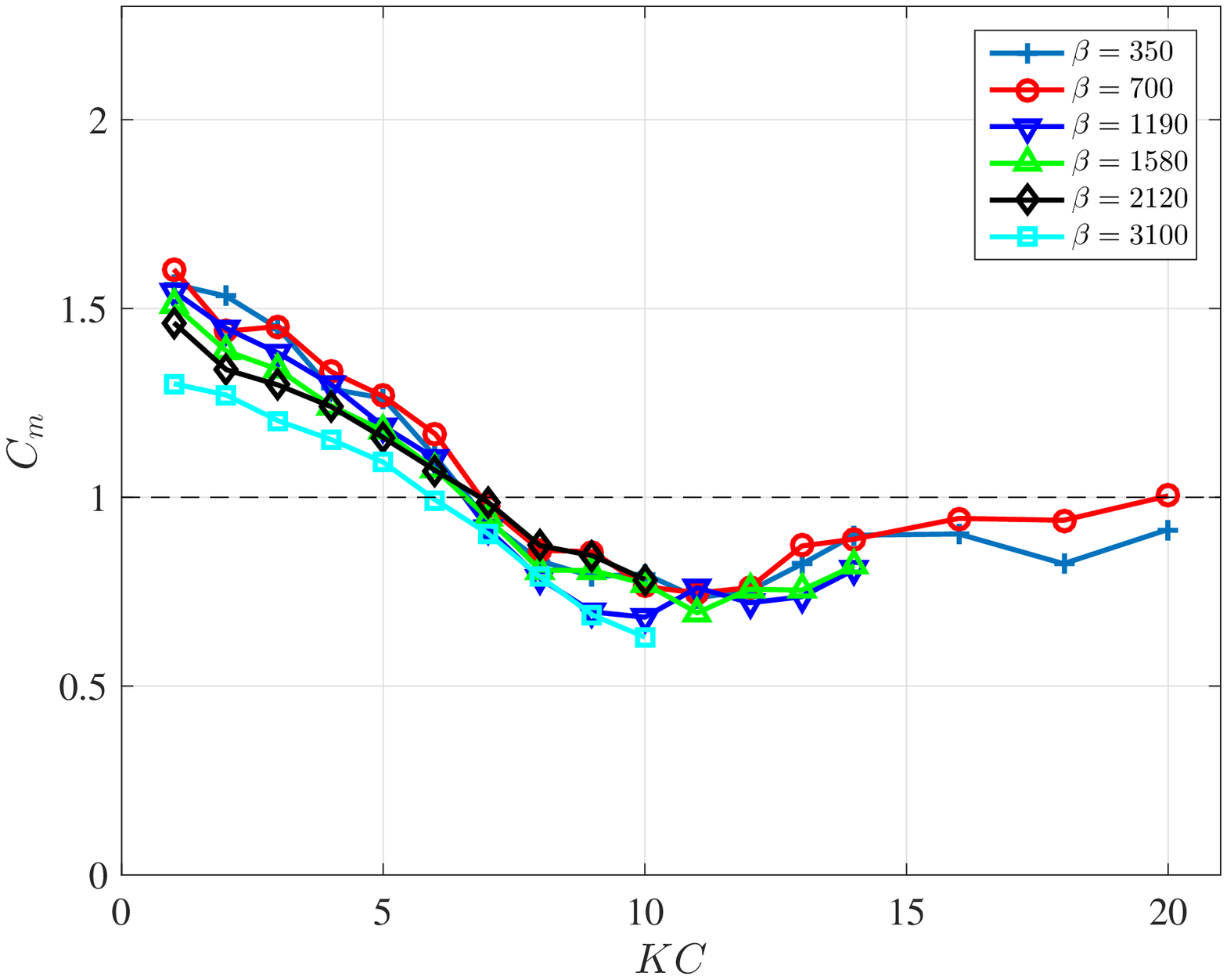}
\end{subfigure}
\begin{subfigure}{0.4\columnwidth}
\includegraphics[width=\columnwidth,keepaspectratio]{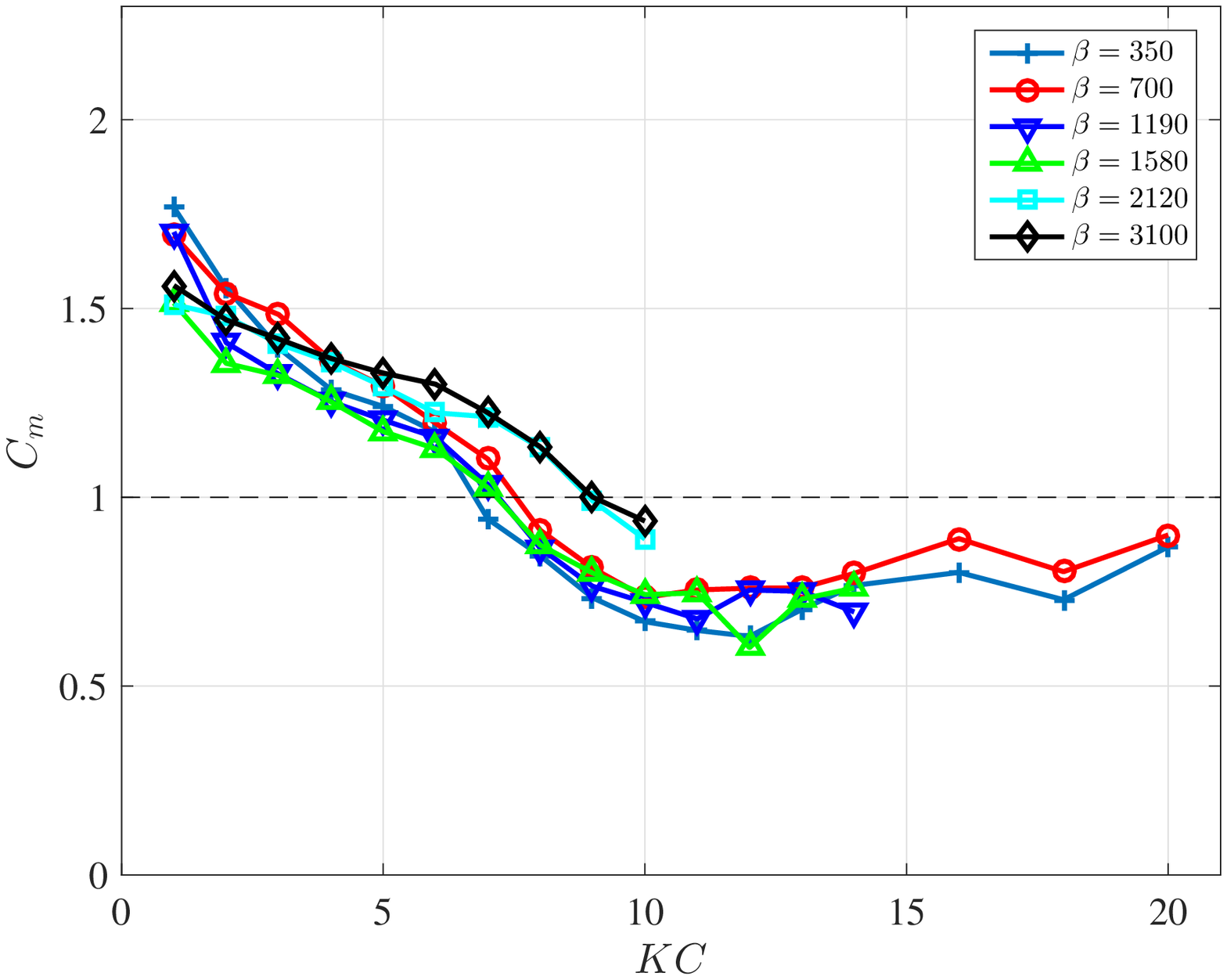}
\end{subfigure}
\caption{Cylinders in side-by-side configuration: trend of $C_m$ along with $KC$ for different $\beta$ at $G/d = 0.5$ (top left), $G/d = 1.0$ (top right), $G/d = 2.0$ (bottom left) and $G/d = 3.0$ (bottom right)}
\label{fig:CmSbS}
\end{figure}

\begin{figure}[!ht]
\centering
\includegraphics[width=0.4\columnwidth,keepaspectratio]{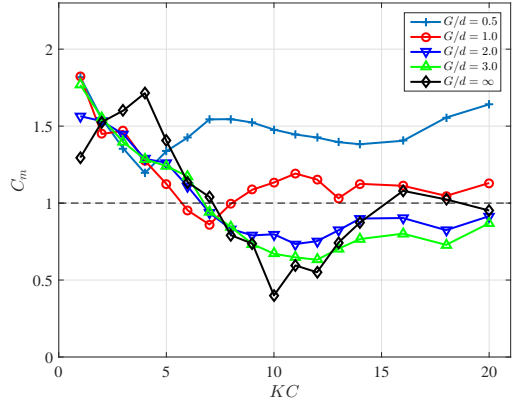}
\caption{Cylinders in side-by-side configuration: trend of $C_m$ along with $KC$ for $\beta = 350$ at different $G/d$}
\label{fig:CmSbSGap}
\end{figure}

Fig. \ref{fig:CmSbS}  plots the added mass coefficient $C_m$ of different $\beta$ for 4 gap ratios $G/d$ of 0.5, 1.0, 2.0 and 3.0. Similar to the single cylinder case, in the current $\beta$ range, $C_m$ of side-by-side dual cylinders for all gap ratios is not sensitive to the change of $\beta$ while strongly dependent on the increasing $KC$. It is found in general for all gap ratios, $C_m$ will first decrease with increasing $KC$ and then converge to a constant value for large $KC$. However, a difference can be found between different gap ratios on how fast $C_m$ decreases to what constant value. As for $G/d = 0.5$, when the interaction between the two cylinders is the strongest, it is observed $C_m$ will reach a minimum around $KC = 4$ and then converge to a value of around $C_m = 1.5$, and yet the $KC$ for the minimum $C_m$ increases with the increasing $G/d$, while the steady value of $C_m$ approaches 1.0 when gap distance gets larger. To reveal the effect of the interaction between two cylinders in a side-by-side configuration, in Fig. \ref{fig:CmSbSGap}, it plots $C_m$ at different $G/d$ (including single cylinder case of $G/d=\infty$) for $\beta = 350$ along with $KC$. As mentioned above, the $C_m$ difference between cylinders of different gap ratios lies in how fast $C_m$ decreases and to what constant value. In addition, it is observed that when gap distance is larger as 2.0 and 3.0, the distribution of $C_m$ along with $KC$ is close to the single cylinder case at same $\beta$, as naturally when the gap distance is larger, the interaction between two cylinders become weaker and weaker and hence simply behave like two independent cylinders.

\begin{figure}[!ht]
\centering
\begin{subfigure}{0.4\columnwidth}
\includegraphics[width=\columnwidth,keepaspectratio]{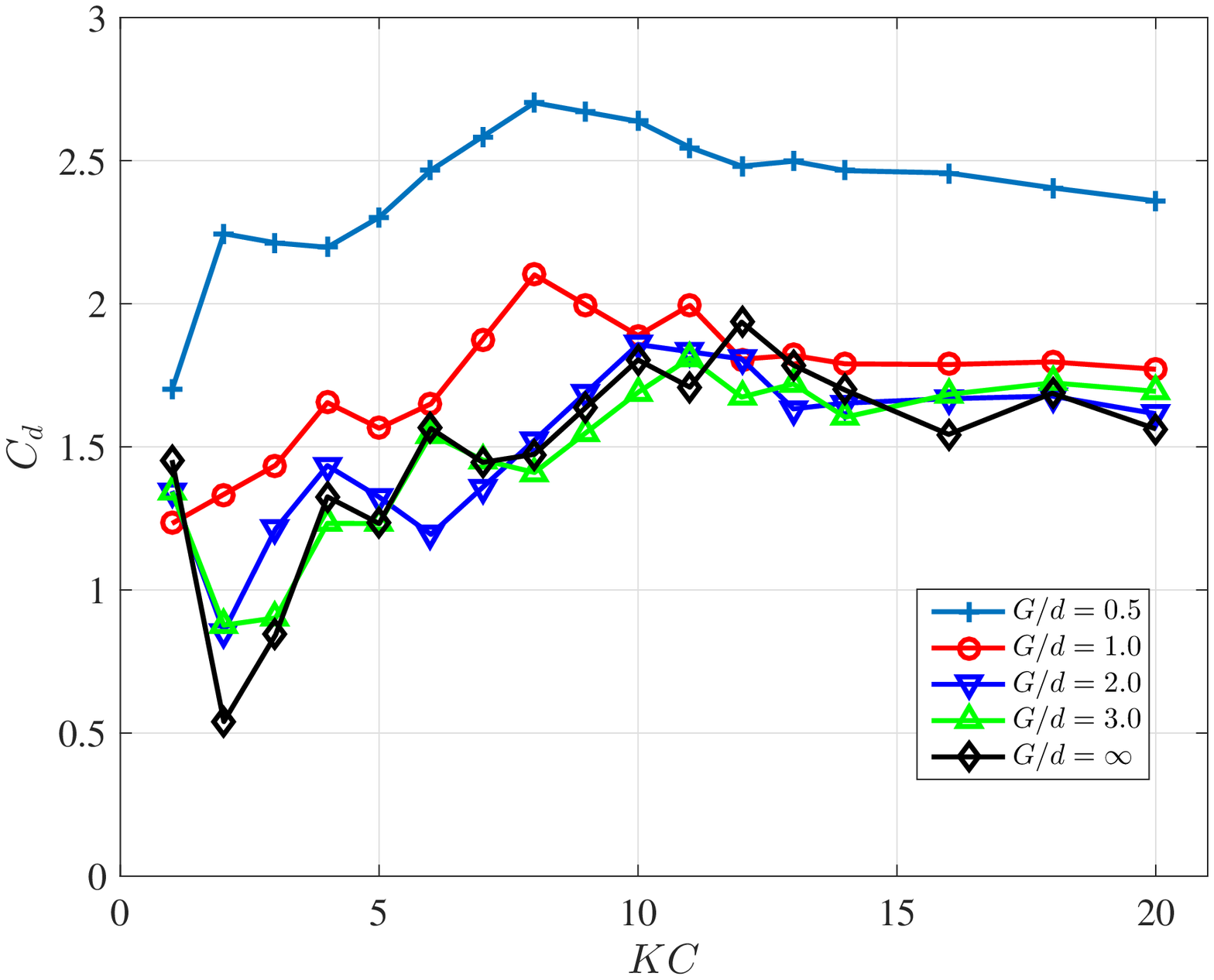}
\end{subfigure}
\begin{subfigure}{0.4\columnwidth}
\includegraphics[width=\columnwidth,keepaspectratio]{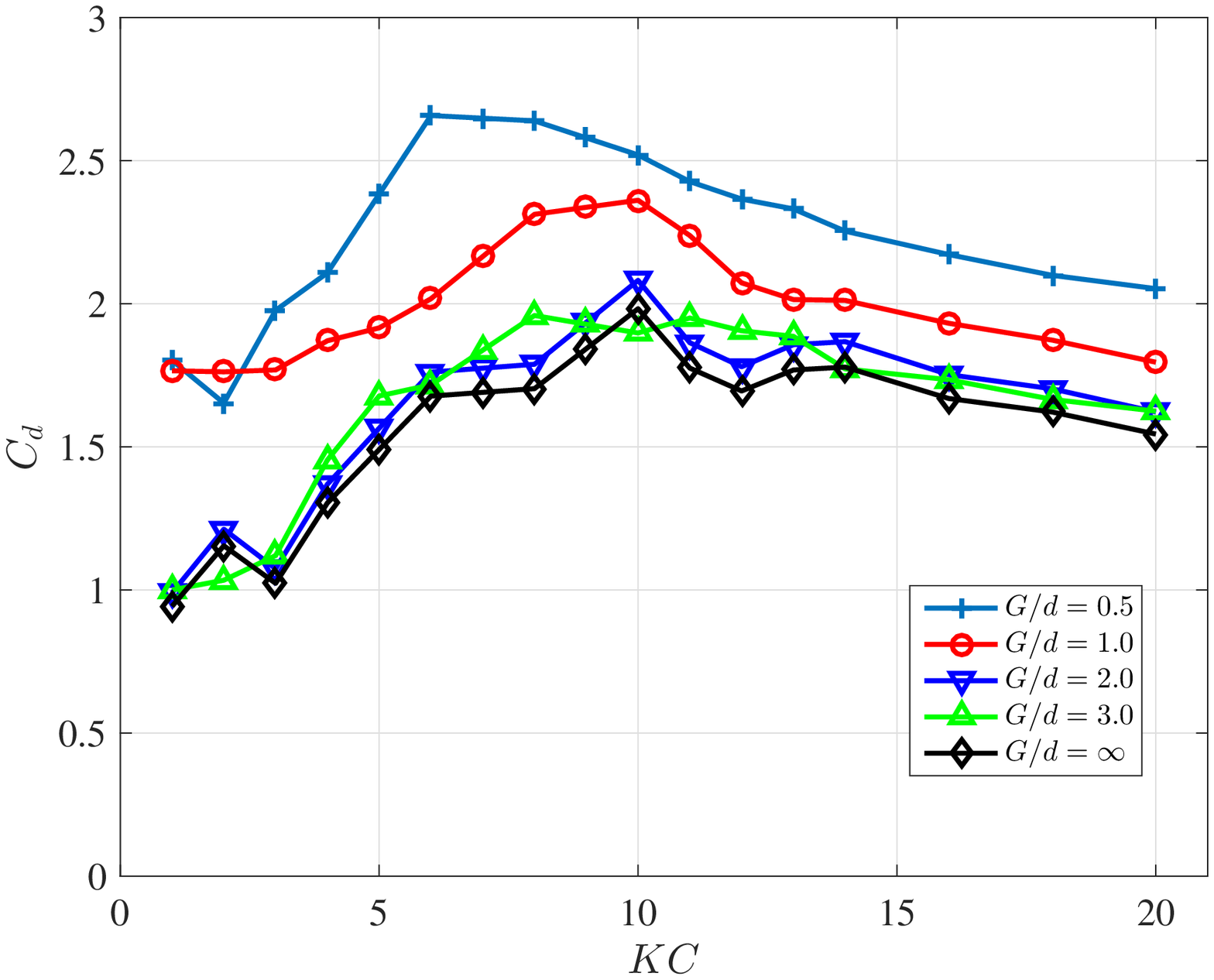}
\end{subfigure}
\caption{Cylinders in side-by-side configuration: trend of $C_d$ along with $KC$ at different $G/d$ for $\beta = 350$ (left) and $\beta = 700$ (right)}
\label{fig:CdSbSGap}
\end{figure}

\begin{figure}[!ht]
\centering
\begin{subfigure}{0.4\columnwidth}
\includegraphics[width=\columnwidth,keepaspectratio]{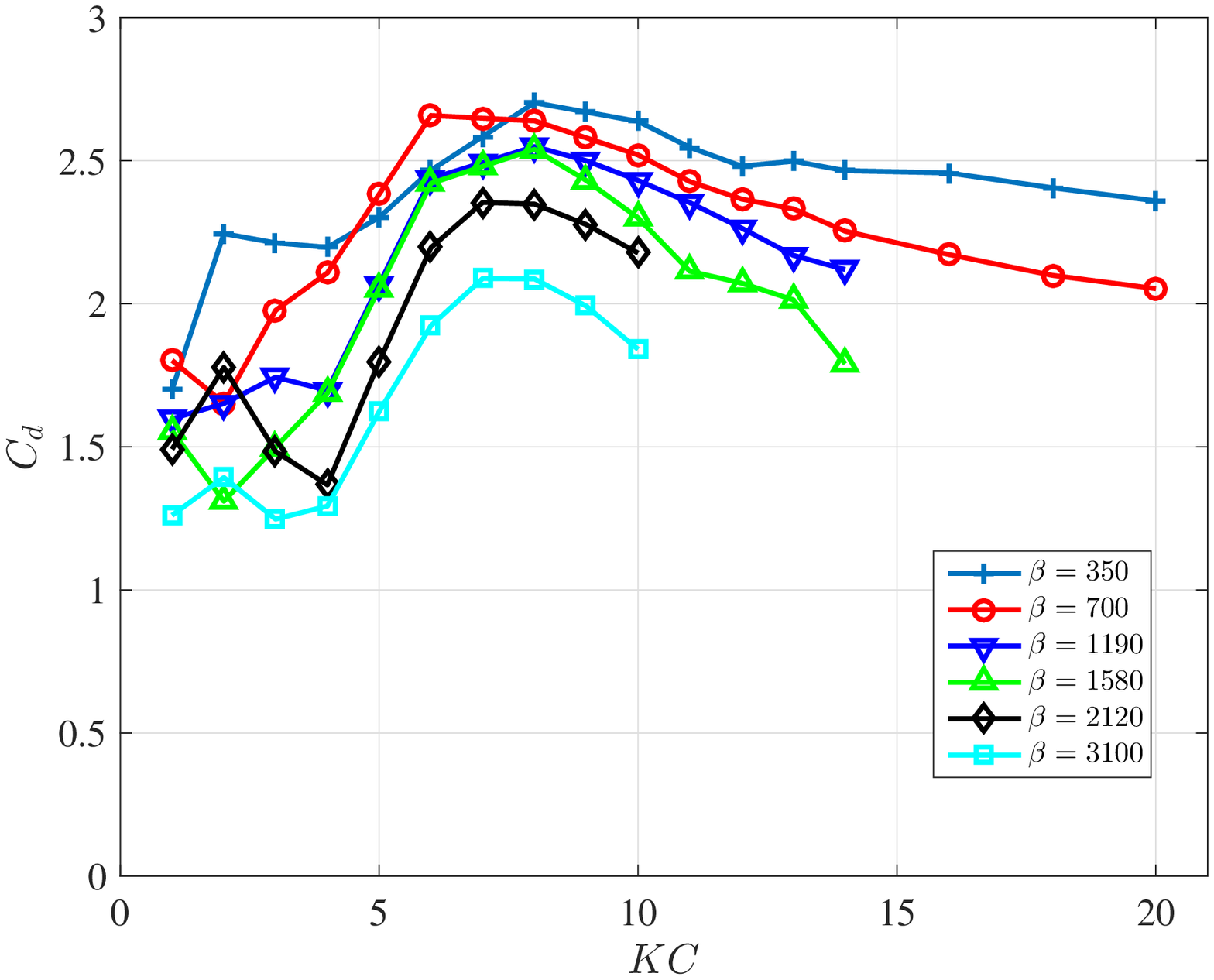}
\end{subfigure}
\begin{subfigure}{0.4\columnwidth}
\includegraphics[width=\columnwidth,keepaspectratio]{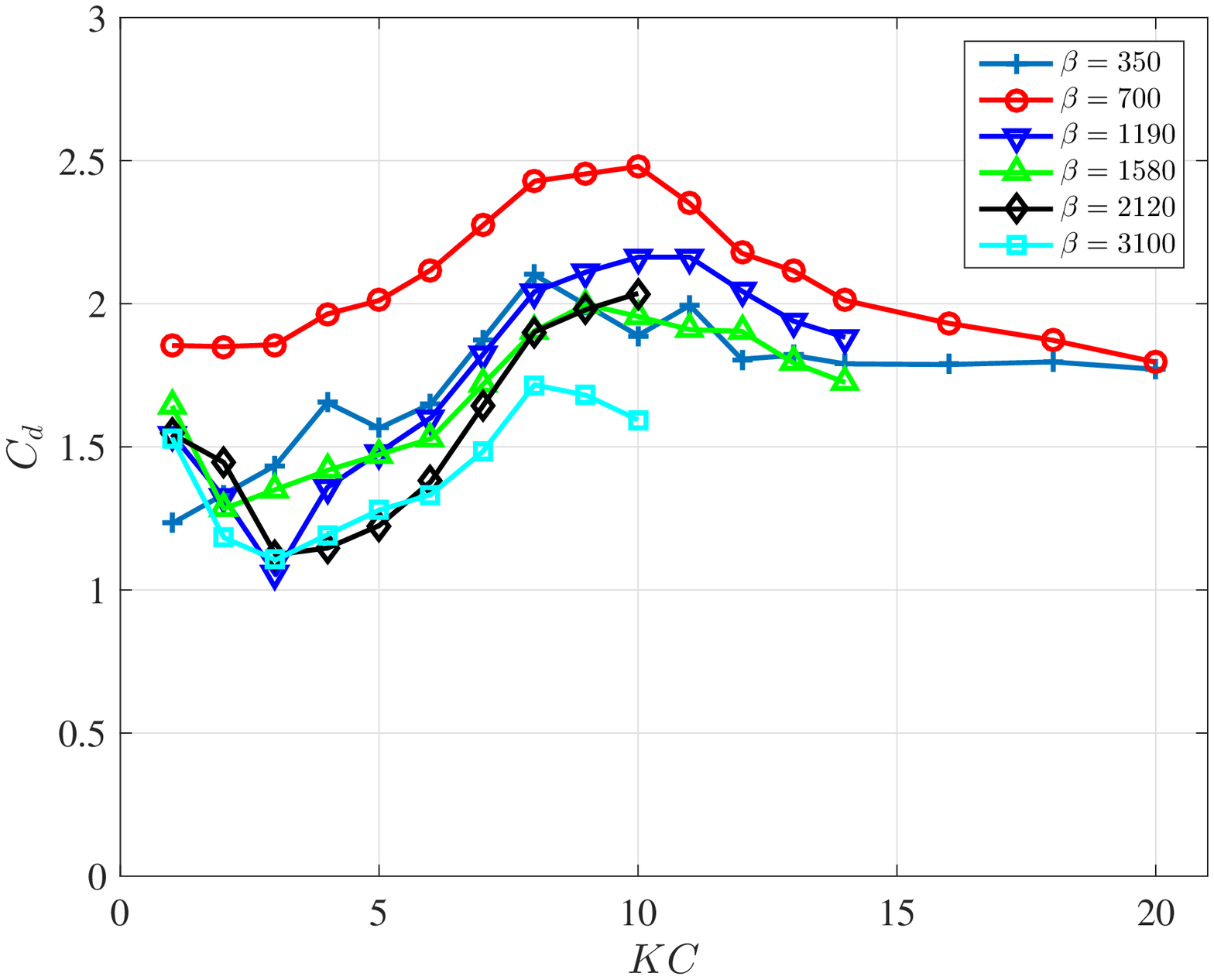}
\end{subfigure}
\begin{subfigure}{0.4\columnwidth}
\includegraphics[width=\columnwidth,keepaspectratio]{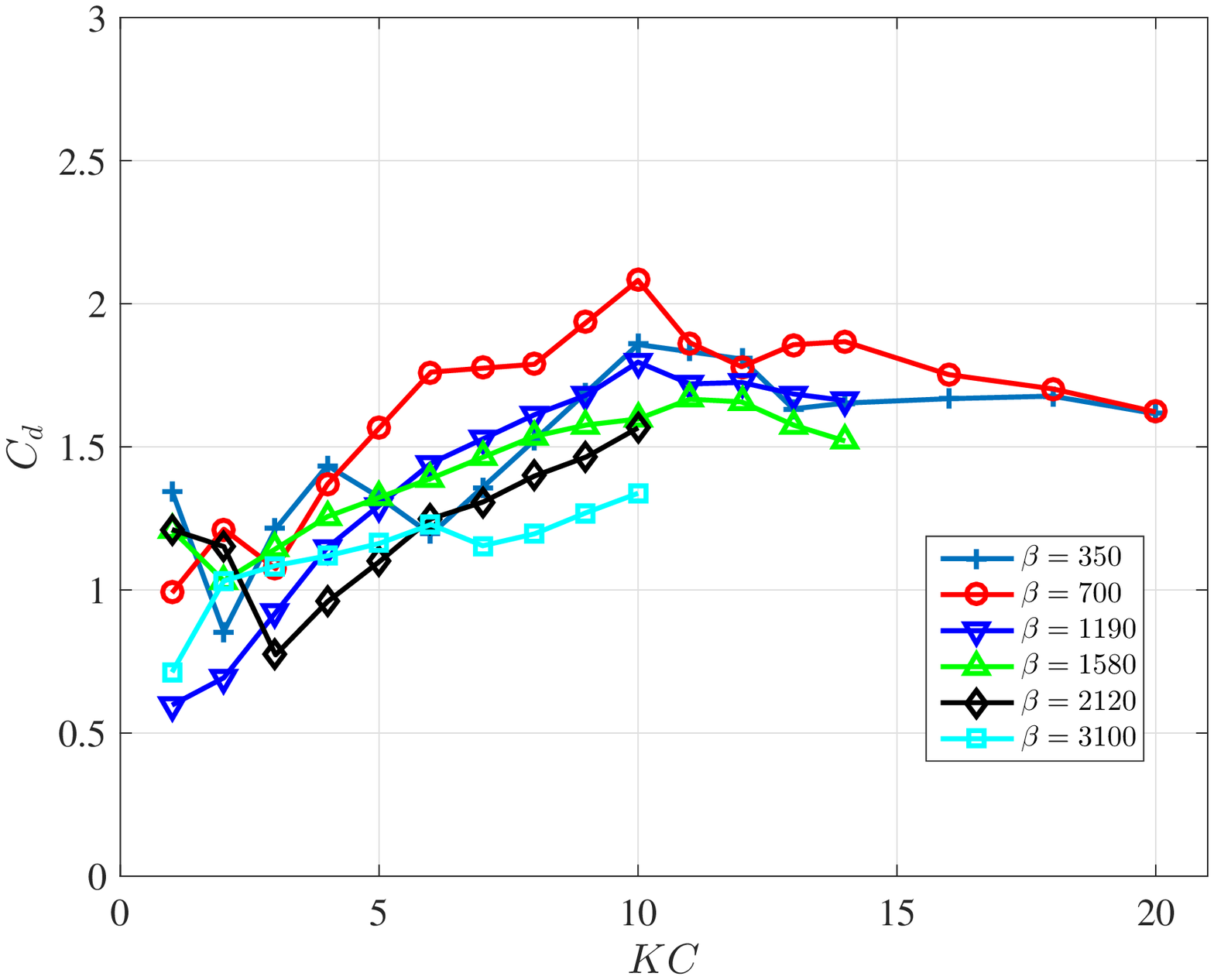}
\end{subfigure}
\begin{subfigure}{0.4\columnwidth}
\includegraphics[width=\columnwidth,keepaspectratio]{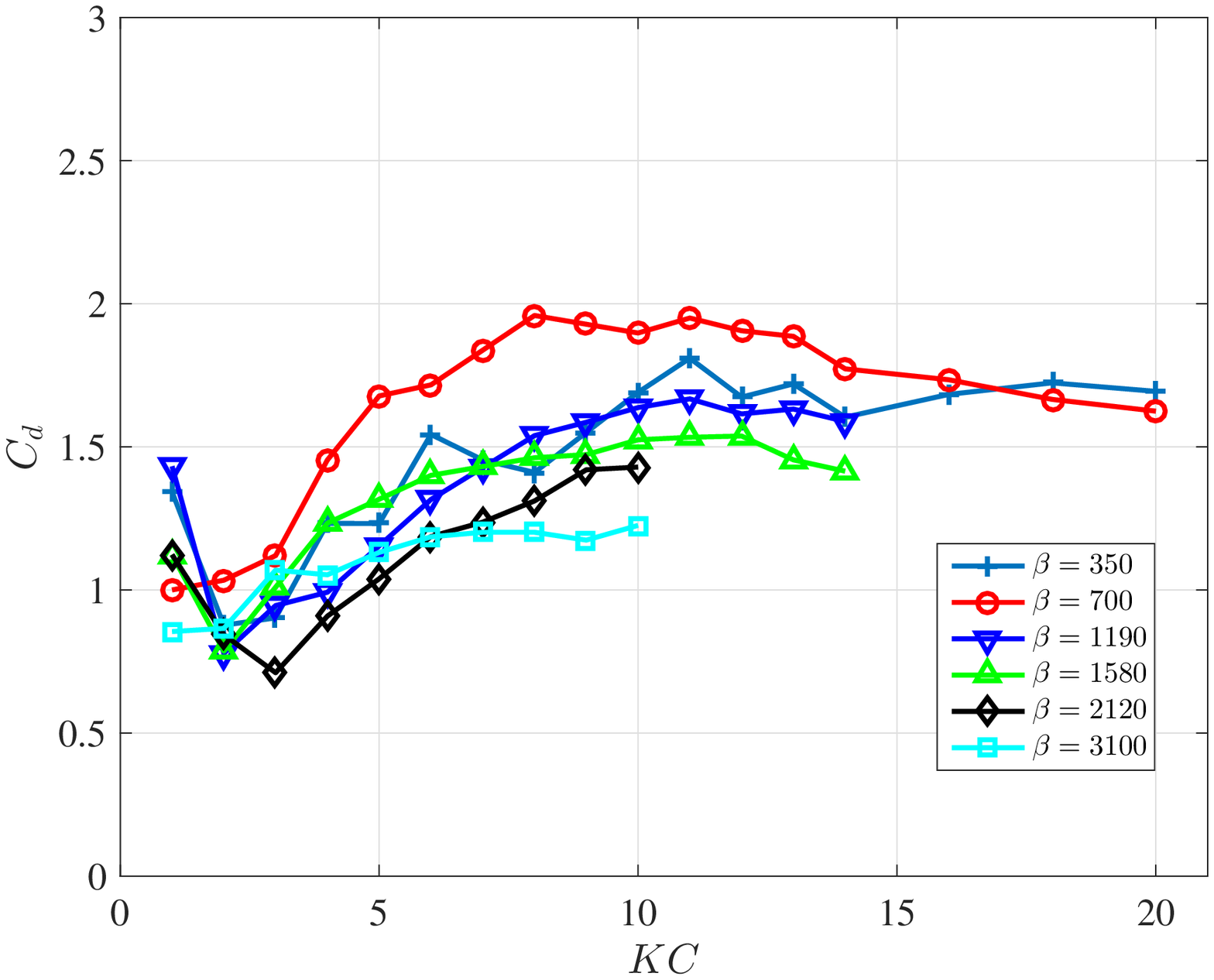}
\end{subfigure}
\caption{Cylinders in side-by-side configuration: trend of $C_d$ along with $KC$ for different $\beta$ at $G/d = 0.5$ (top left), $G/d = 1.0$ (top right), $G/d = 2.0$ (bottom left) and $G/d = 3.0$ (bottom right)}
\label{fig:CdSbS}
\end{figure}

In Fig. \ref{fig:CdSbS}, it plots the drag coefficient $C_d$ of different $\beta$ for 4 gap ratios $G/d$ of 0.5, 1.0, 2.0 and 3.0. In general, $C_d$ has a similar trend with KC for side-by-side dual cylinders, compared to the single cylinder case that $C_d$ increases with increasing $KC$ and then decreases slightly to a constant value. However, one of the most striking results is that for all $\beta$, with the decrease of the $G/d$, we observe a coherent enhancement for Cd. Here in Fig. \ref{fig:CdSbSGap}, it plots two cases of $\beta = 350$ and $\beta = 700$ for different gap ratios G/d (including single cylinder case of $G/d=\infty$.) along with $KC$. It is revealed that compared to the single cylinder, $C_d$ of $G/d = 0.5$ and $G/d = 1.0$ are largely amplified, especially for the case of $G/d = 0.5$, its $C_d$ can reach almost twice that of the single cylinder. At the same time, when gap distances get larger as to 2 and 3 times of the cylinder diameter, it is found the results of their $C_d$ are really close to the single cylinder case, and hence the interaction between the two cylinders are weak, same as what is observed from $C_m$ results.

\begin{figure}[!ht]
\centering
\begin{subfigure}{0.4\columnwidth}
\includegraphics[width=\columnwidth,keepaspectratio]{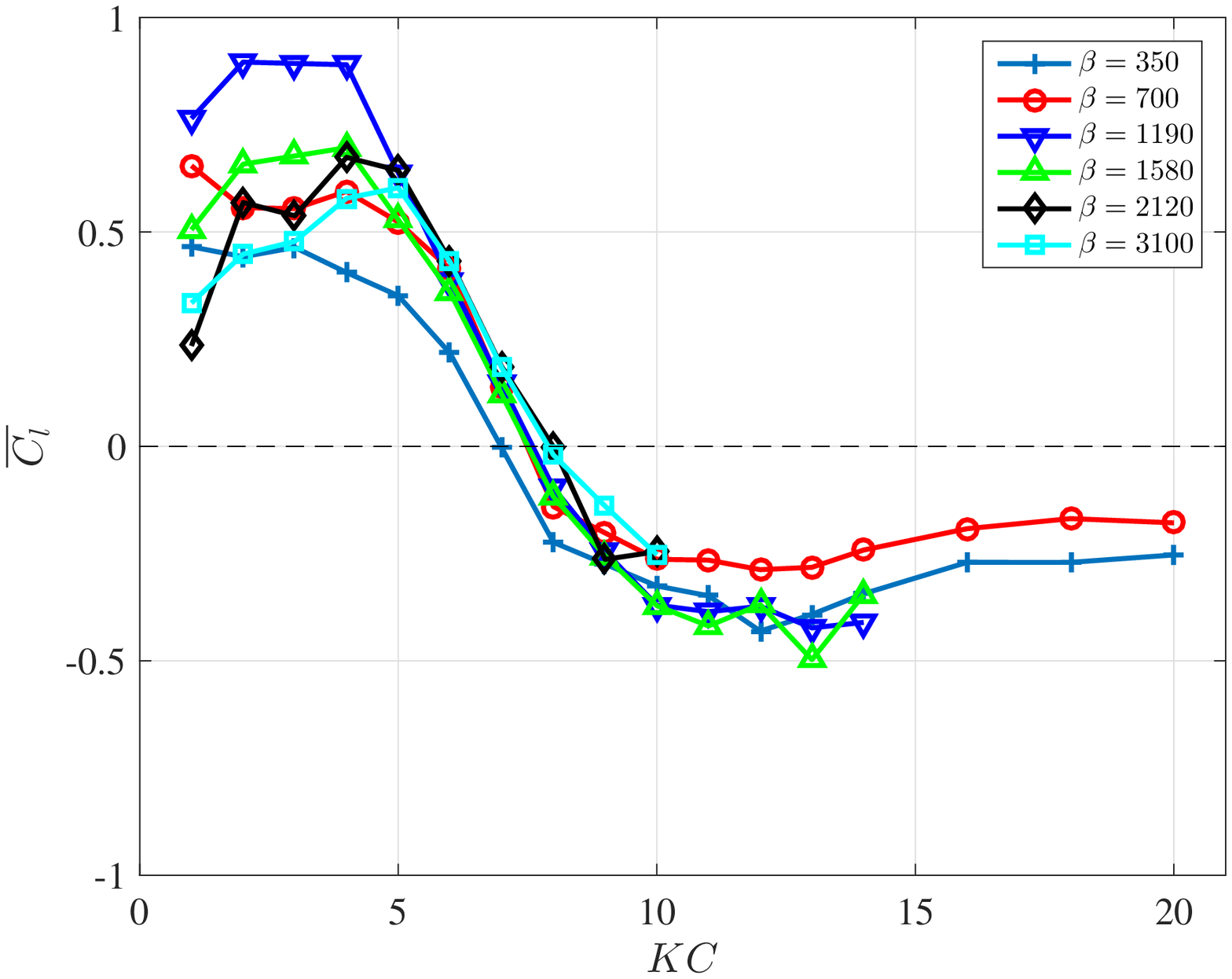}
\end{subfigure}
\begin{subfigure}{0.4\columnwidth}
\includegraphics[width=\columnwidth,keepaspectratio]{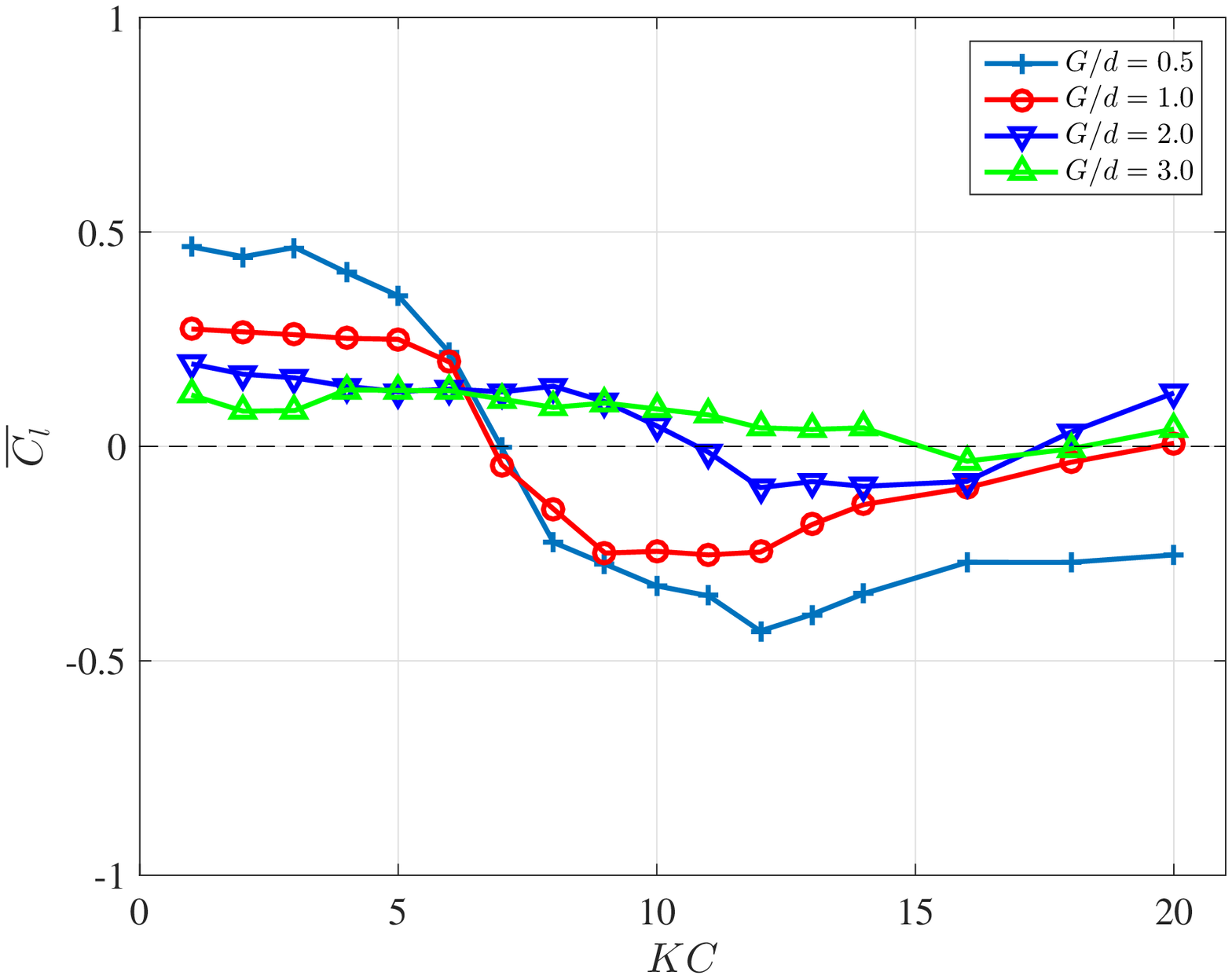}
\end{subfigure}
\caption{Cylinders in side-by-side configuration: trend of $\overline{C_l}$ along with $KC$ for different $\beta$ at $G/d = 0.5$ (left) and for $\beta = 350$ at different $G/d$ (right)}
\label{fig:ClavgSbS}
\end{figure}

\begin{figure}[!ht]
\centering
\begin{subfigure}{0.44\columnwidth}
\includegraphics[width=\columnwidth,keepaspectratio]{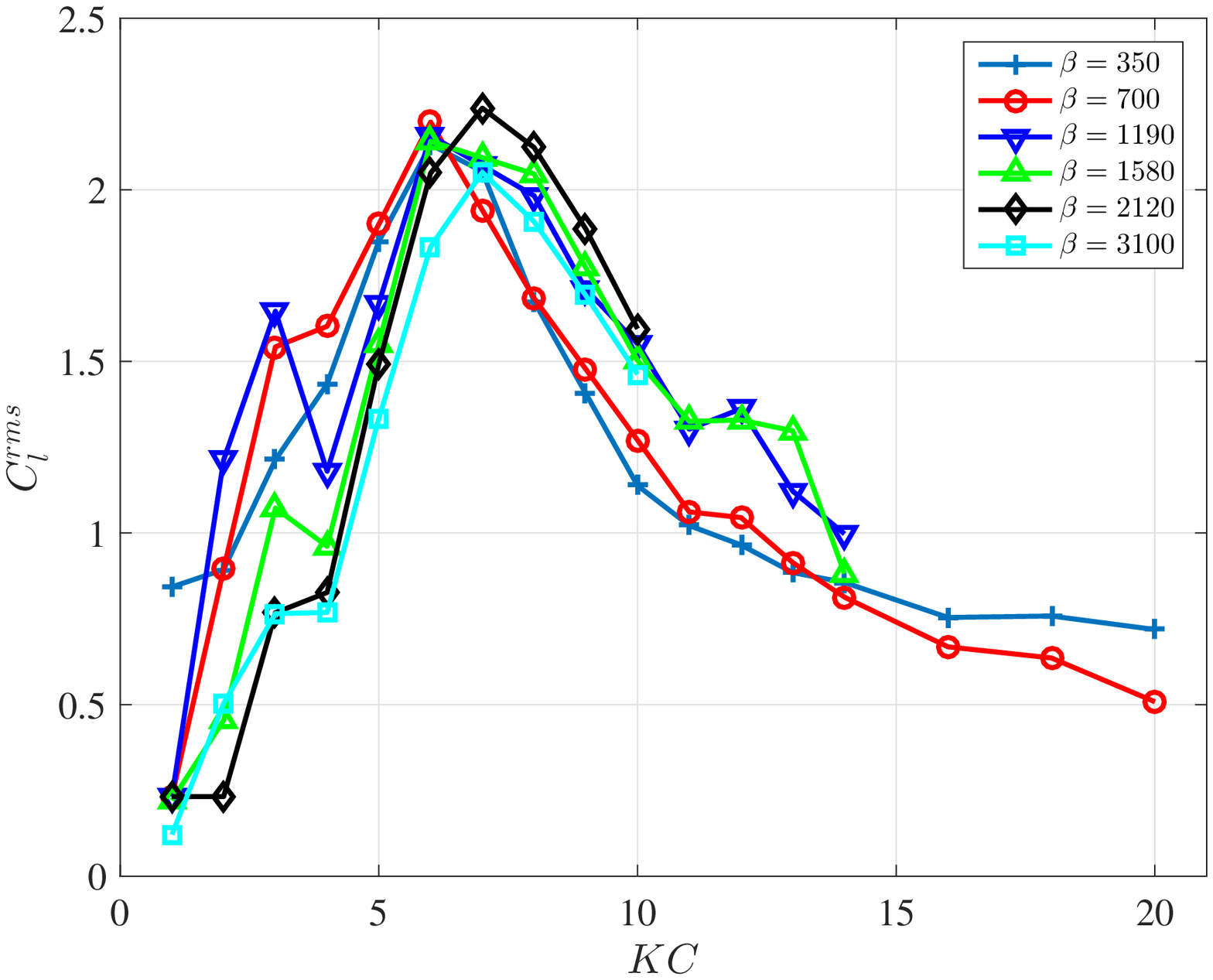}
\end{subfigure}
\begin{subfigure}{0.4\columnwidth}
\includegraphics[width=\columnwidth,keepaspectratio]{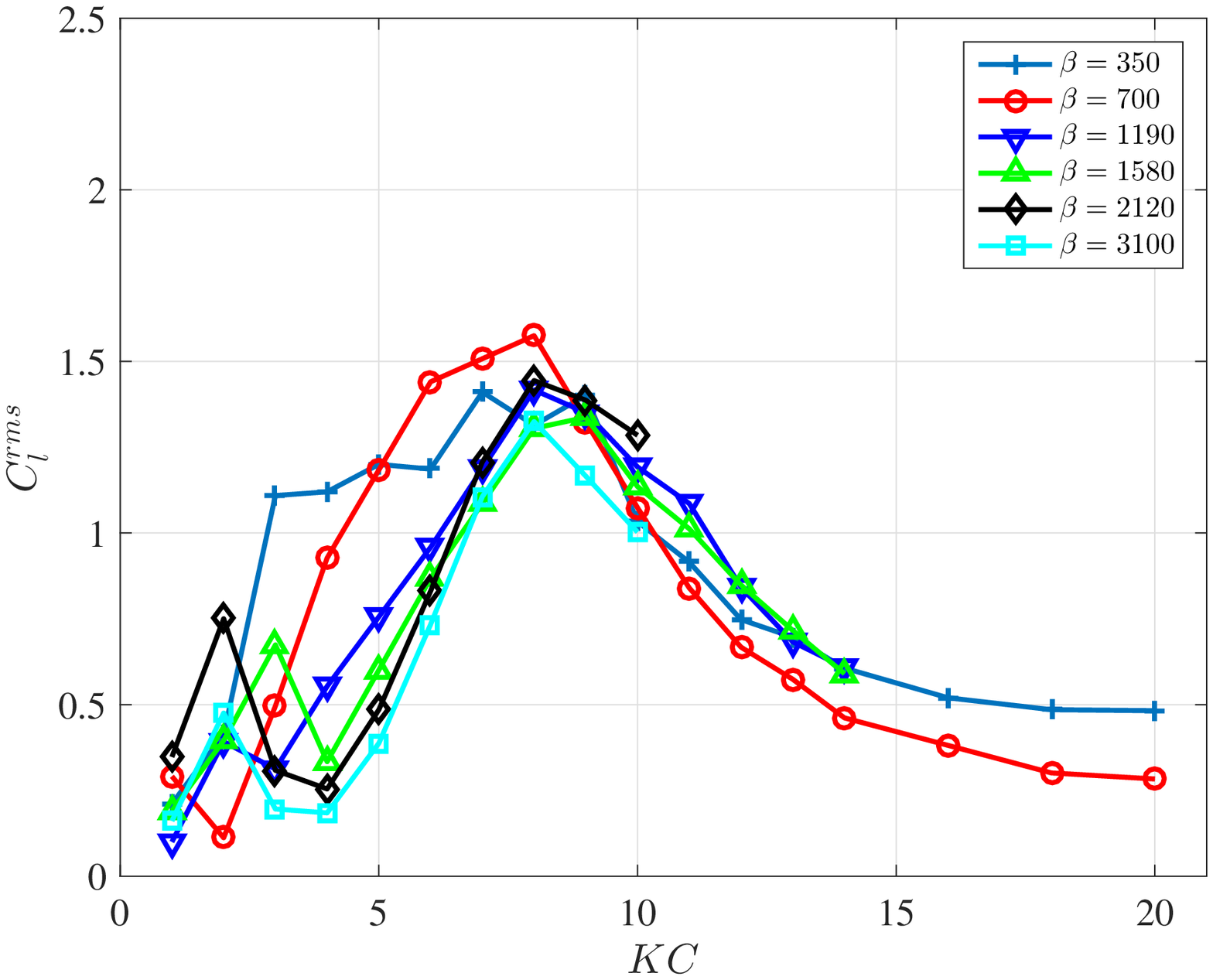}
\end{subfigure}
\begin{subfigure}{0.4\columnwidth}
\includegraphics[width=\columnwidth,keepaspectratio]{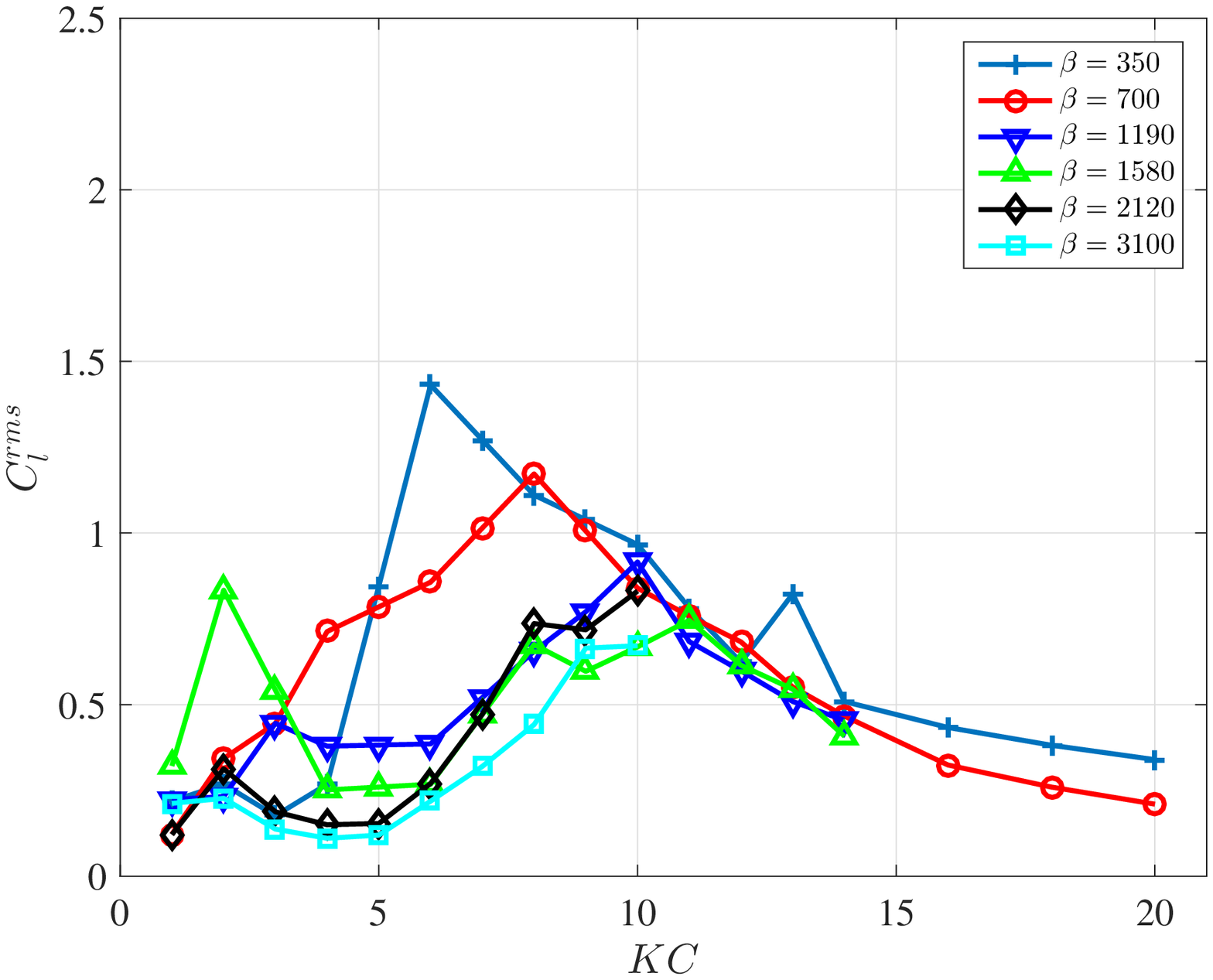}
\end{subfigure}
\begin{subfigure}{0.4\columnwidth}
\includegraphics[width=\columnwidth,keepaspectratio]{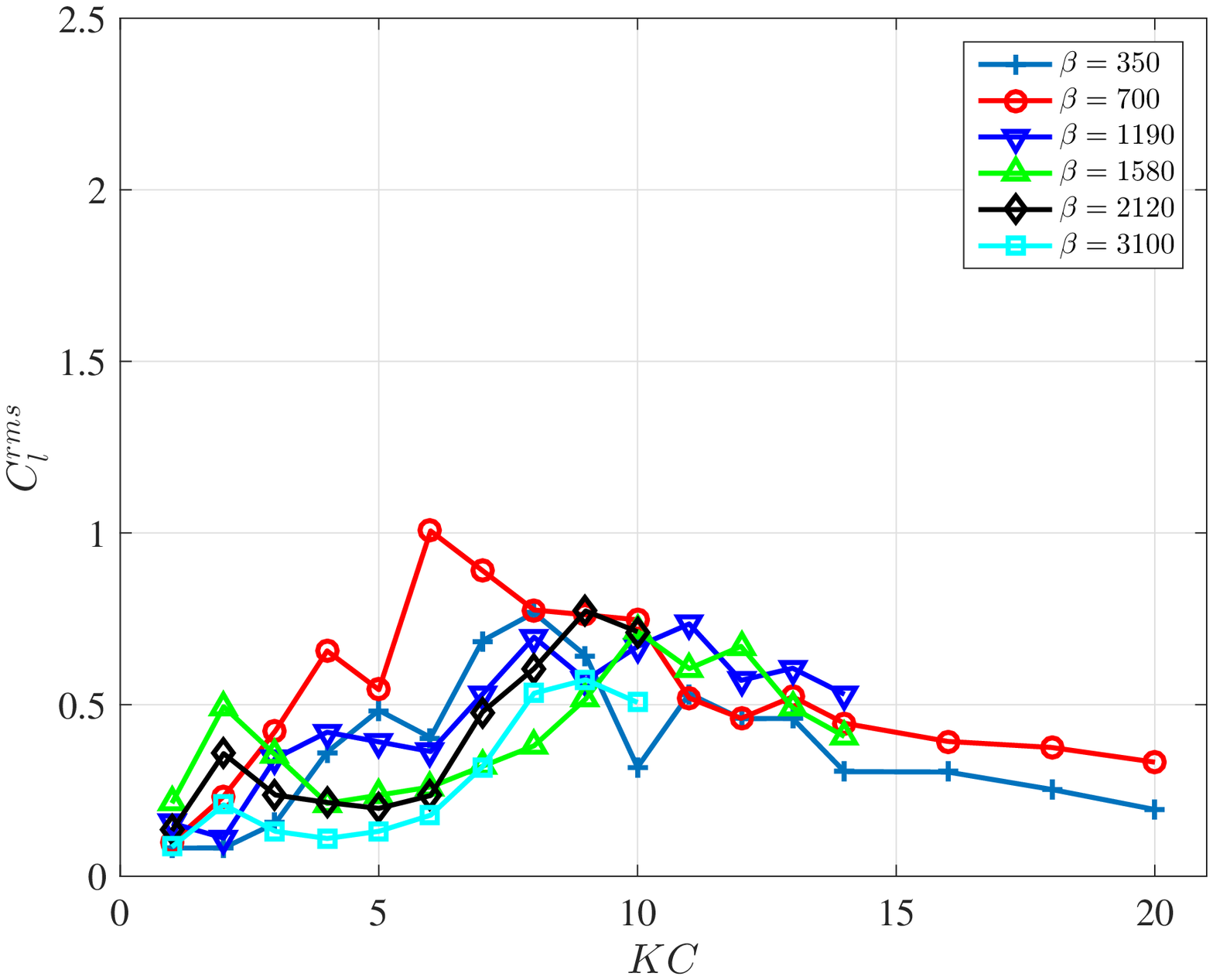}
\end{subfigure}
\caption{Cylinders in side-by-side configuration: trend of $C^{rms}_l$ along with $KC$ for different $\beta$ at $G/d = 0.5$ (top left), $G/d = 1.0$ (top right), $G/d = 2.0$ (bottom left) and $G/d = 3.0$ (bottom right)}
\label{fig:ClstdSbS}
\end{figure}

With the configuration of the side-by-side dual cylinder, it may induce a mean lift coefficient on each cylinder pointing towards or away from each other \cite{williamson1985fluid}. In Fig. \ref{fig:ClavgSbS}(left), it plots the mean lift coefficient $C_l$ on one cylinder for different $\beta$ at $G/d = 0.5$. It shows some interesting phenomena that cylinders will first experience a mean lift force pushing two cylinders close to each other at smaller $KC$ at $G/d = 0.5$, with an increase of $KC$, and this mean force will switch direction around $KC = 7$, pushing the two cylinders away from each other. This holds for all $\beta$. When the gap distance between two cylinders becomes larger, the mean lift coefficient hence will also get smaller, and this is revealed in Fig. \ref{fig:ClavgSbS}(right) of $C_l$ on one cylinder for $\beta = 350$ at $G/d$. It can be observed that for $G/d = 1.0$, its $C_l$ will hold a similar trend but a smaller value as that of $G/d = 0.5$, and when gap becomes large as $G/d = 2.0$, the mean lift coefficient $C_l$ is close to 0 for all $KC$.

\begin{figure}[!ht]
\centering
\includegraphics[width=0.4\columnwidth,keepaspectratio]{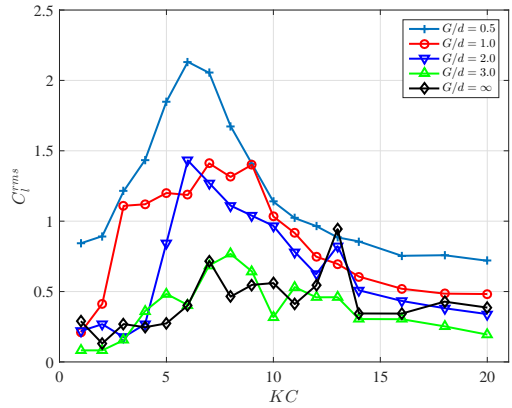}
\caption{Cylinders in side-by-side configuration: trend of $C^{rms}_l$ along with $KC$ for $\beta = 350$ at different $G/d$}
\label{fig:ClstdSbSGap}
\end{figure}

Furthermore, the lift coefficient RMS $C^{rms}_l$ is shown in Fig. \ref{fig:ClstdSbS} for all $\beta$ at different $G/d$. The result shows that though there is difference among different $\beta$ and $G/d$, the overall trend holds for $C^{rms}_l$ that it will first rise up then drop down with the increasing $KC$. Moreover, it is observed that with the smaller gap ratio and hence the stronger interaction between two cylinders, the larger $C^{rms}_l$ will be, as it is plotted in Fig. \ref{fig:ClstdSbSGap} of $C^{rms}_l$ for $\beta = 350$ at different $G/d$ along with $KC$. Unlike $C_m$, $C_d$ and $\overline{C_l}$, even for $G/d=2.0$, there is still a large difference between $C^{rms}_l$ of the side-by-side dual cylinders and the single cylinder. 

\subsection{Numerical Simulation: Flow Visualization}
\label{sec4_2}

Numerical simulation, using the same mesh setup as in the single cylinder case, was conducted on the dual cylinders in side-by-side configuration for $KC$ of 4 and 8, $\beta = 20$ and G/d of 0.5, 0.6, 0.7, 0.75, 0.9, 1.0, 1.25, 1.5, 1.6, 1.75, 2.0 and 3.0 respectively.

In the last subsection, we observe that when the two cylinders get closer, the drag coefficient is enhanced, which indicates a stronger wake interaction between two cylinders. As the Morison Equation states, the drag coefficient component is a dissipative term while the added mass coefficient component is a conservative term, which can be manifest itself through the following integral in one oscillation period,

\begin{equation}
    \int^{T}_0 F_x \cdot \dot{X}dt=\int^{T}_0(\frac{1}{2} \rho SC_d|\dot{X}|\dot{X} + \rho \nabla C_{m} \ddot{X})\cdot \dot{X}dt=\frac{4}{3\omega}\rho SC_dA^3.
\end{equation}

\begin{figure}[!ht]
\centering
\begin{subfigure}{0.32\columnwidth}
\includegraphics[width=\columnwidth,keepaspectratio]{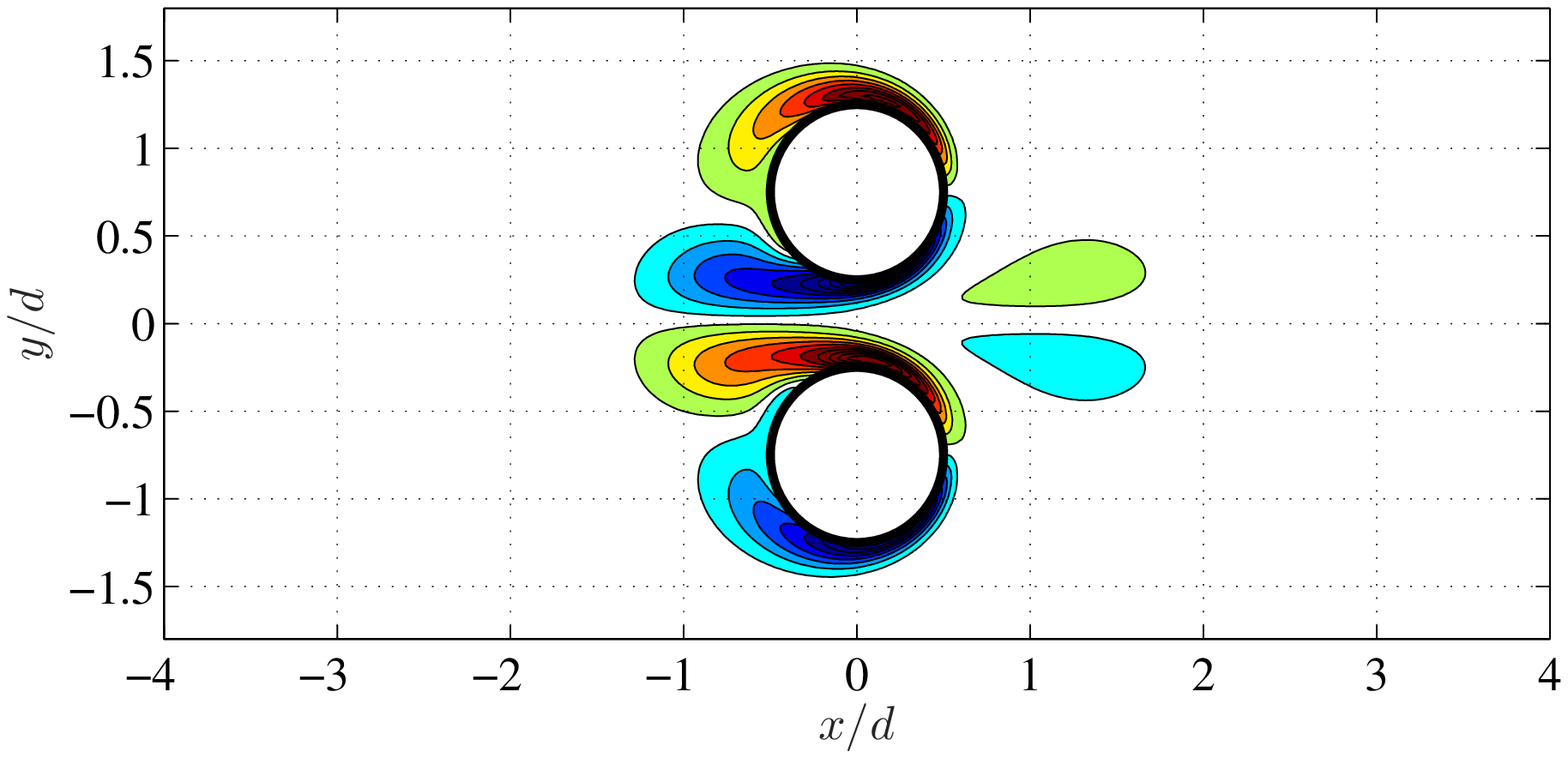}
\end{subfigure}
\begin{subfigure}{0.32\columnwidth}
\includegraphics[width=\columnwidth,keepaspectratio]{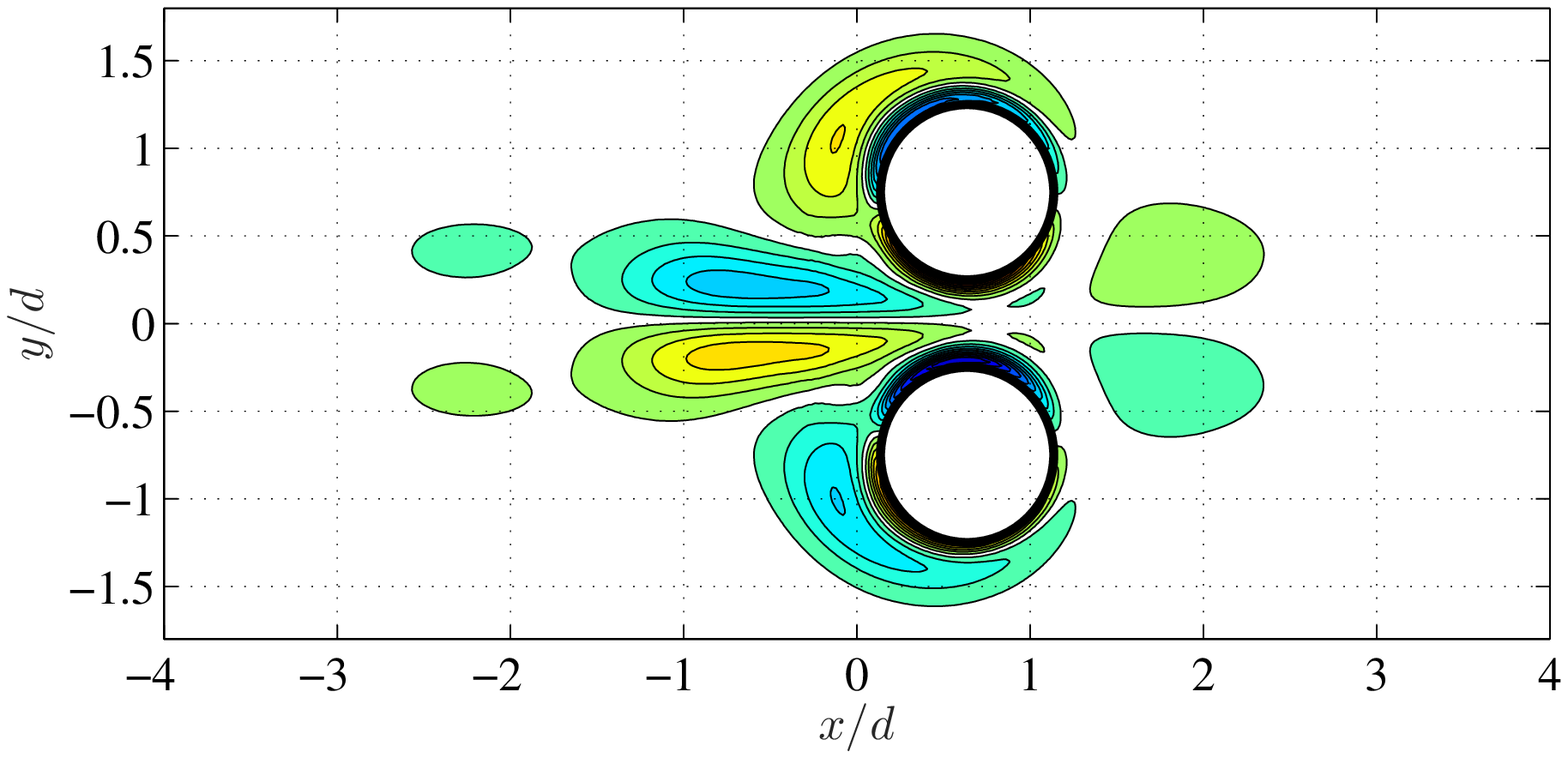}
\end{subfigure}
\begin{subfigure}{0.32\columnwidth}
\includegraphics[width=\columnwidth,keepaspectratio]{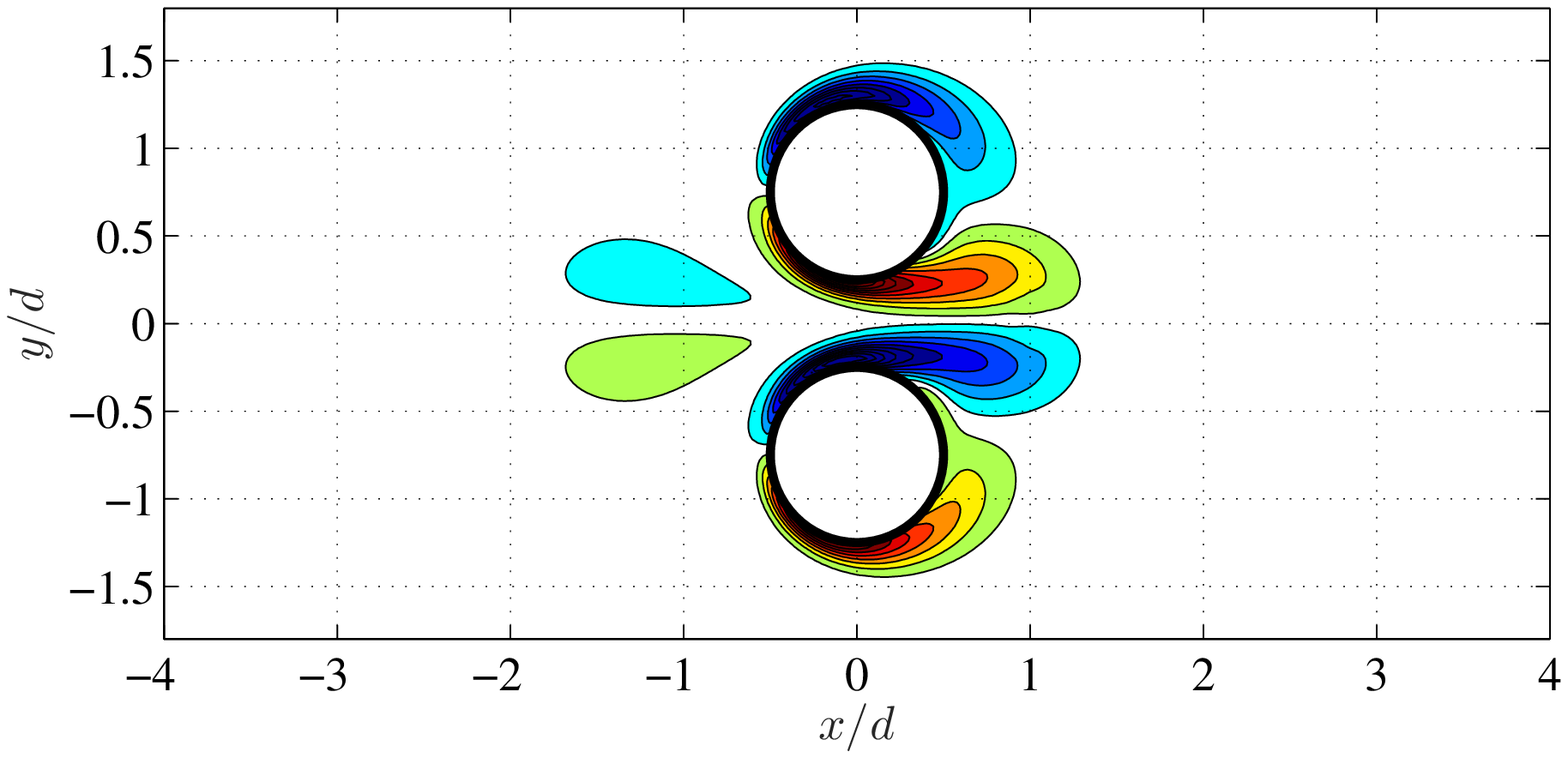}
\end{subfigure}
\begin{subfigure}{0.32\columnwidth}
\includegraphics[width=\columnwidth,keepaspectratio]{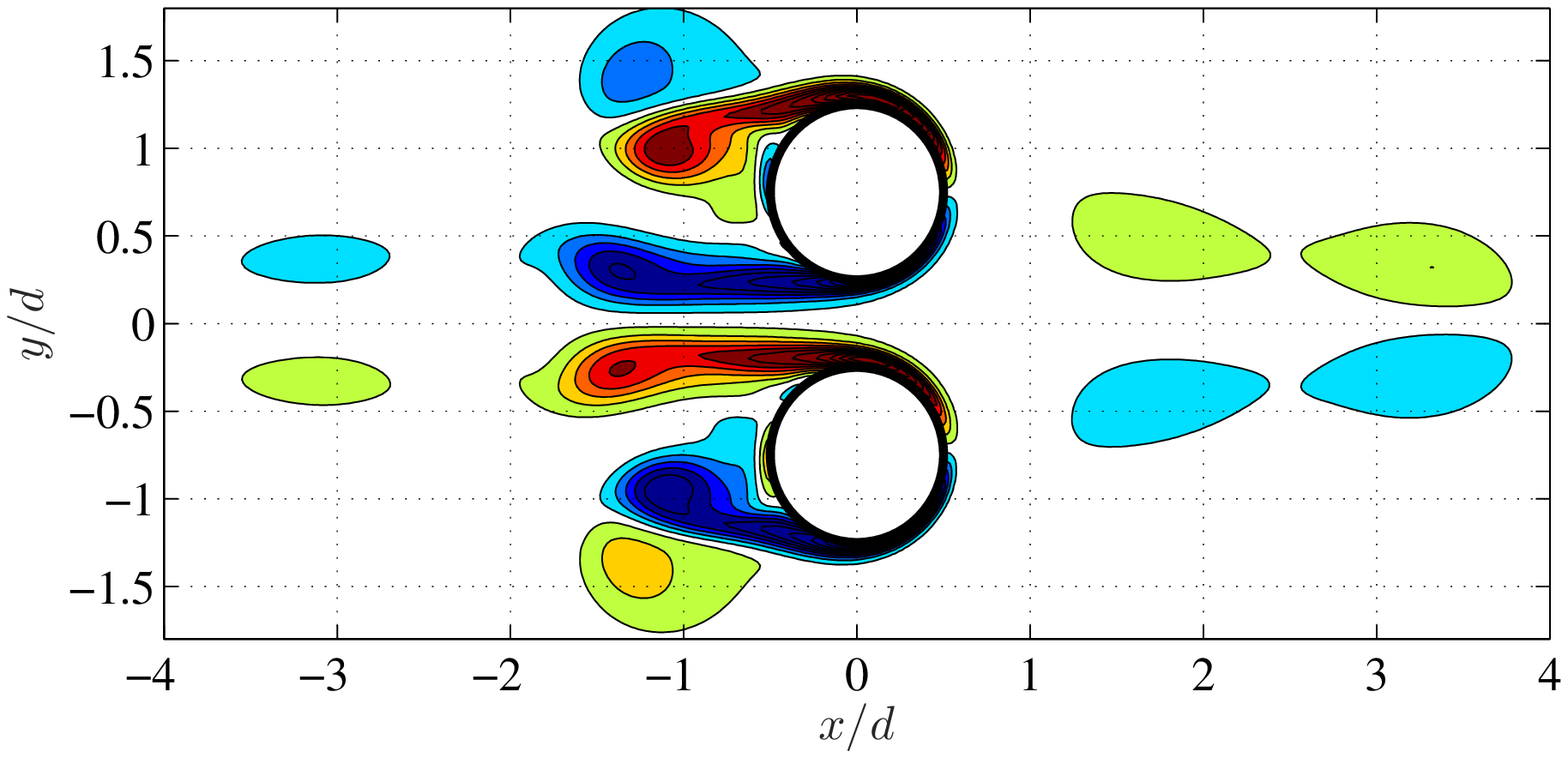}
\end{subfigure}
\begin{subfigure}{0.32\columnwidth}
\includegraphics[width=\columnwidth,keepaspectratio]{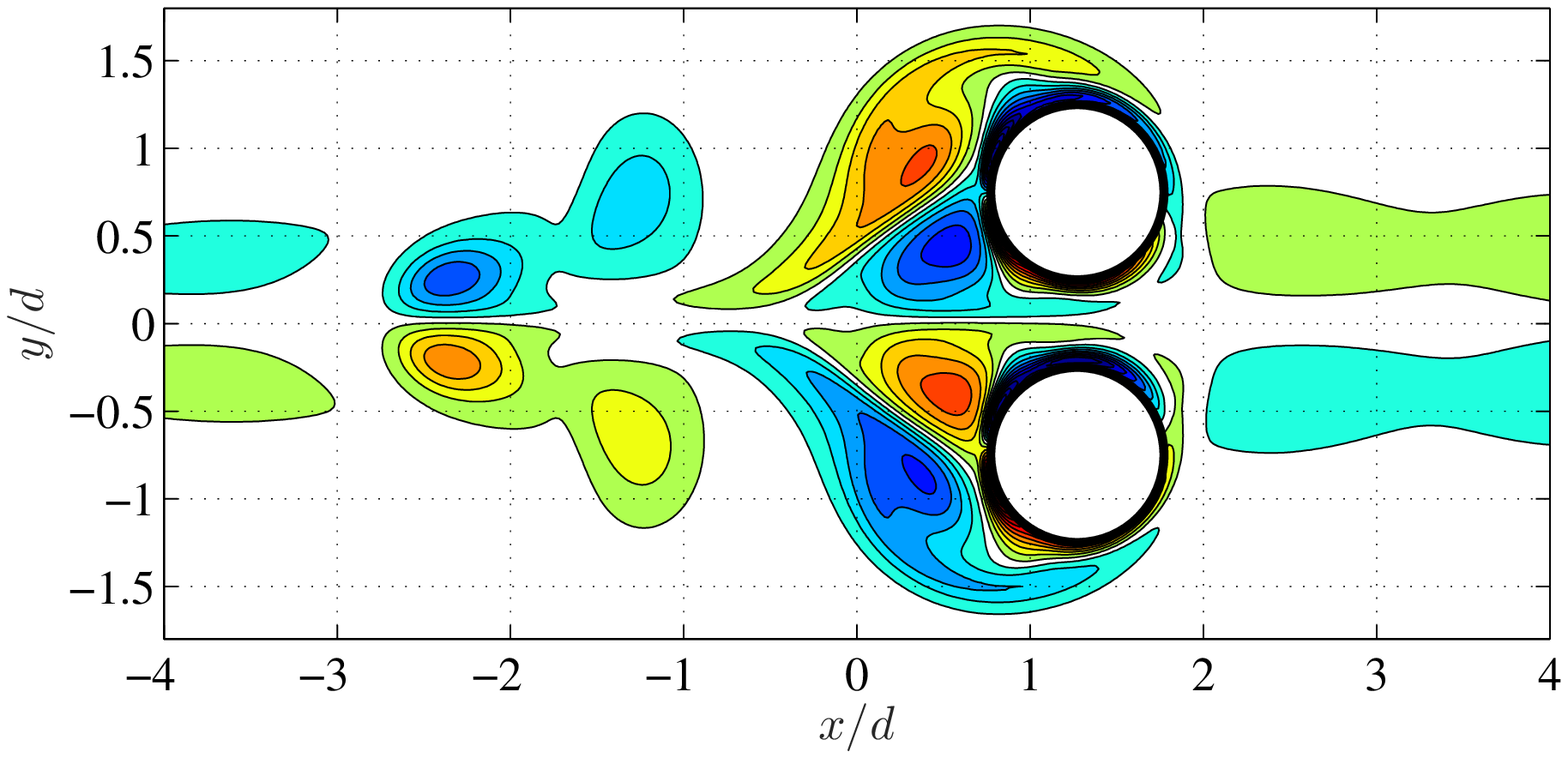}
\end{subfigure}
\begin{subfigure}{0.32\columnwidth}
\includegraphics[width=\columnwidth,keepaspectratio]{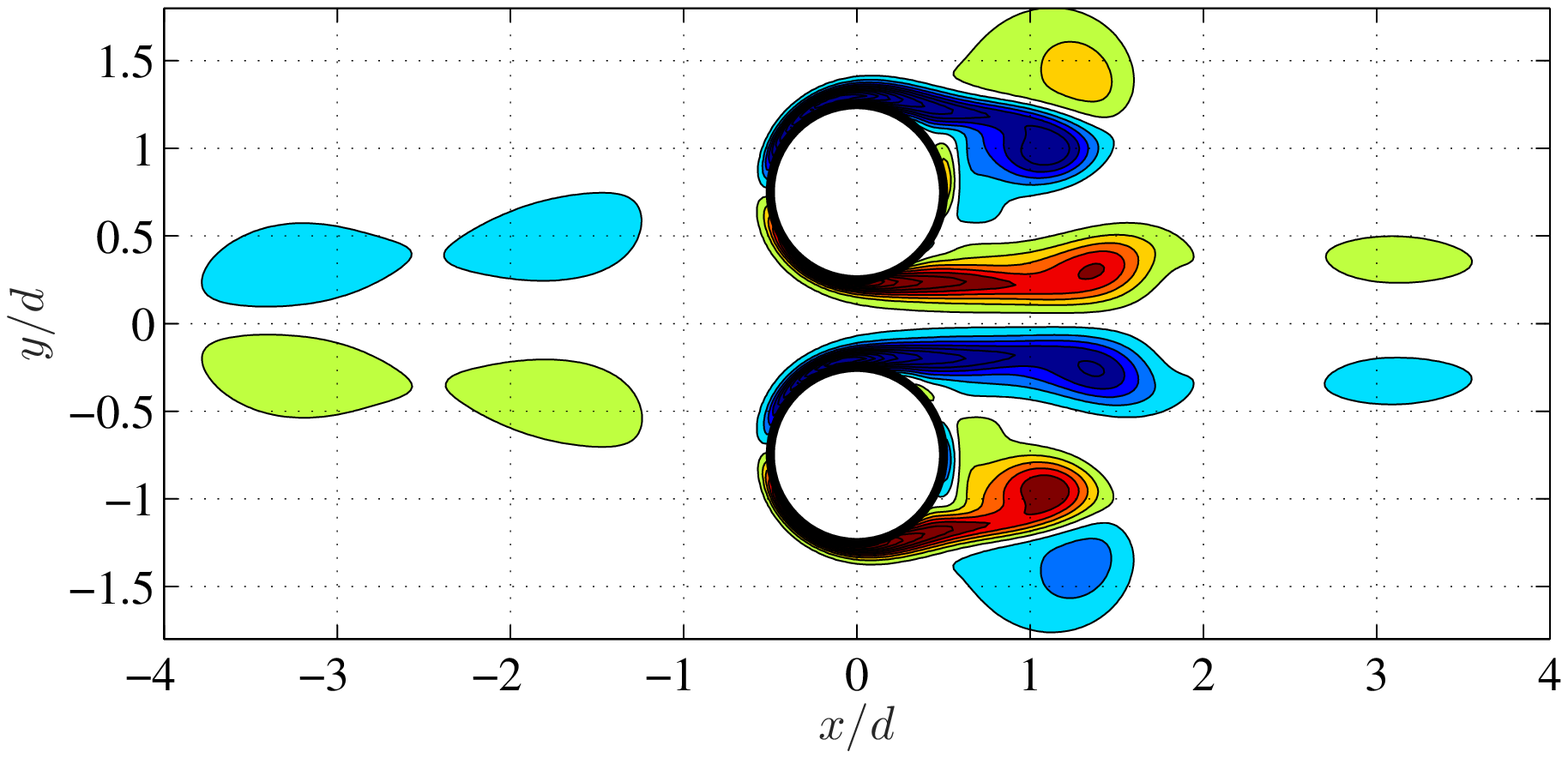}
\end{subfigure}
\begin{subfigure}{0.55\columnwidth}
\includegraphics[width=\columnwidth,keepaspectratio]{Fig/Single/Flow/cbar.eps}
\end{subfigure}
\caption{Snapshot of vorticity plot for side-by-side cylinders ($\beta = 20$, $G/d = 0.5$) at $KC=4$ (first row, $C_m = 1.61$, $C_d = 3.24$) and at $KC=8$ (second row, $C_m = 1.76$, $C_d = 2.79$) at $t/\tau$ of 0 (left), $\frac{1}{4}$ (middle) and $\frac{1}{2}$ (right).}
\label{fig:vortSbSG05}
\end{figure}

The flow visualization is shown in Fig. \ref{fig:vortSbSG05} of the cylinder pair in a side-by-side configuration for (first row) and  (second row) reveals that compared to the single cylinder at same $KC = 4$ and $KC = 8$, the shedding vortex is strengthened due to the existence of the energetic flow jet induced by the gap between two cylinders. Tracking the vortex pair left the cylinder, we can see that the vortex pair can maintain its shape and also achieve high self-induced velocity away from the cylinder for a much longer time and distance, compared to the faster-decayed vortex wake in the single cylinder case. Moreover, therefore, the vortex pair generated between the gap enhances the energy transfer/dissipation from the cylinder to the fluid into the far flow field. Thus this explains the phenomenon of the enhancement of the drag coefficient, which is directly associated with the energy transfer between the fluid and structure.

\section{Dual Cylinder in Tandem Configuration}
\label{sec5}
Experimental and numerical results of the oscillating tandem cylinders in still water will be presented in this section with a comparison with a single cylinder case.

\subsection{Experimental Results: Hydrodynamic Coefficients}
\label{sec5_1}
\begin{figure}[!ht]
\centering
\begin{subfigure}{0.4\columnwidth}
\includegraphics[width=\columnwidth,keepaspectratio]{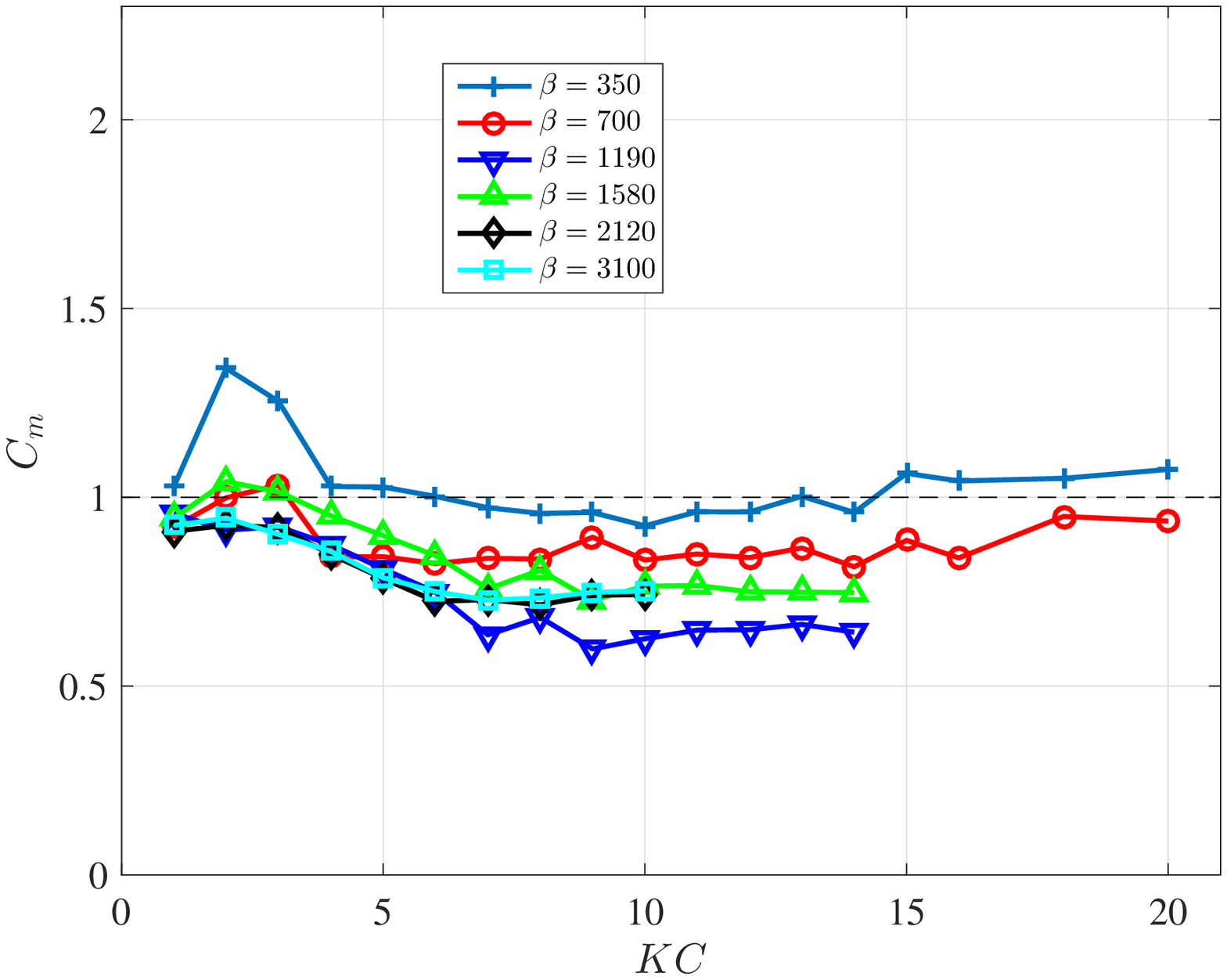}
\end{subfigure}
\begin{subfigure}{0.4\columnwidth}
\includegraphics[width=\columnwidth,keepaspectratio]{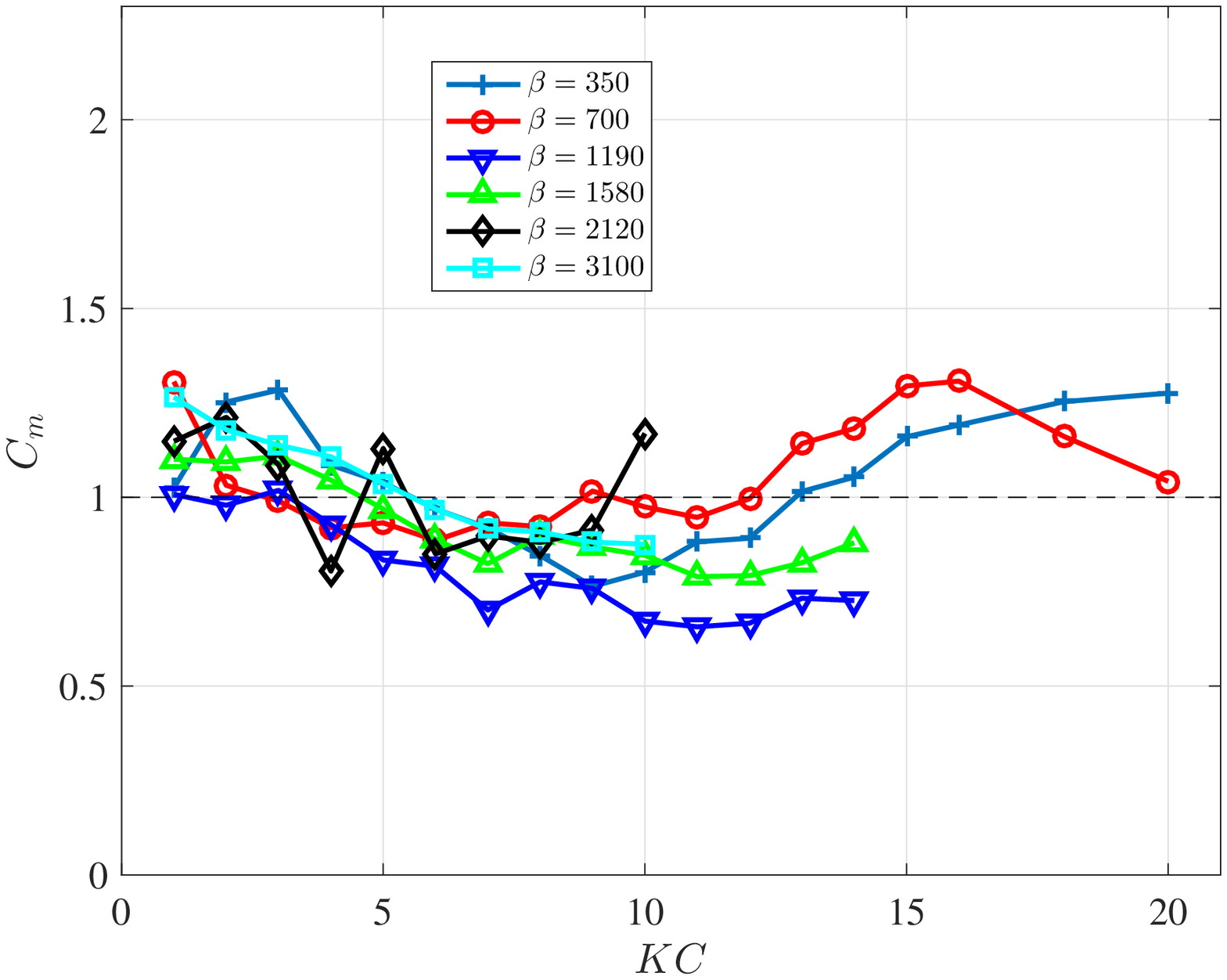}
\end{subfigure}
\begin{subfigure}{0.4\columnwidth}
\includegraphics[width=\columnwidth,keepaspectratio]{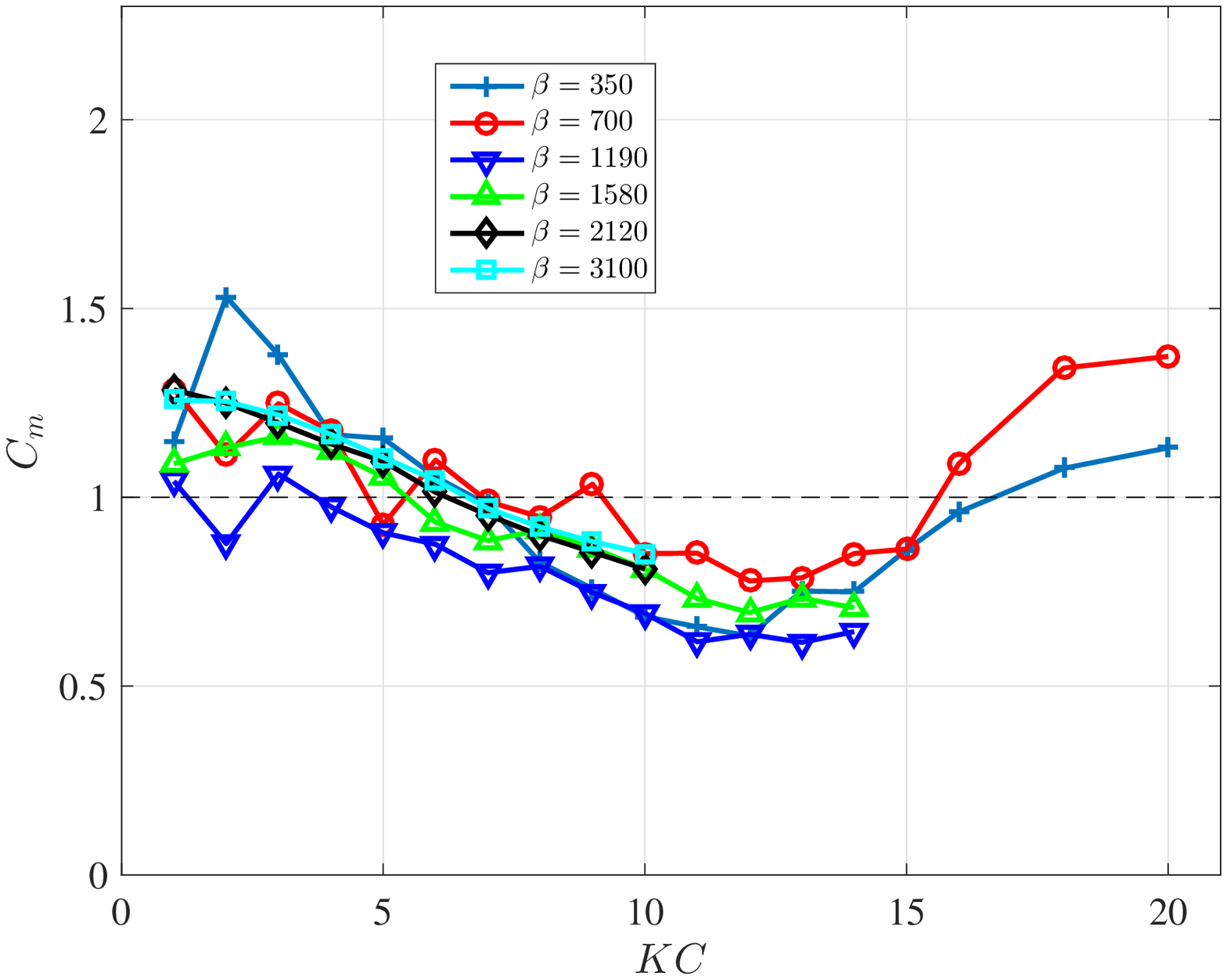}
\end{subfigure}
\begin{subfigure}{0.4\columnwidth}
\includegraphics[width=\columnwidth,keepaspectratio]{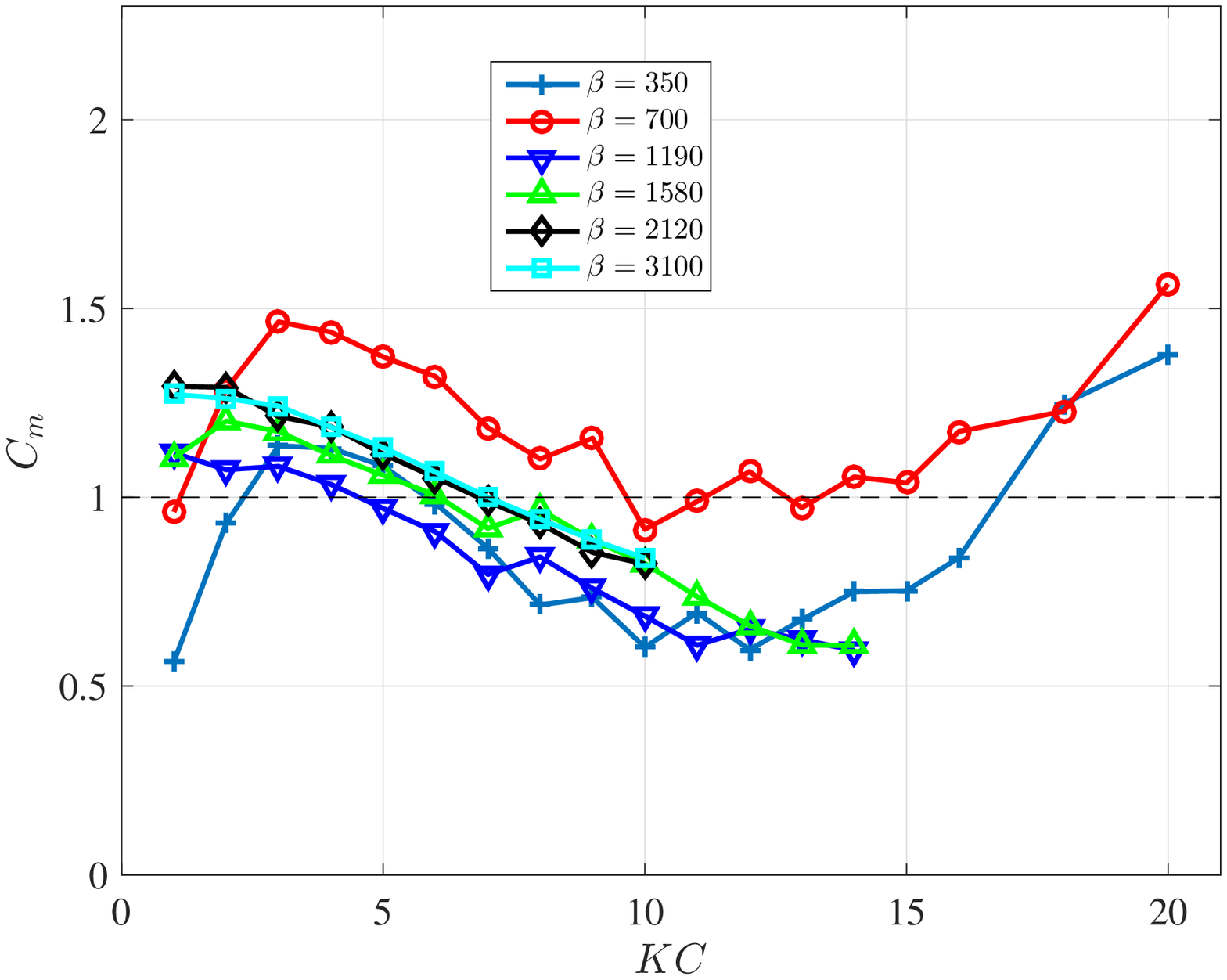}
\end{subfigure}
\caption{Cylinders in tandem configuration: trend of $C_m$ along with $KC$ for different $\beta$ at $G/d = 0.5$ (top left), $G/d = 1.0$ (top right), $G/d = 2.0$ (bottom left) and $G/d = 3.0$ (bottom right)}
\label{fig:CmTandem}
\end{figure}

\begin{figure}[!ht]
\centering
\includegraphics[width=0.4\columnwidth,keepaspectratio]{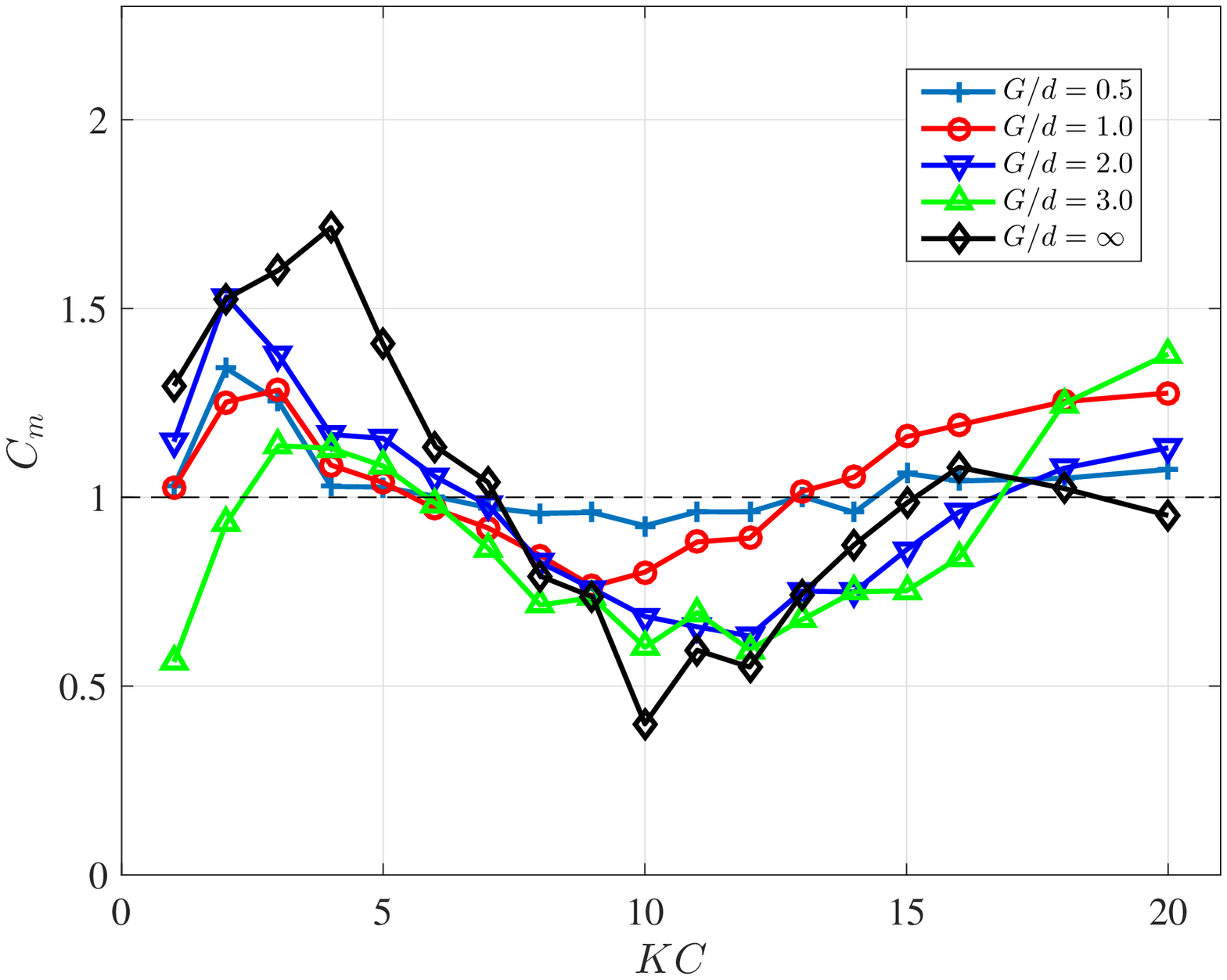}
\caption{Cylinders in tandem configuration: trend of $C_m$ along with $KC$ for $\beta = 350$ at different $G/d$}
\label{fig:CmTandemGap}
\end{figure}

\begin{figure}[!ht]
\centering
\begin{subfigure}{0.4\columnwidth}
\includegraphics[width=\columnwidth,keepaspectratio]{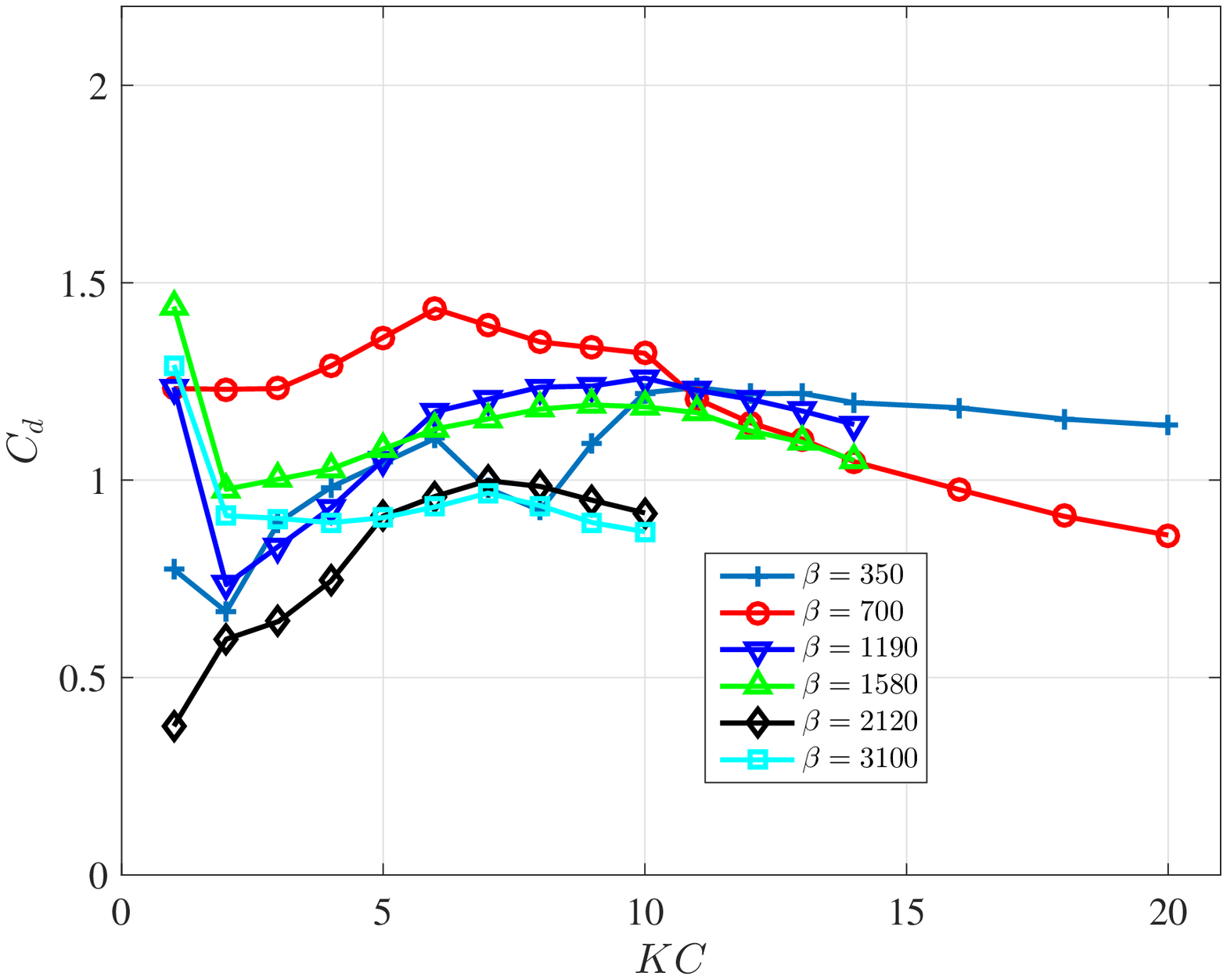}
\end{subfigure}
\begin{subfigure}{0.4\columnwidth}
\includegraphics[width=\columnwidth,keepaspectratio]{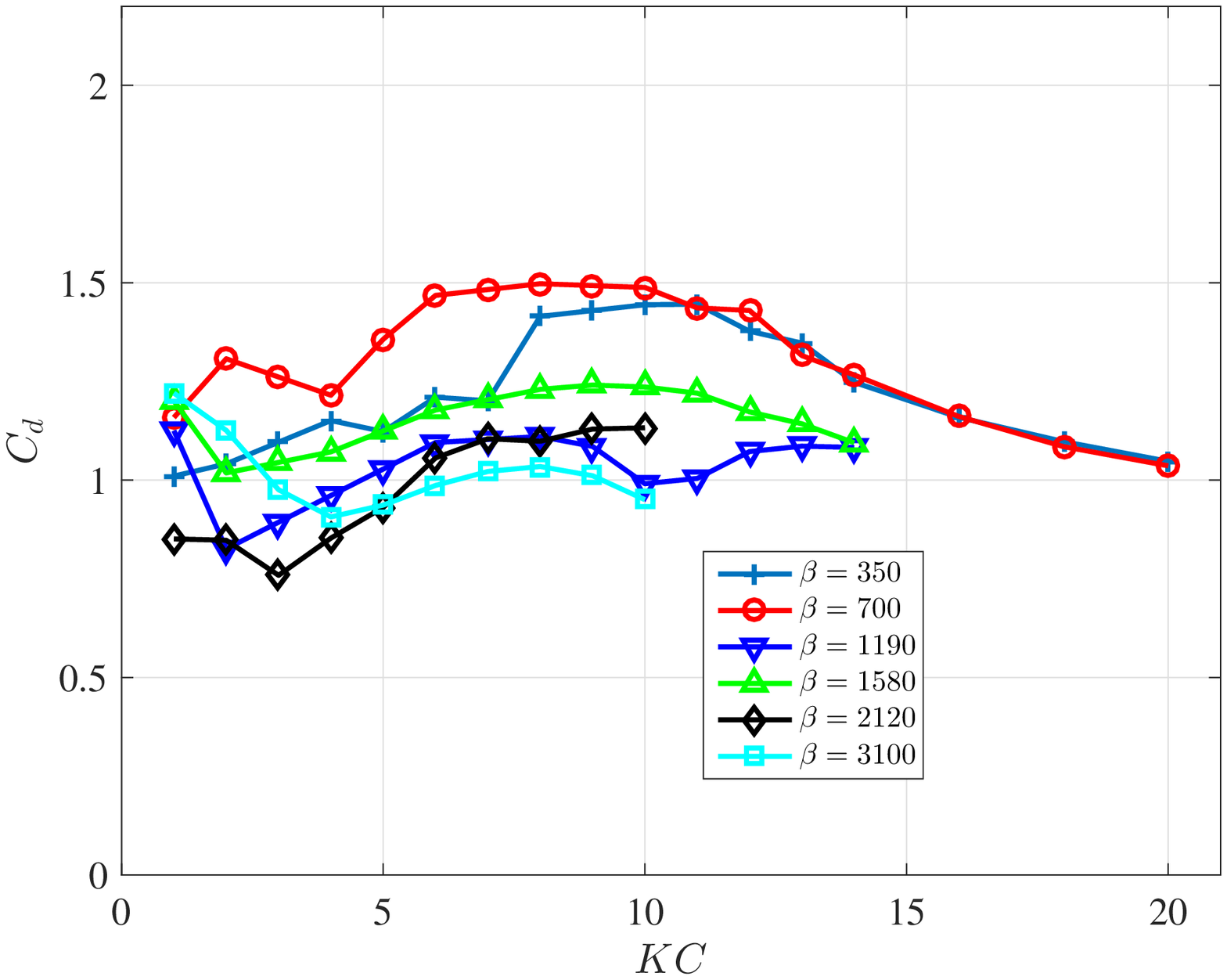}
\end{subfigure}
\begin{subfigure}{0.4\columnwidth}
\includegraphics[width=\columnwidth,keepaspectratio]{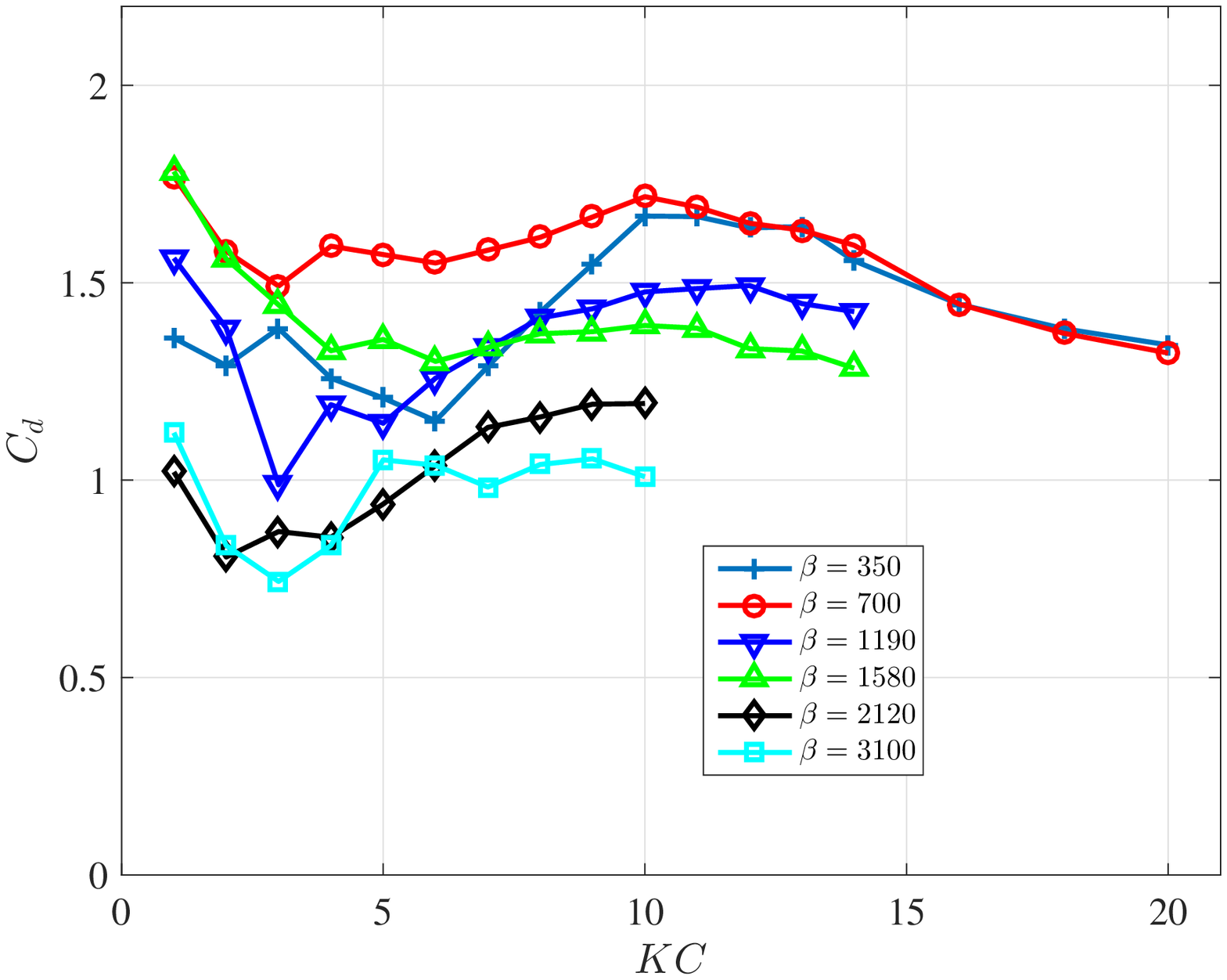}
\end{subfigure}
\begin{subfigure}{0.4\columnwidth}
\includegraphics[width=\columnwidth,keepaspectratio]{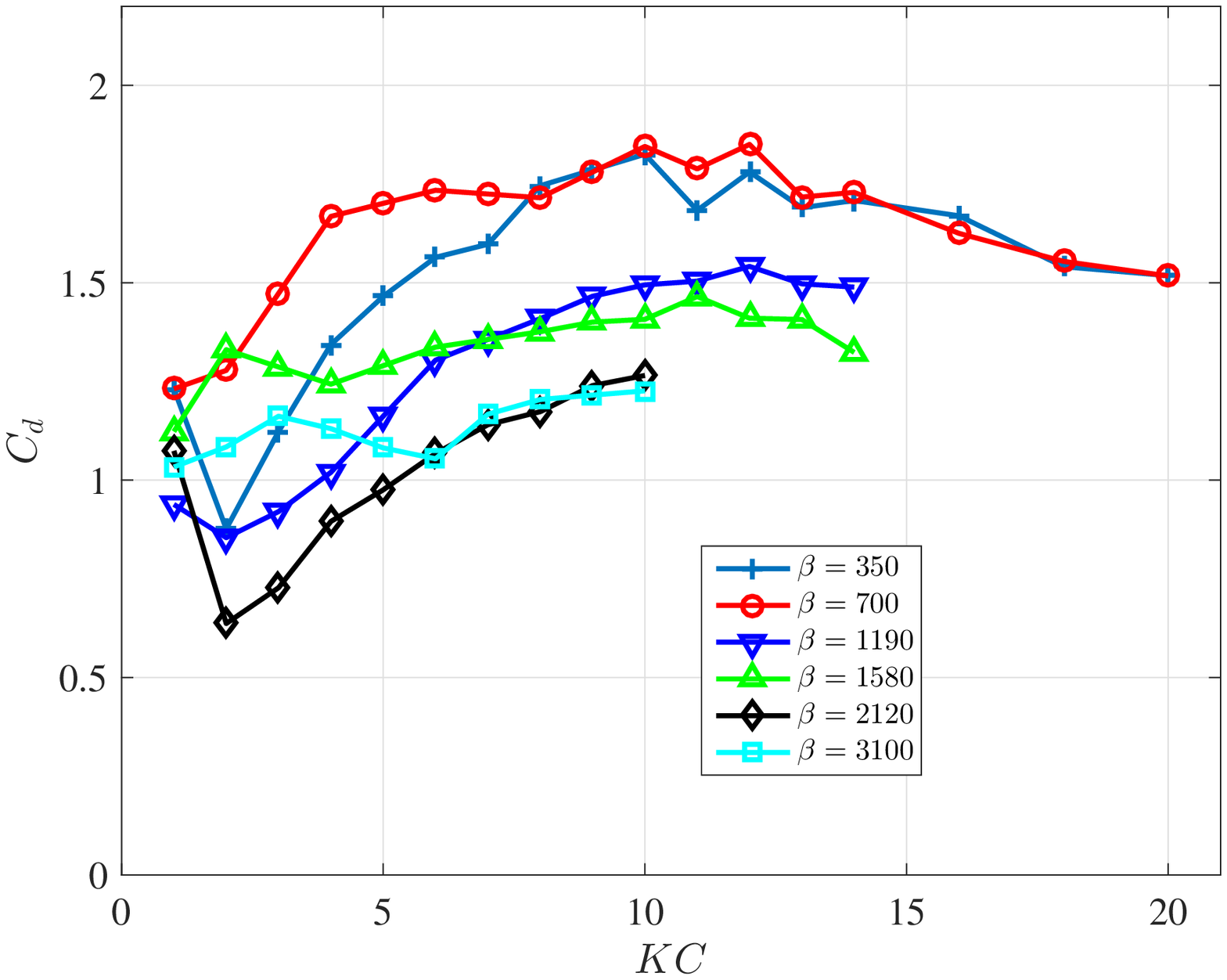}
\end{subfigure}
\caption{Cylinders in tandem configuration: trend of $C_d$ along with $KC$ for different $\beta$ at $G/d = 0.5$ (top left), $G/d = 1.0$ (top right), $G/d = 2.0$ (bottom left) and $G/d = 3.0$ (bottom right)}
\label{fig:CdTandem}
\end{figure}

The added mass coefficient $C_m$ for tandem dual cylinder is first plotted in Fig. \ref{fig:CmTandem} for 4 gap ratios $G/d$ of 0.5, 1.0, 2.0 and 3.0. The result shows that the overall $C_m$ of tandem cylinders will be smaller than that of the cylinders in a side-by-side configuration. Furthermore, it is observed that when the gap distance is smaller and naturally the interaction between two cylinders will be stronger, $C_m$ does not change too much with $KC$, and hence is more like a constant value. Meanwhile, when the gap ratio $G/d$ gets larger, $C_m$ is again dependent on $KC$. In Fig. 15, it plots $C_m$ at different $G/d$ (including single cylinder case of $G/d=\infty$) for $\beta = 350$ along with $KC$. It shows the gap effect on $C_m$ for a tandem oscillating cylinder that when $G/d$ gets larger, the difference of $C_m$ between the single cylinder and tandem cylinders becomes smaller. In addition, $C_m$ first increases, then decreases and finally rises again with an increasing $KC$ number.

\begin{figure}[!ht]
\centering
\includegraphics[width=0.4\columnwidth,keepaspectratio]{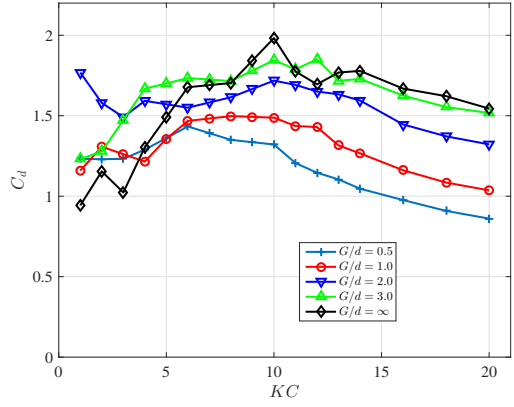}
\caption{Cylinders in tandem configuration: trend of $C_d$ along with $KC$ for $\beta = 700$ at different $G/d$}
\label{fig:CdTandemGap}
\end{figure}

Fig. \ref{fig:CdTandem} plots the drag coefficient $C_d$ for 4 gap ratios $G/d$ of 0.5, 1.0, 2.0 and 3.0. The result shows that $C_d$ will first increase then decrease with an increasing $KC$ for all gap ratios, and however, the $KC$ corresponding to the maximum $C_d$ varies with different gap ratios, that is when the gap is smaller, the maximum $C_d$ $KC$ becomes smaller. To further reveal the gap effect on $C_d$, in Fig. \ref{fig:CdTandemGap}, it plots $C_d$ at different $G/d$ (including single cylinder case of $G/d=\infty$) for $\beta = 700$ along with $KC$. Again we can see, when $G/d$ is larger, $C_d$, of a single cylinder and that of tandem cylinders are alike. At the same time, it is found that, at a larger $KC$, $C_d$ will decrease with a decrease of $G/d$. Compared to the results of the cylinders in a side-by-side configuration, when the gap is smaller, and interaction is stronger, $C_d$ for both tandem and side-by-side configuration will be largely altered compared to that in a single cylinder case. However, the side-by-side configuration will enhance $C_d$, while the tandem configuration will result in a drop in $C_d$.

\begin{figure}[!ht]
\centering
\begin{subfigure}{0.4\columnwidth}
\includegraphics[width=\columnwidth,keepaspectratio]{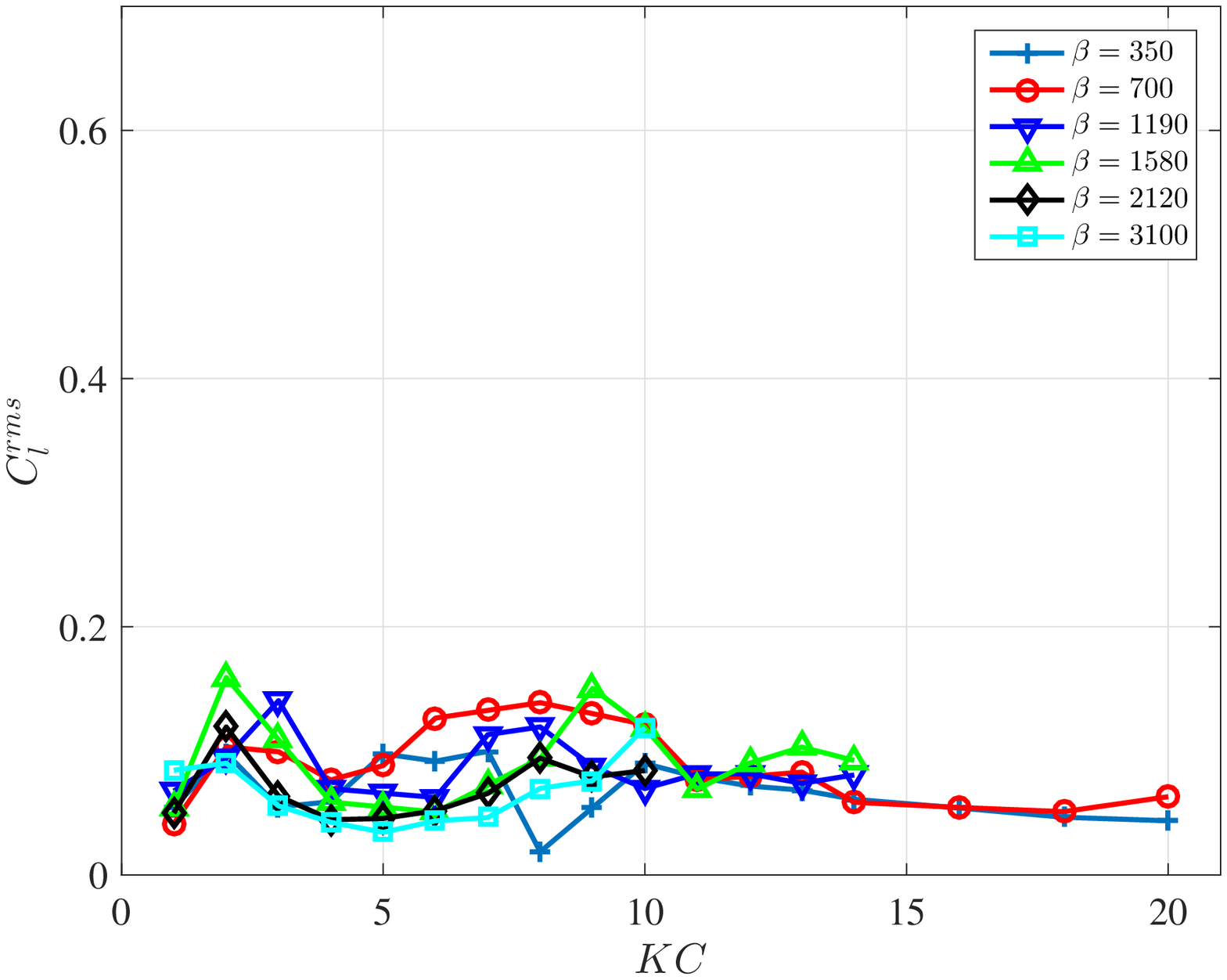}
\end{subfigure}
\begin{subfigure}{0.4\columnwidth}
\includegraphics[width=\columnwidth,keepaspectratio]{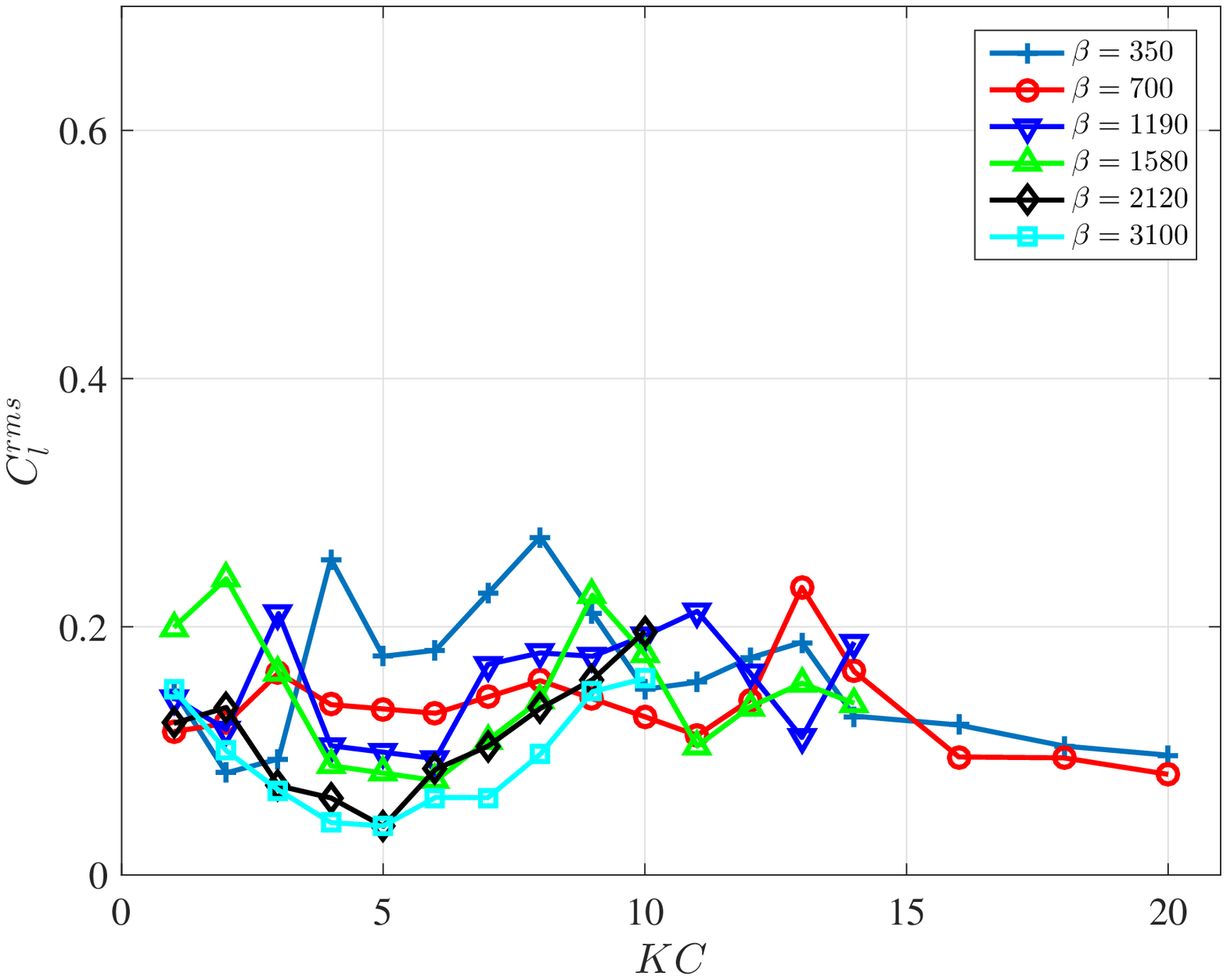}
\end{subfigure}
\begin{subfigure}{0.4\columnwidth}
\includegraphics[width=\columnwidth,keepaspectratio]{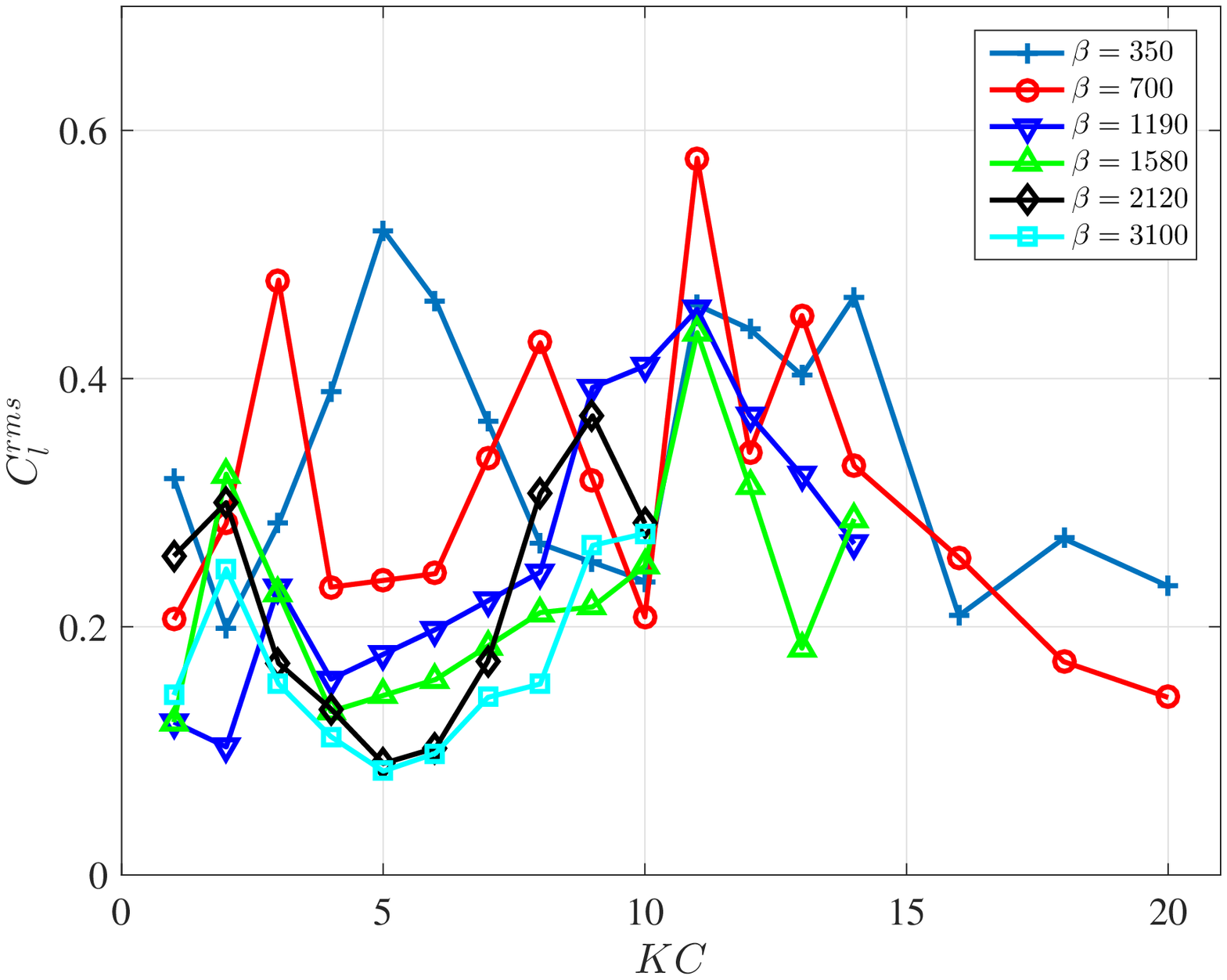}
\end{subfigure}
\begin{subfigure}{0.4\columnwidth}
\includegraphics[width=\columnwidth,keepaspectratio]{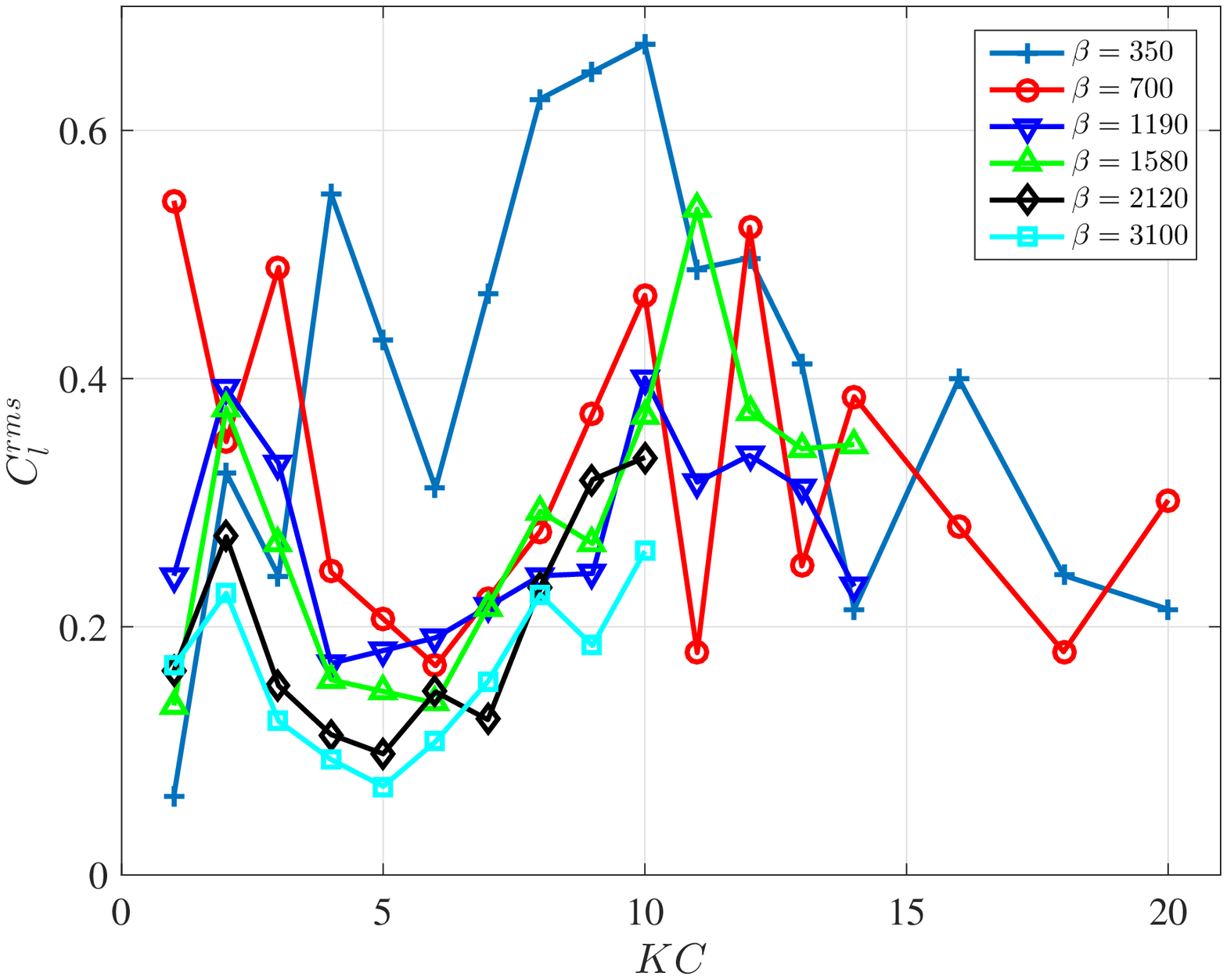}
\end{subfigure}
\caption{Cylinders in tandem configuration: trend of $C^{rms}_l$ along with $KC$ for different $\beta$ at $G/d = 0.5$ (top left), $G/d = 1.0$ (top right), $G/d = 2.0$ (bottom left) and $G/d = 3.0$ (bottom right)}
\label{fig:ClstdTandem}
\end{figure}

\begin{figure}[!ht]
\centering
\begin{subfigure}{0.4\columnwidth}
\includegraphics[width=\columnwidth,keepaspectratio]{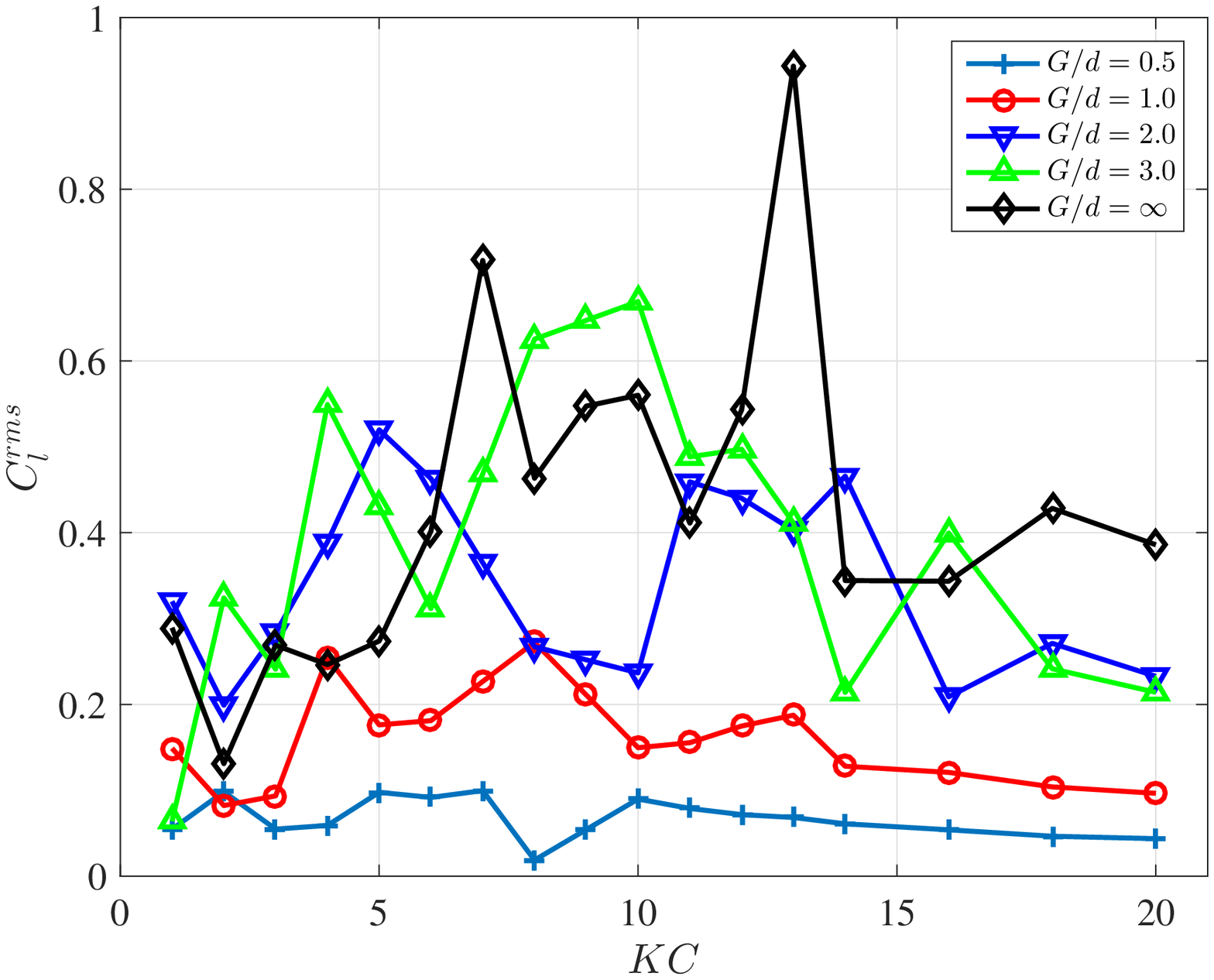}
\end{subfigure}
\begin{subfigure}{0.4\columnwidth}
\includegraphics[width=\columnwidth,keepaspectratio]{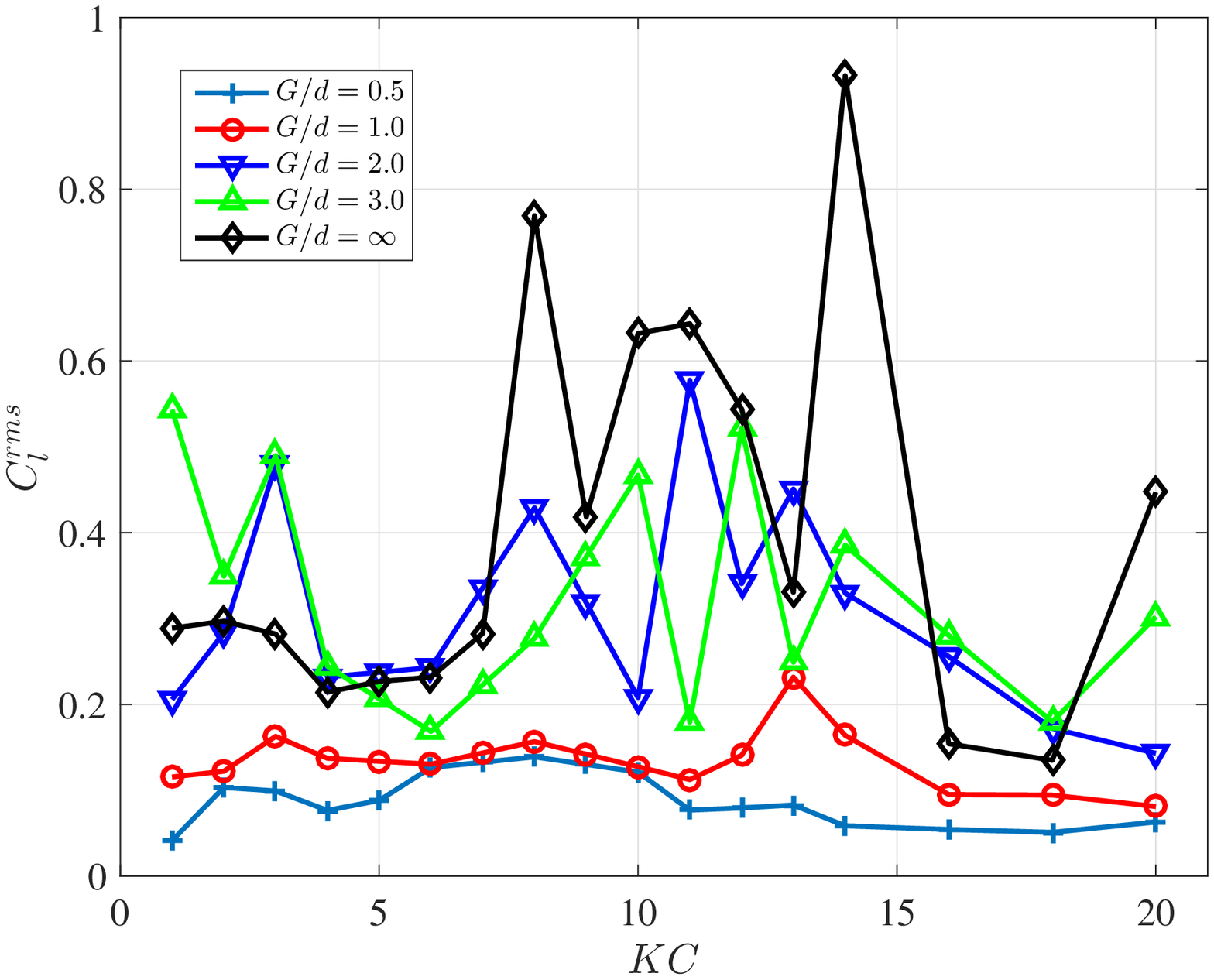}
\end{subfigure}
\caption{Cylinders in tandem configuration: trend of $C^{rms}_l$ along with $KC$ at different $G/d$ for $\beta = 350$ (left) and $\beta = 700$ (right)}
\label{fig:ClstdTandemGap}
\end{figure}

For cylinders in a tandem configuration, in the current experiment, we find that it will have a zero mean lift coefficient $\overline{C_l}$, compared to the side-by-side configuration. Moreover, the lift coefficient RMS $C^{rms}_l$ is observed to have some more interesting phenomena and hence is plotted in Fig. \ref{fig:ClstdTandem} for 4 gap ratios $G/d$ of 0.5, 1.0, 2.0 and 3.0. Though there is no definite trend for $C^{rms}_l$ that can be concluded for $KC$ and $\beta$, it is obvious that $C^{rms}_l$ is smaller when $G/d$ is small. To better reveal the effect of $G/d$, it is plotted in Fig. \ref{fig:ClstdTandemGap} of $C^{rms}_l$ at different $G/d$ (including single cylinder case of $G/d=\infty$) for $\beta=350$ and $\beta=700$. Clearly it is revealed that, compared to the single cylinder as well as the larger $G/d$ cases, $C^{rms}_l$ of $G/d=0.5$ is much smaller over the entire $KC$ range.

\subsection{Numerical Simulation: Flow Visualization}
\label{sec5_2}
\begin{figure}[!ht]
\centering
\begin{subfigure}{0.32\columnwidth}
\includegraphics[width=\columnwidth,keepaspectratio]{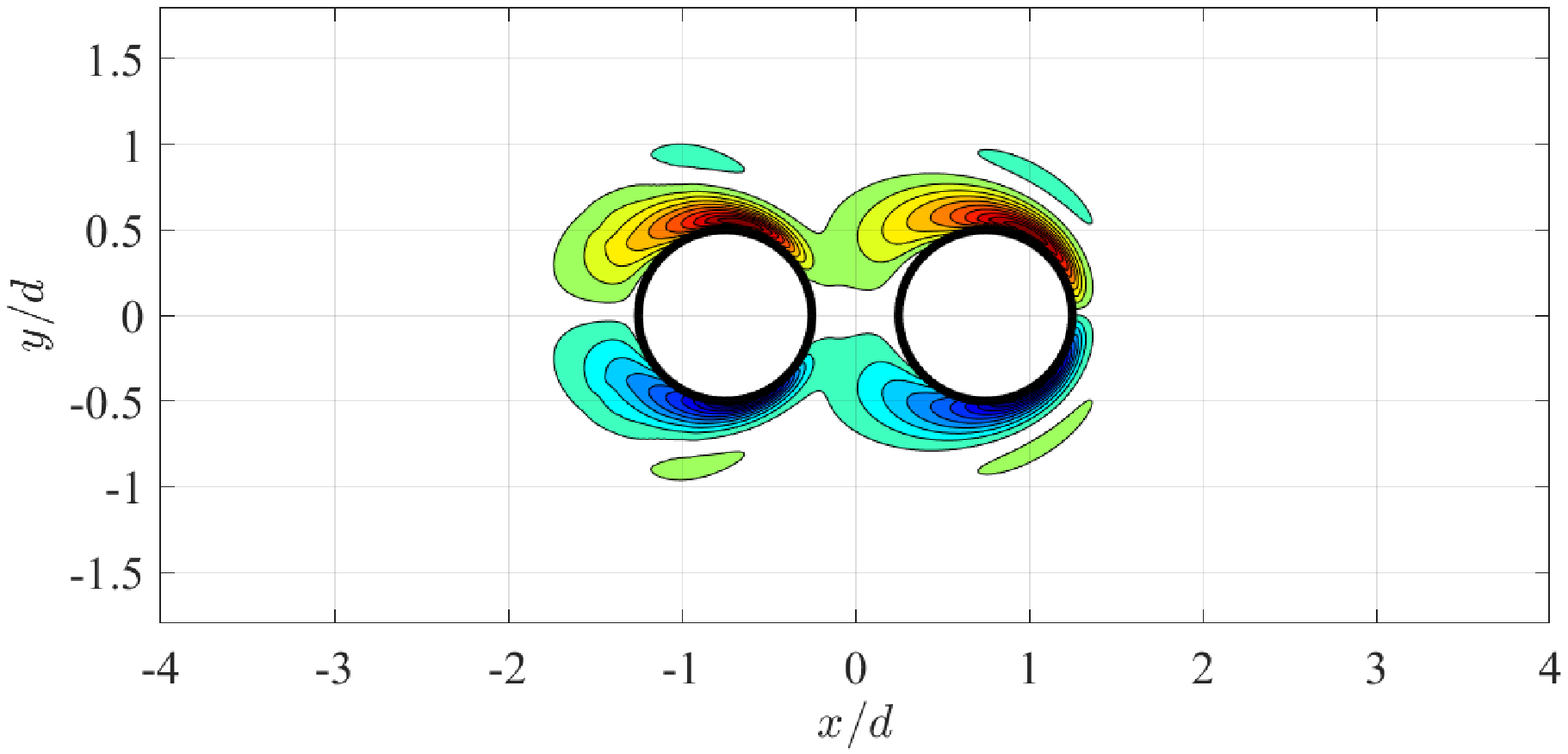}
\end{subfigure}
\begin{subfigure}{0.32\columnwidth}
\includegraphics[width=\columnwidth,keepaspectratio]{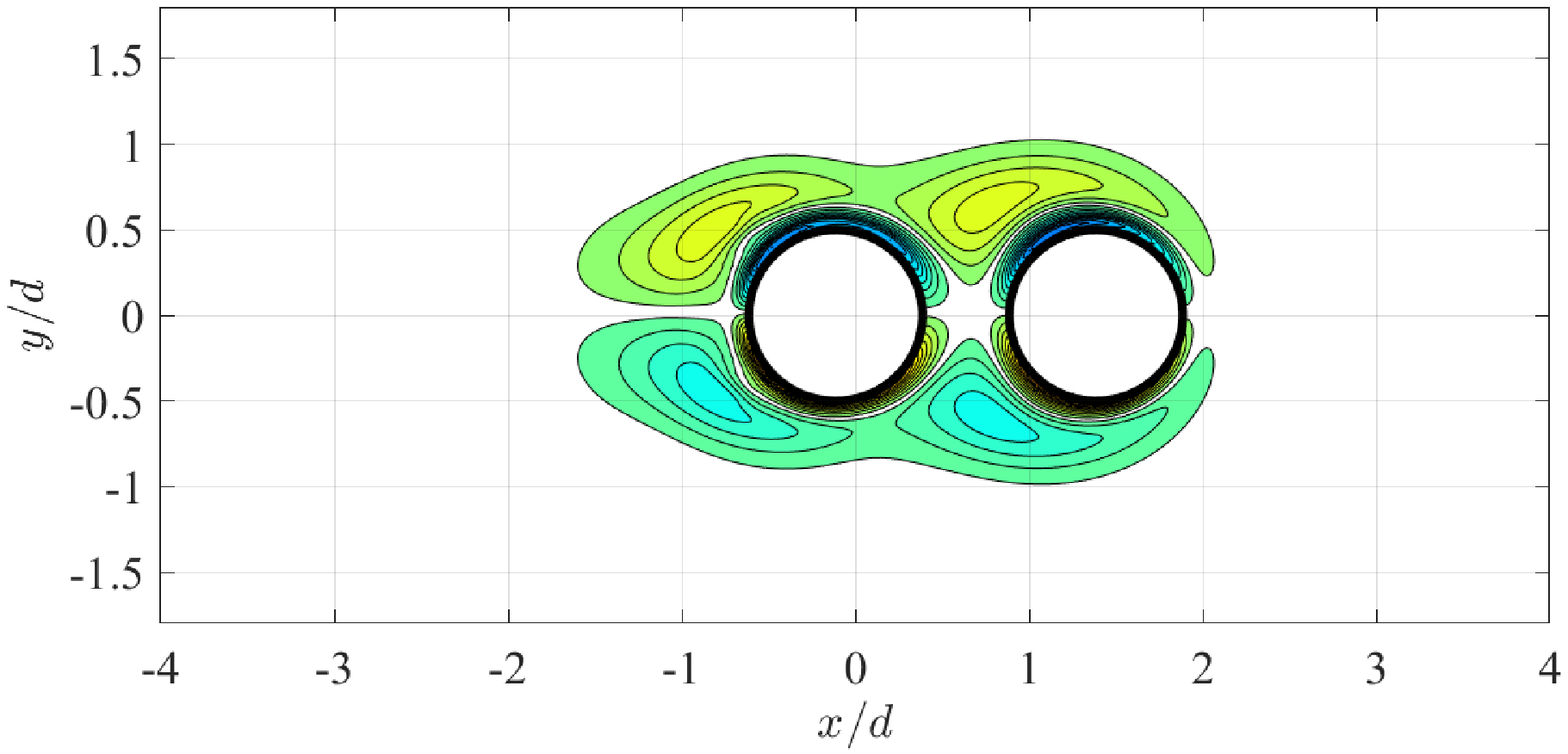}
\end{subfigure}
\begin{subfigure}{0.32\columnwidth}
\includegraphics[width=\columnwidth,keepaspectratio]{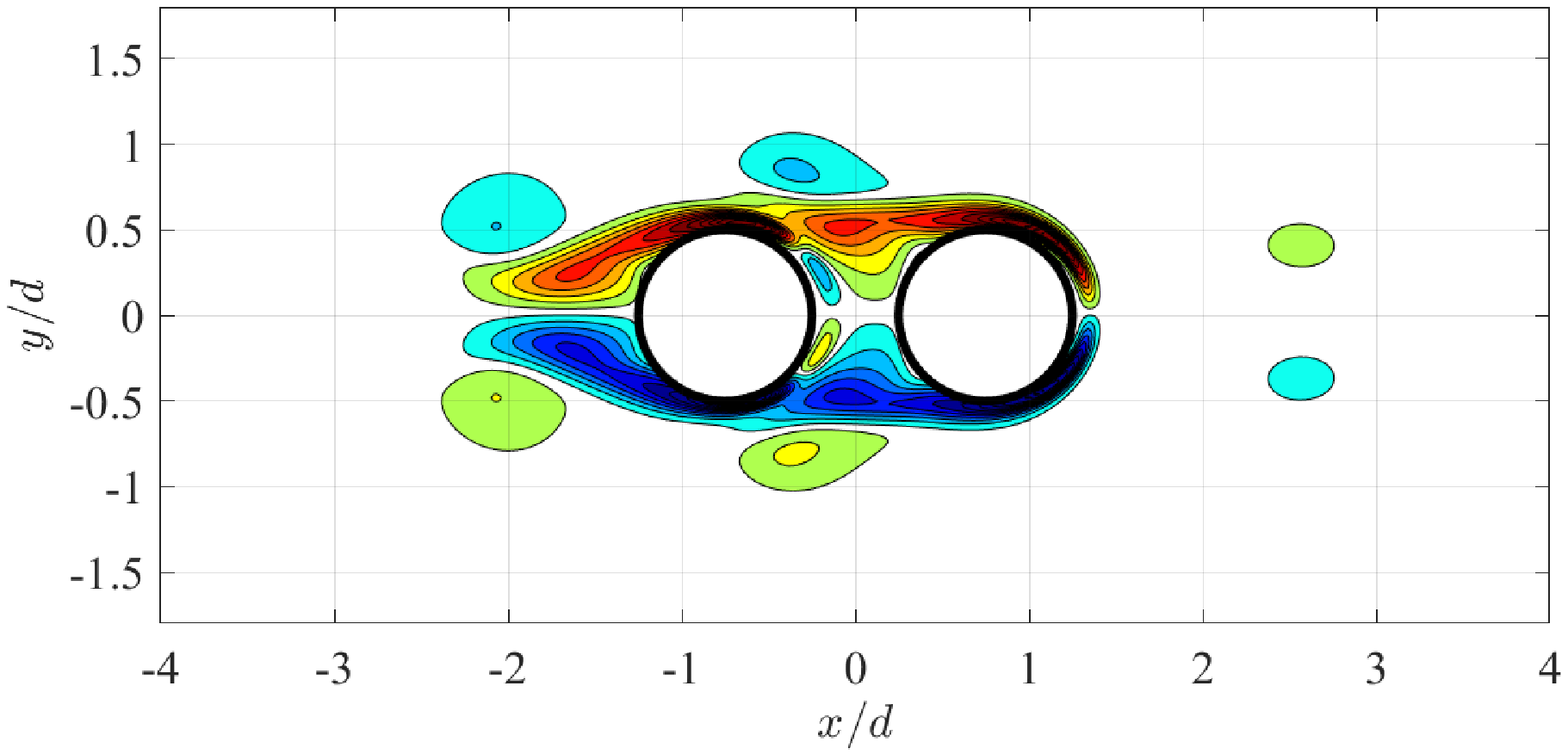}
\end{subfigure}
\begin{subfigure}{0.32\columnwidth}
\includegraphics[width=\columnwidth,keepaspectratio]{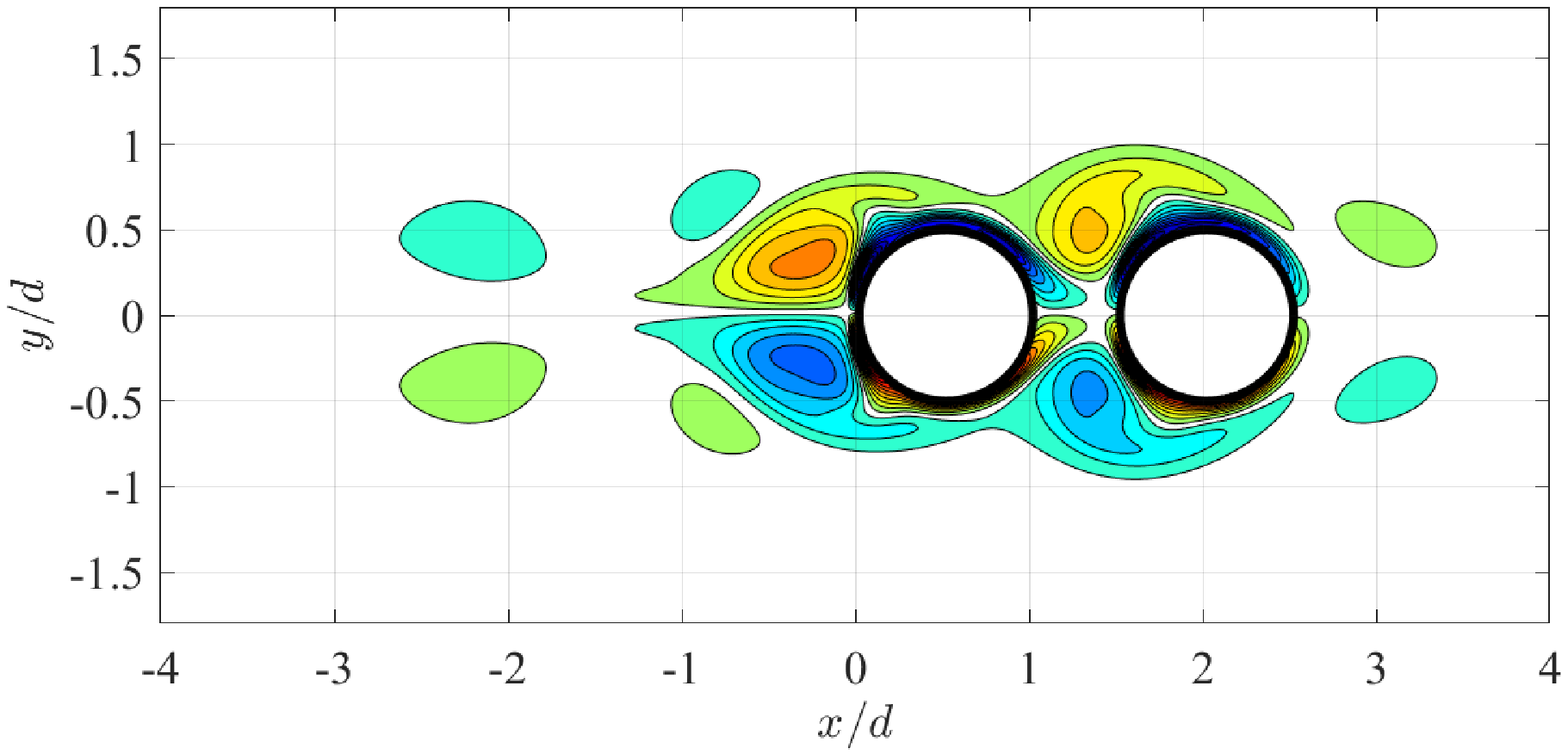}
\end{subfigure}
\begin{subfigure}{0.32\columnwidth}
\includegraphics[width=\columnwidth,keepaspectratio]{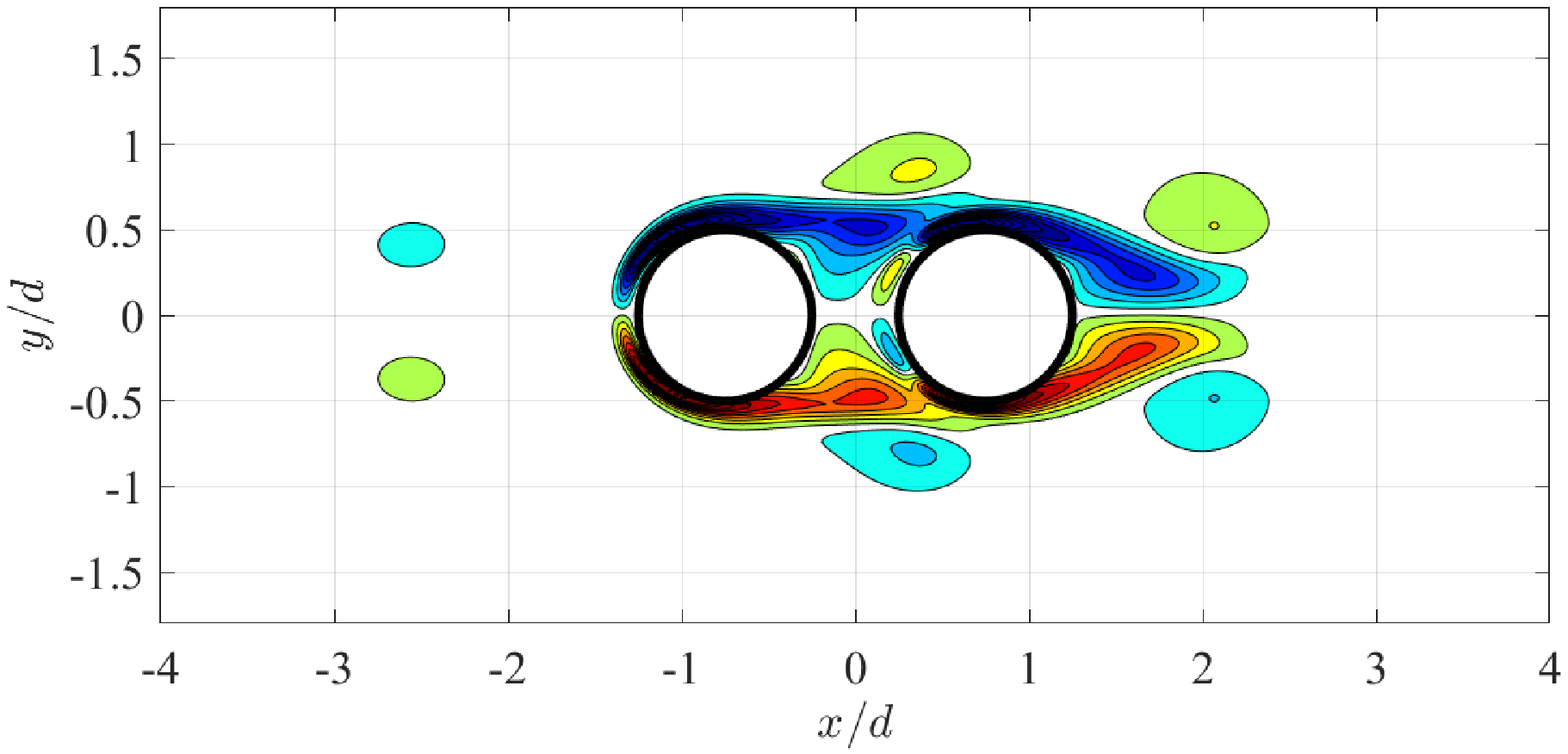}
\end{subfigure}
\begin{subfigure}{0.32\columnwidth}
\includegraphics[width=\columnwidth,keepaspectratio]{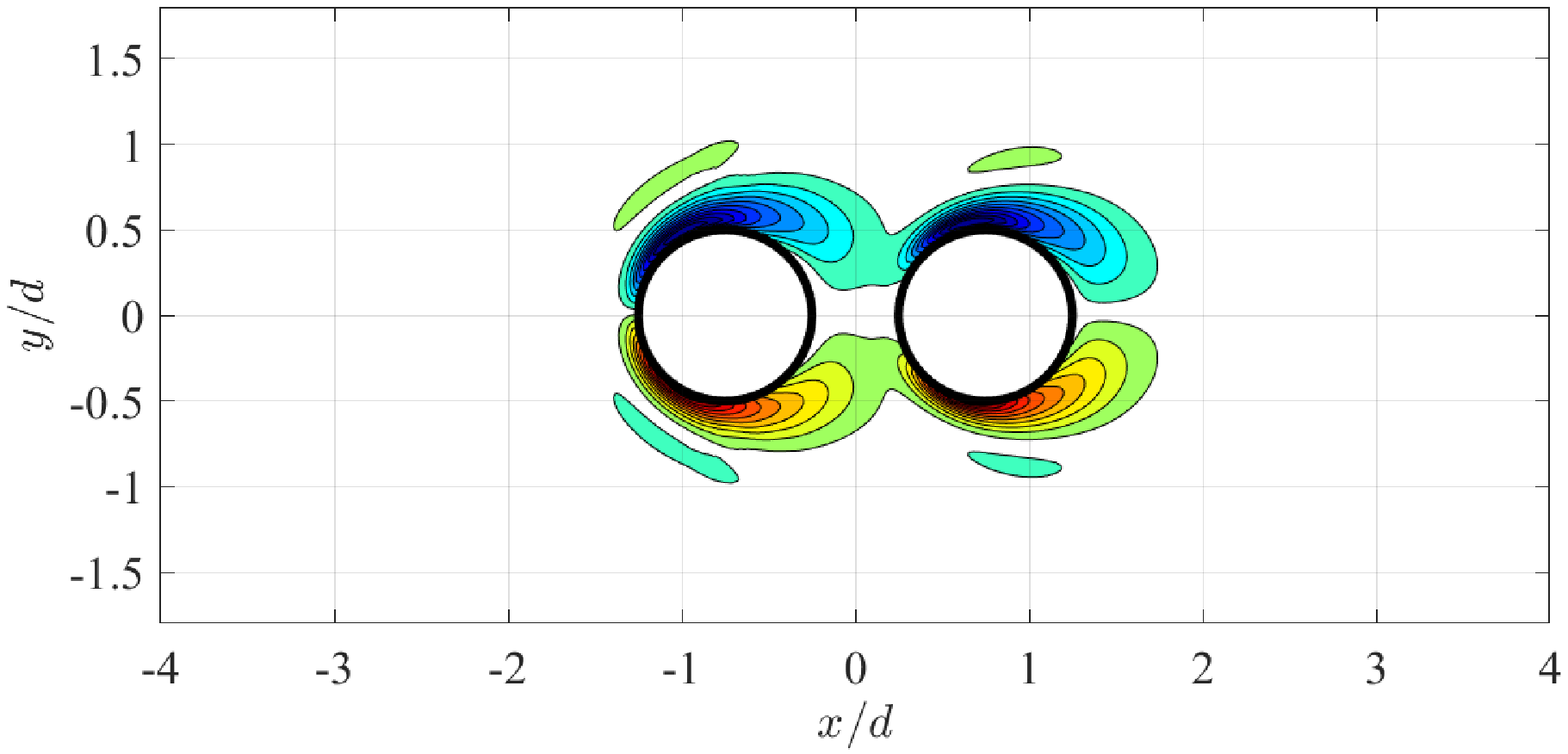}
\end{subfigure}
\begin{subfigure}{0.55\columnwidth}
\includegraphics[width=\columnwidth,keepaspectratio]{Fig/Single/Flow/cbar.eps}
\end{subfigure}
\caption{Snapshot of vorticity plot for tandem cylinders ($\beta = 20$, $G/d = 0.5$) at $KC=4$ (first row, $C_m = 1.18$, $C_d = 1.44$) and at $KC=8$ (second row, $C_m = 1.25$, $C_d = 1.23$) at $t/\tau$ of 0 (left), $\frac{1}{4}$ (middle) and $\frac{1}{2}$ (right).}
\label{fig:vortTandemG05}
\end{figure}

The simulated flow visualization of the cylinder pair in a tandem configuration at $G/d = 0.5$ is plotted in Fig. \ref{fig:vortTandemG05} for $KC = 4$ (first row) and $KC = 8$ (second row). The result shows that at $G/d = 0.5$, the shedding or attached vortices from the cylinder merge with the vortices of the same sign from the other cylinder, which creates a symmetric vortical wake pattern. Such a wake pattern explains the reduction of the drag coefficient of the cylinder pair in a tandem configuration, shown in Fig. \ref{fig:CdTandem}. At a small $G/d$, due to the blocking effect of the other cylinders, strength of the vortex generated is weaker, and therefore the drag coefficient is smaller than that of a single cylinder. In the meantime, due to the symmetric wake as a result of the merge and thus elongation of the vortices on either side of the cylinder pair in a tandem configuration, the RMS of the lift forces coefficient is small.

\section{Conclusion}
\label{sec6}

In this paper, we focus on the hydrodynamic problems of a cylinder pair in the oscillatory flow with a side-by-side and a tandem configuration. The experiment findings reveal that over a wide range of $KC$ and $\beta$, with a decrease in the gap ratio, the interaction between two cylinders gets stronger for both side-by-side and tandem cylinder pairs. The experiment also shows that with the decreasing gap ratio, the drag coefficient and the RMS of the lift coefficient will be enhanced for the cylinder pair in a side-by-side configuration, while those two coefficients will be reduced for the cylinder pair in a tandem configuration. In order to clearly reveal the flow physics of the hydrodynamic force variation due to the interaction between the dual cylinders, flow visualization is provided by numerical simulation at the same $KC$ but smaller $\beta$. The result demonstrates a strong coupling effect between fluid forces and the generated vortices around cylinders. The smaller gap of the cylinder pair in a side-by-side configuration induces a gap flow that results in a stronger vortex shedding and hence increases the mean energy outflow rate and the drag coefficient. On the contrary, the smaller gap of the cylinder pair in a tandem configuration leads to a symmetric flow pattern of the merge and elongation of the vortices of the same signs and hence decreasing the drag coefficient and the RMS of the lift coefficient.

\section*{Acknowledgments}
\label{sec7}
Author would like to acknowledge Dr. Zhicheng Wang's constructive discussion that facilitated this paper




\bibliographystyle{elsarticle-num}

\bibliography{main}

\end{document}